\documentclass[
 reprint,onecolumn,
 amsmath,amssymb,
 aps,
]{revtex4}

\usepackage{epsfig,bm,epsf,graphics}
\usepackage{graphicx}
\usepackage{dcolumn}
\usepackage{bm}

\begin{document}
\preprint{APS/123-QED}

\title{Dynamics of two-group conflicts: a statistical physics model}

\author{H. T. Diep$^a$\footnote{diep@u-cergy.fr}, Miron Kaufman$^b$\footnote{m.kaufman@csuohio.edu, corresponding author}
and Sanda Kaufman$^c$\footnote{s.kaufman@csuohio.edu}}
\affiliation{%
$^a$Laboratoire de Physique Th\'eorique et Mod\'elisation,
Universit\'e de Cergy-Pontoise, CNRS, UMR 8089\\
2, Avenue Adolphe Chauvin, 95302 Cergy-Pontoise Cedex, France.\\
 }%
 \affiliation{%
$^b$Department of Physics, Cleveland State University, Cleveland, OH 44115, USA\\
 }%
\affiliation{%
$^c$Levin College of Urban Affairs, Cleveland State University, Cleveland, OH 44115, USA\\
 }%


\begin{abstract}
We propose a "social physics" model for two-group conflict. We consider two disputing groups. Each individual $i$ in each of the two groups has a preference $s_i$ regarding the way in which the conflict should be resolved. The individual preferences span a range between $+M$ (prone to protracted conflict) and $-M$ (prone to settle the conflict). The noise in this system is quantified by a "social temperature." Individuals interact within their group and with individuals of the other group. A pair of individuals ($i,j$) within a group contributes -$s_i*s_j$ to the energy. The inter-group energy of individual $i$ is taken to be proportional to the product between $s_i$ and the mean value of the preferences from the other group's members. We consider an equivalent-neighbor Renyi - Erdos network where everyone interacts with everyone. We present some examples of conflicts that may be described with this model.

\vspace{0.5cm}
\begin{description}
\item[PACS numbers: 89.65.-s 	Social and economic system, 89.75.-k 	 Complex systems,]
\item[05.90.+m 	Other topics in statistical physics, thermodynamics, and nonlinear dynamical systems]
\end{description}
\end{abstract}

\pacs{Valid PACS appear here}
\maketitle


\section{Introduction}
Social conflicts emerge among groups of individuals. For example, in 2016, the elections in the United States and the Brexit referendum illustrate such conflicts. So do debates around whether a country should sign climate change accords, whether the pipeline from Canada to Texas should be built, or whether various pieces of land should be developed or conserved. Social conflicts have been extensively studied theoretically and practically by scholars in sociology \cite{Coser}, social psychology \cite{Druckman} \cite{Pruitt} as well as negotiation and decision making \cite{Schelling} \cite{Simons}. There is broad consensus in the literature that social conflicts are complex, and therefore their outcomes are difficult to predict. For example, the outcome of the Brexit referendum surprised many of those who had ventured to predict it; 50 days before the US elections, opinion polls fluctuate significantly from one day to the next.  Results of the Paris climate change accord will not be known for decades.  Nevertheless, conflicting parties need the ability to prepare and strategize in order to navigate through uncertainty and be effective in attaining their objectives. Climate change is a rather polarizing social conflict in the US. Nevertheless, decision makers at the local and state levels have to make ongoing decisions and engage in adaptive management while uncertain about how the conflict will be resolved and how it will affect their regions.
To manage social conflicts, we can borrow tools from other settings such as planning and policy making which also require decisions and strategizing about complex, interrelated and unpredictable systems in the absence of sufficient information. In such settings, prediction can be replaced by anticipation. That is, instead of basing decisions on a predicted future, anticipation entails generating and exploring  possible scenarios.  Decision makers can then devise strategies likely to yield desired results across a range of scenarios \cite{Kaufman-Kaufman2013}, \cite{Kaufman-Kaufman2015}. Such strategies are considered robust in the sense that they do not depend on the advent of a specific future. Anticipative scenarios are helpful, and even critical in informing parties to complex social conflicts. In the Brexit example, each of the sides could have improved their respective strategies by considering a range of scenarios; similarly, in the 2016 elections, each of the two major parties can devise response strategies based on anticipated moves of the opponent. In conflicts surrounding climate change, parties can develop strategies for the possibility that all countries abide by their commitments, as well as responses to the cases where key actors default or fulfill commitments only partially and even contemplate the situations where nature responds differently than expected to climate actions or lack thereof. Note that such uses of anticipative scenarios can be one-sided, when one or more disputants use them to construct their strategies; or they can even be used by interveners as persuasion tools to manage the conflict, by getting parties to make decisions that use the same information base (such an approach is called data mediation).  Such an approach is more appropriate to conflicts around climate change, for example, where consensus agreements can emerge, rather than in the elections example. The Brexit case is interesting in that it was handled by a referendum as a binary choice (as in elections) but it could also have been resolved by addressing the disputants' concerns with various anticipated scenarios of consequences of staying, or of leaving the EU. In these and other examples

Physics can contribute conceptually and through modeling \cite{Galam}, \cite{Diep2014} to the task of analyzing, modeling and generating possible future states of complex systems such as those involved in social conflicts. For example, the author of Ref.\onlinecite{Galam2016} has explored the effects of various voting rules on the 2016 US election outcomes. At the conceptual level, the notion of using toy models to explore states of systems closely parallels the scenario-generating activities involved in anticipating social futures. At the modeling level, the family of multiplex networks can be used to represent parsimoniously various patterns of interconnections between individuals within and between groups. In what follows, we consider the interactions of two groups experiencing conflict around decisions with respect to some specific set of issues. We use the 2016 elections in the United States to illustrate how anticipatory scenarios can be used to understand a social conflict and devise resolution strategies.

We consider two groups in conflict. In each group, each individual has a preference or attitude regarding whether or not to engage in negotiations to resolve their conflict. Preferences in Group $A$ range from $-M_A$ to $M_A$. The number of preferences, or the number of "states", is $q_A$.  $M_A$ reflects a preference for protracted conflict, stemming from extreme adherence to the group's ideology, and desire to defend it by any means. This type of attitude leads to being against any concessions. Its polar opposite, -$M_A$, is equivalent to being prone to any compromise in negotiations. The midpoint of this range is 0 and represents adherence to the values of one's group combined with willingness to find a way out of the conflict with the opposing group. Individual preferences in Group $B$ range similarly from -$M_B$ to $M_B$.
For example in the conflict between the Democratic and Republican parties, members' views range from very strong adherence to progressivism or conservatism and brooking no compromise ($M$), to centrist ones (0) adhering to party values but open to negotiations with members of the opposing party, to quasi-independence and even ability to switch allegiance (-$M$). Think of "Reagan democrats" and "RINOs" as holding the -$M$ values. In gun control debates for instance, some Republican politicians (at or close to $M$) claim any control measure contravenes to the Second Amendment and is therefore nonnegotiable; some (in the vicinity of 0) would negotiate with their Democrat counterparts for limited measures they consider consistent with the Second Amendment; and some (at the -$M$ end) would be willing to accept any measure that might reduce gun violence. Some Democrat politicians ($M$) do not countenance any military intervention in Syria, based on lessons learned in the Iraq war; others (0) would negotiate for very limited intervention as in Libya, and some (-$M$) are willing to side with their Republican counterparts in a vigorous intervention plan.  The variability or noise in individual preferences is quantified by a social temperature. Low temperature situations are more settled, while high temperature situations are in flux.  Gun control is an example of a low social temperature conflict as the competing camps are unlikely to be changed by external events. The current US elections, where things are in flux and the camps are shifting is an example of a high social temperature conflict.

Members within each group interact with each other as well as with members of the opponent group. We can conceive of each group as a network of members; the networks can interact with each other forming a multiplex \cite{Gao}.
A multiplex model of the translational and rotational degrees of freedom was used \cite{GalamGabay}to describe plastic crystals.  Recently a similar model \cite{Alvarez} was used to describe social processes.  Each individual acts with a certain intensity to persuade others in the group to his/her point of view, and is in turn subject to others' persuasion efforts. To begin, we assume that each individual interacts with every other individual inside and between the groups. This corresponds to the Renyi-Erdös equivalent neighbor network, a network with links of equal strength between all nodes \cite{Kaufman1989},\cite{Cohen},\cite{Fernandez}.  We will also analyse the same model using Monte Carlo simulations \cite{Binder}.

The paper is organized as follows. In section \ref{Model} we describe our model using the mean-field method borrowed from  statistical physics.  Dynamical equations are established and the behavior of the groups in interaction is shown in section \ref{MF}. Discussion on the meaning of the results in real conflicts is given in section \ref{MFD}. Monte Carlo simulations are shown in section \ref{MC} in various situations. Section \ref{MCMF} is devoted to a comparison between mean-field and MC results with various parameter assumptions. Our concluding remarks are presented in section \ref{Concl}.

\section{Mean-Field Model}\label{Model}

Our model of individual interactions within each group and between the groups yields group preference averages s at any time $t$. For each of two networks (groups) their average values are $S_A$ and $S_B$ respectively. The in-group intensity of advocacy (negative energy) of an individual from Group $A$ is $J_1*s*S_A$, while the corresponding intensity of advocacy of a Group $B$ individual is:  $J_2*s*S_B$, where $S_A$ is the average of all individual preferences in Group $A$ and $S_B$ is the average of all individual preferences in Group $B$. When an individual interacts with members of the opposing group, his/her inter-group intensity of interaction (negative energy) is taken to be proportional to the product between that individual's preference $s$ and the mean value of the preferences of the other group's members: $K_{12}s*S_B$ for an individual in Group $A$, and $K_{21}s*S_A$ for an individual in Group $B$.  In this system, the variability (noise) in individual preferences $s$ in a group is quantified by a "social temperature."  We capture this noise using the Boltzmann probability distribution.  We generate a dynamic model of the evolution of preferences by assuming that the intensity of interaction involves the product of preference at current time to preference at an earlier time, i.e. introduce a lag time.

On the Renyi-Erdos (equivalent neighbor) network, the mean of preferences $s$ of each group is proportional to the exponential of the intensity of interactions (negative energy):
\begin{eqnarray}
S_A(t+1)&=&\frac{\sum_{s=-M_A}^{M_A}se^{s[J_1S_A(t)+K_{12}S_B(t)]}}
{\sum_{s=-M_A}^{M_A}e^{s[J_1S_A(t)+K_{12}S_B(t)]}}\label{eq1}\\
S_B(t+1)&=&\frac{\sum_{s=-M_B}^{M_B}se^{s[J_2S_B(t)+K_{21}S_A(t)]}}
{\sum_{s=-M_B}^{M_B}e^{s[J_2S_B(t)+K_{21}S_A(t)]}}\label{eq2}
\end{eqnarray}
We introduce a lag time as we assume the preference $s$ at time $t + 1$ interacts with the averages $S_A$ and $S_B$ at an earlier time $t$.  The time is measured in units of the delay time.
The sums on the right hand sites of Eqs. (\ref{eq1})-(\ref{eq2}) involve the Brillouin function \cite{Diep2015}:
\begin{equation}
B(x,y,J,K,M)=(M+\frac{1}{2 })\mbox{cotanh}[(M+\frac{1}{2 })(Jx+Ky)]-\frac{1}{2 }\mbox{cotanh}[\frac{1}{2 }(Jx+Ky)]\label{eq3}
\end{equation}
Equations (\ref{eq1})-(\ref{eq2}) can be written as:
\begin{eqnarray}
S_A(t+1)&=&B(S_A(t),S_B(t),J_1,K_{12},M_A)\label{eq3a}\\
S_B(t+1)&=&B(S_B(t),S_A(t),J_2,K_{21},M_B) \label{eq3b}
\end{eqnarray}
An analysis of the linearized equations (\ref{eq3a})-(\ref{eq3b}), valid for small $S_A$ and $S_B$, gives the region of the parameter space where an ordered phase $(|S_A| > 0$, $|S_B| > 0)$ can exist beside the disordered phase ($S_A = S_B = 0)$:

\begin{equation}\label{eq4}
\frac{J_1-J_{1c}}{K_{12}}\  \frac{J_2-J_{2c}}{K_{21}}=1						
\end{equation}
where $J_{1c}$ and $J_{2c}$ are the critical values of the couplings when the two networks are decoupled: if the  states of an individual are $-M_A,-M_A+1,-M_A+2,...,M_A-1,M_A$ for $A$ and the similar for $B$ then
$J_{1c}=3/M_A(M_A+1)$  ; $J_{2c}=3/M_B(M_B+1)$.	
The eigenvalues of the Jacobian   $\partial [J_1(t+1),J_2(t+1)]/\partial [J_1(t),J_2(t)]$  are:
\begin{equation}
\lambda_{1,2}=\frac{1}{2 }\large[\frac{J_1}{J_{1c}}+\frac{J_2}{J_{2c}}\pm\sqrt{(\frac{J_1}{J_{1c}}-\frac{J_2}{J_{2c}})^2+4 \frac{K_{12}}{J_{1c}}\ \frac{  K_{21}}{J_{2c}}}\large]		
\end{equation}
On the manifold of equation (\ref{eq4}) both eigenvalues are equal to 1.  The eigenvalues can become complex provided $K_{12}*K_{21} < 0$.  In such cases the average $S_A$ and $S_B$ exhibit oscillations as function of time.  The period of such oscillations is given by

\begin{equation}
\mbox {Period}=\frac{2\pi}{\arctan(\lambda_I/\lambda_R)}
\end{equation}
where $\lambda_R$ and $\lambda_I$ are the real and imaginary parts of the eigenvalue.  When the eigenvalue absolute value is less than unity the average $S$ evolves in time to a non-zero value while when the eigenvalue absolute value is less than unity the average $S$ evolves in time to zero.

\section{Mean-Field Results}\label{MF}

To get acquainted with the model that involves four couplings and two numbers of states, we consider next only $M_A = M_B = 3$, and present graphs of the time series of the averages $S_A$ and $S_B$.

\begin{figure}[ht!]
\centering
\includegraphics[width=8cm,angle=0]{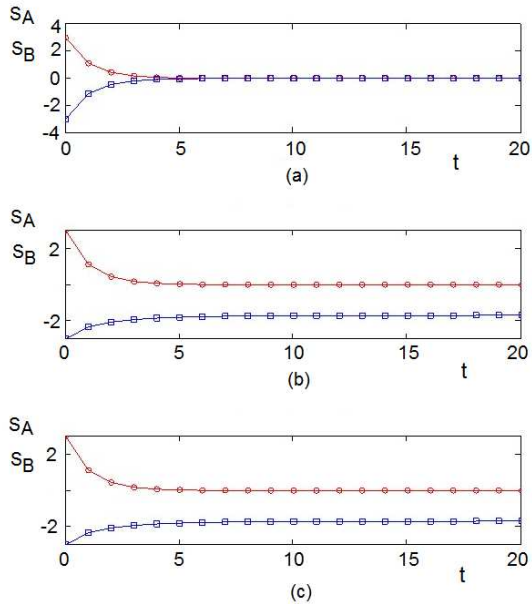}
\caption{Decoupled networks $K_{12}=K_{21}=0$. (a) intra-group couplings weak  with $J_1=J_2=0.1$, $S_A$ and $S_B$ converge to zero; (b) strong and weak intra-coupling  with $J_1=0.1$ and $J_2=0.3$, $S_A$ converges to non-zero value and $S_B$ to zero; (c) strong intra-couplings $J_1=J_2=0.3$, $S_A$ and $S_B$ converge to non-zero values.\label{ffig1}}
\end{figure}

In Fig. \ref{ffig1} we show decoupled networks with:  weak intra-couplings (below the critical value 0.25); one strong intra-group coupling (above the critical value 0.25) and one weak intra-group; and strong intra-group couplings.

\begin{figure}[ht!]
\centering
\includegraphics[width=8cm,angle=0]{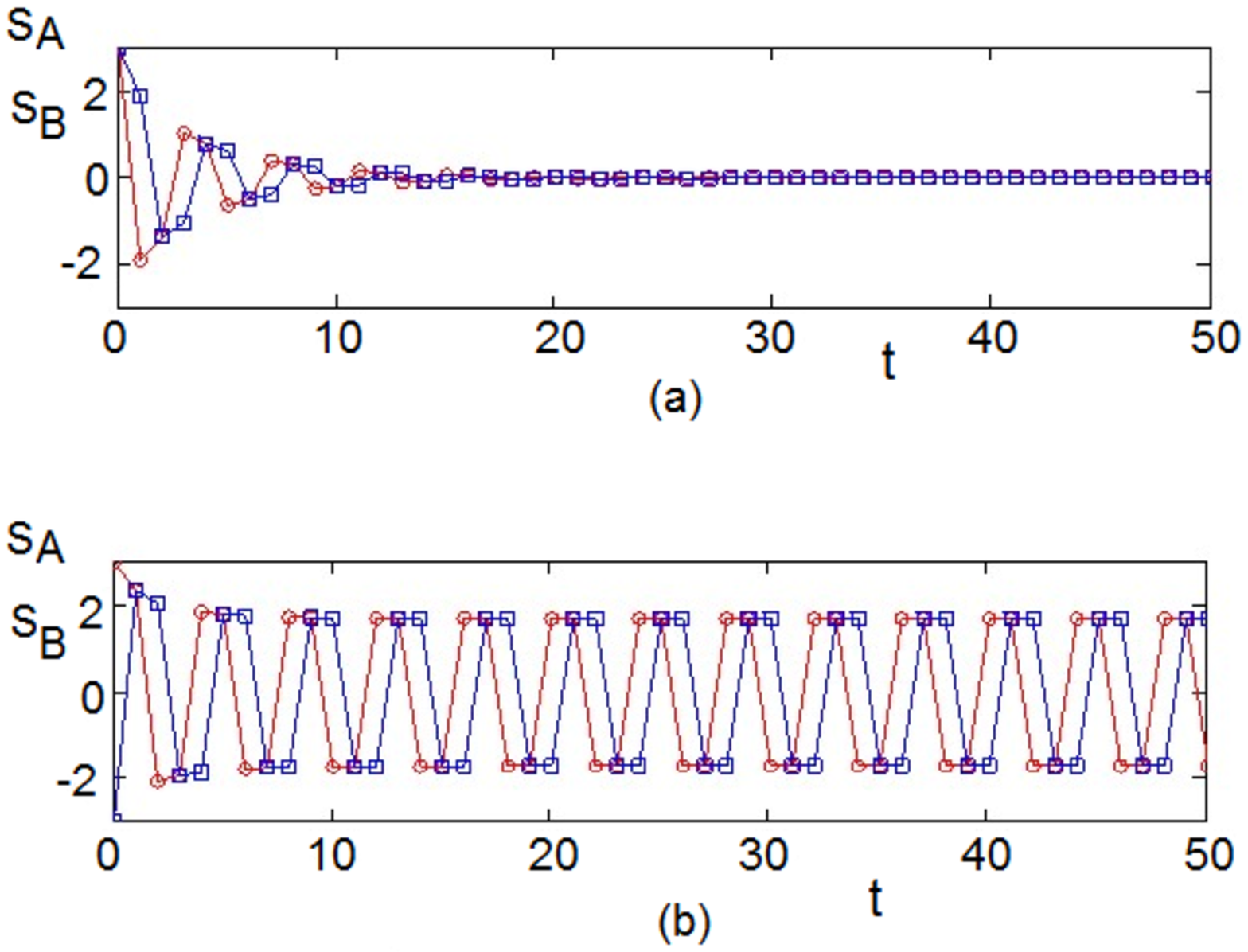}\\
\includegraphics[width=8cm,angle=0]{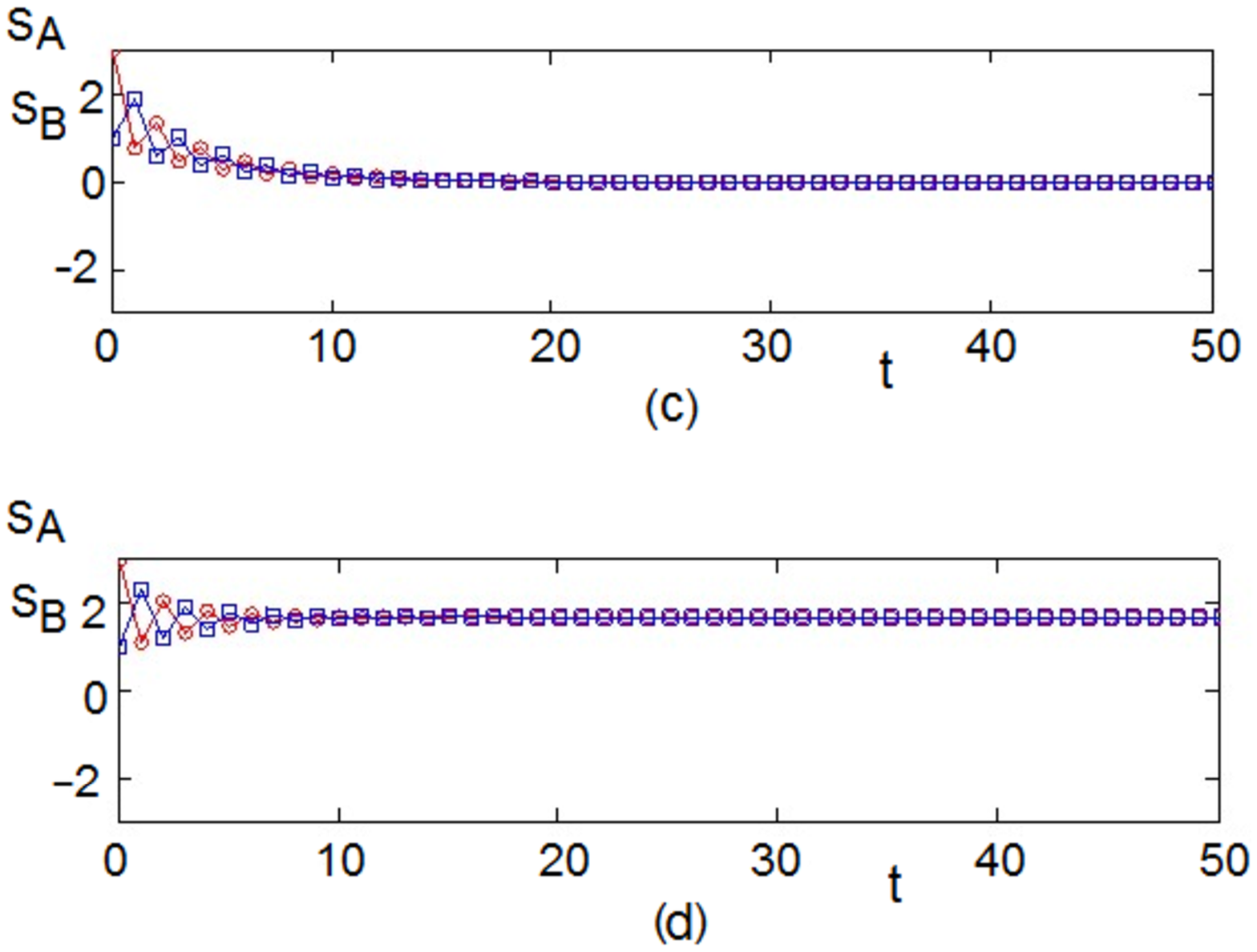}\\
\includegraphics[width=8cm,angle=0]{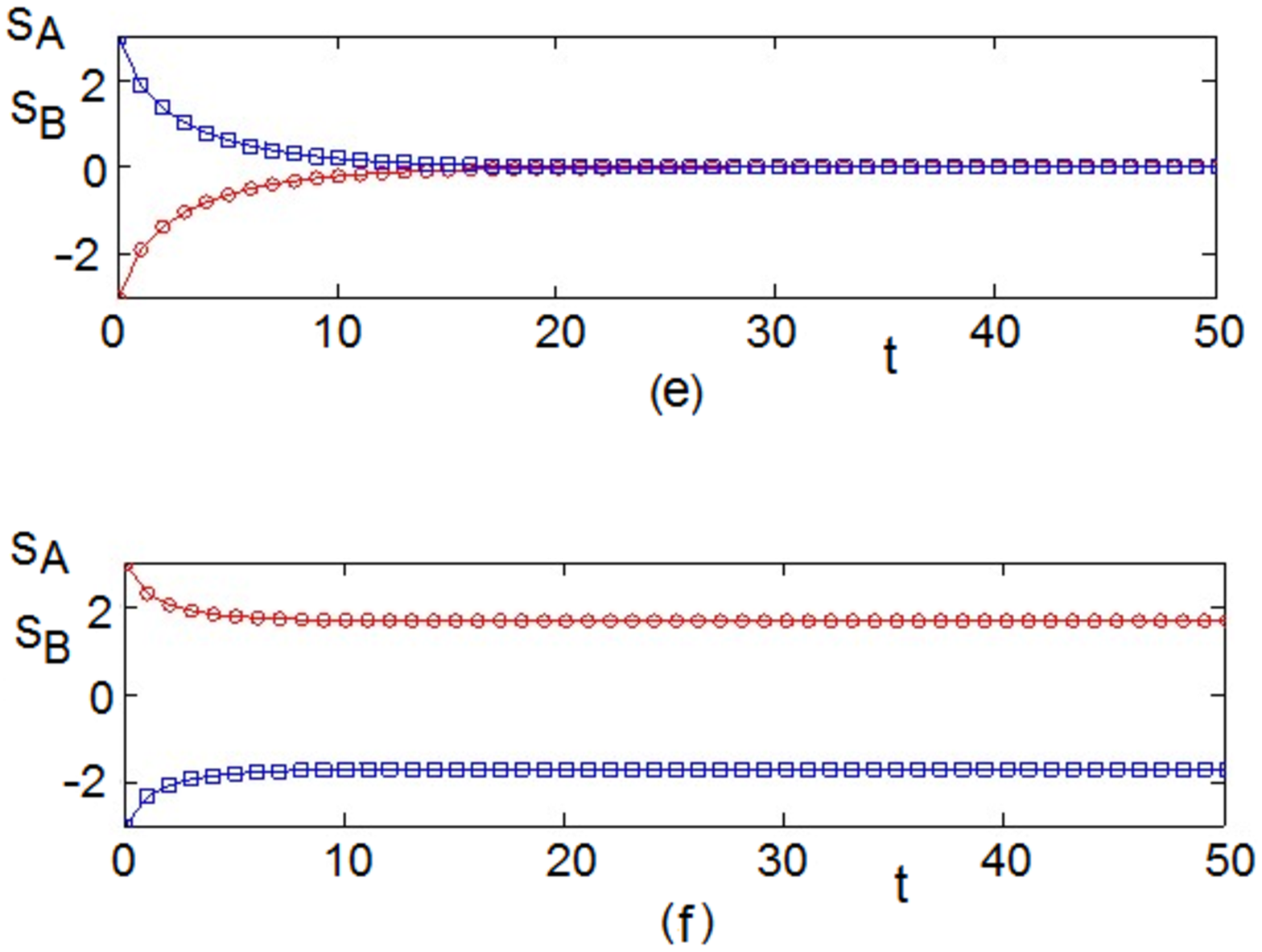}
\caption{zero intra-group interactions  $J_1=J_2=0$, (a) $-K_{12} = K_{21} = 0.2$ damped oscillations; (b) $-K_{12} = K_{21}= 0.3 $ oscillations; (c) $K_{12} = K_{21}= 0.2$ decay to zero; (d) $K_{12} = K_{21}= 0.3$ convergence to non-zero value; (e) $K_{12} = K_{21}= -0.2$ decay to zero; (f) $K_{12} = K_{21}= -0.3$ convergence to non-zero value.\label{ffig2}}
\end{figure}

In the opposite regime, for zero intra-group couplings, we get oscillatory behavior when $K_{12}$ and $K_{21}$ have opposite signs as shown in Fig. \ref{ffig2}.

\begin{figure}[ht!]
\centering
\includegraphics[width=8cm,angle=0]{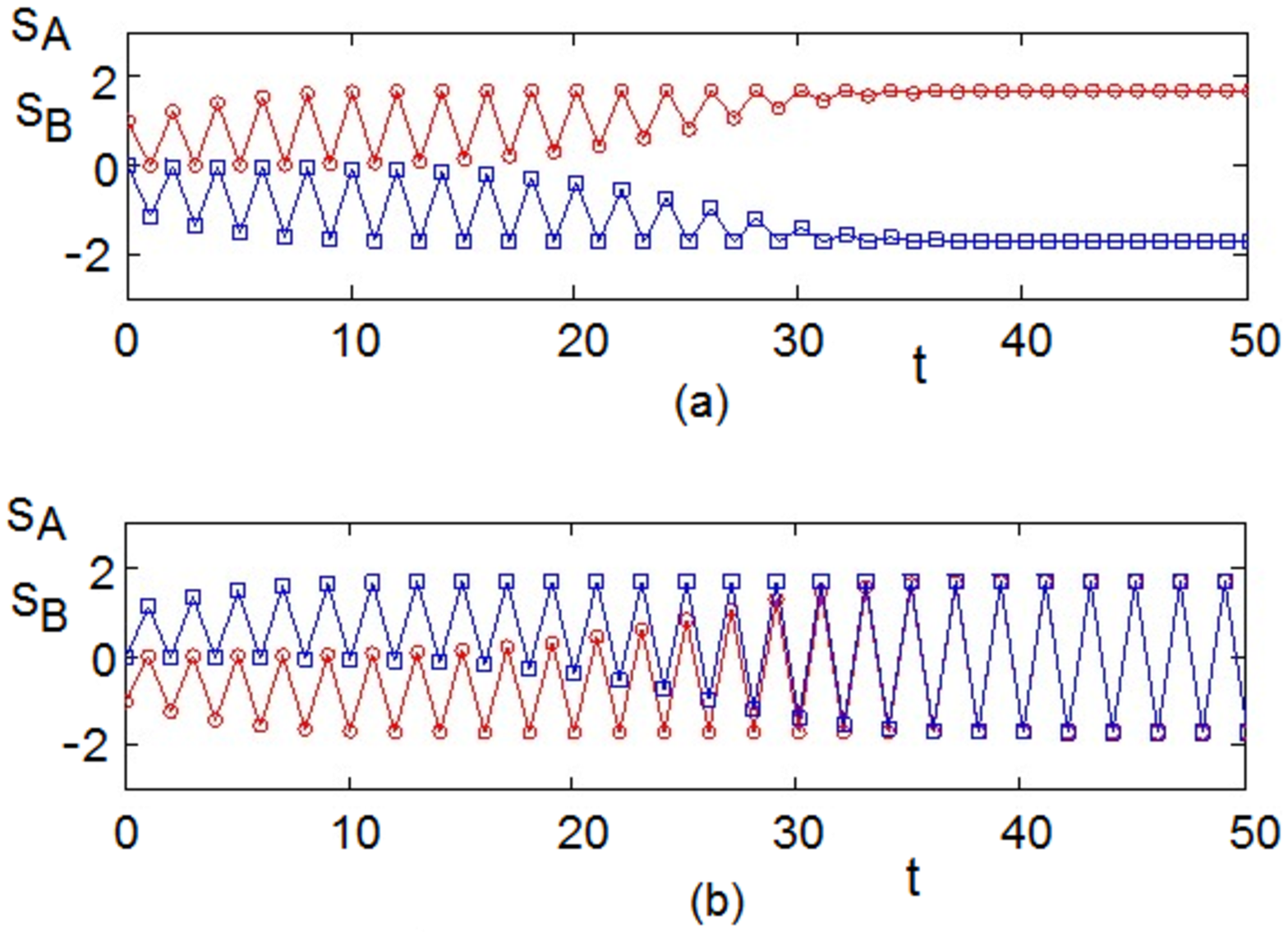}
\caption{ with $J_1=J_2=0$, $K_{12} = K_{21} = -0.3$ as in Fig. \ref{ffig2}f with different initial conditions. Through different patterns systems evolve to same nonzero values.\label{ffig3}}
\end{figure}

The qualitative pattern of time evolution may also depend on initial conditions. In Fig. \ref{ffig3} we show the time evolution for $K_{12} = K_{21} = -0.3$, same values as in Fig. \ref{ffig1}f but with different initial conditions.

\begin{figure}[ht!]
\centering
\includegraphics[width=8cm,angle=0]{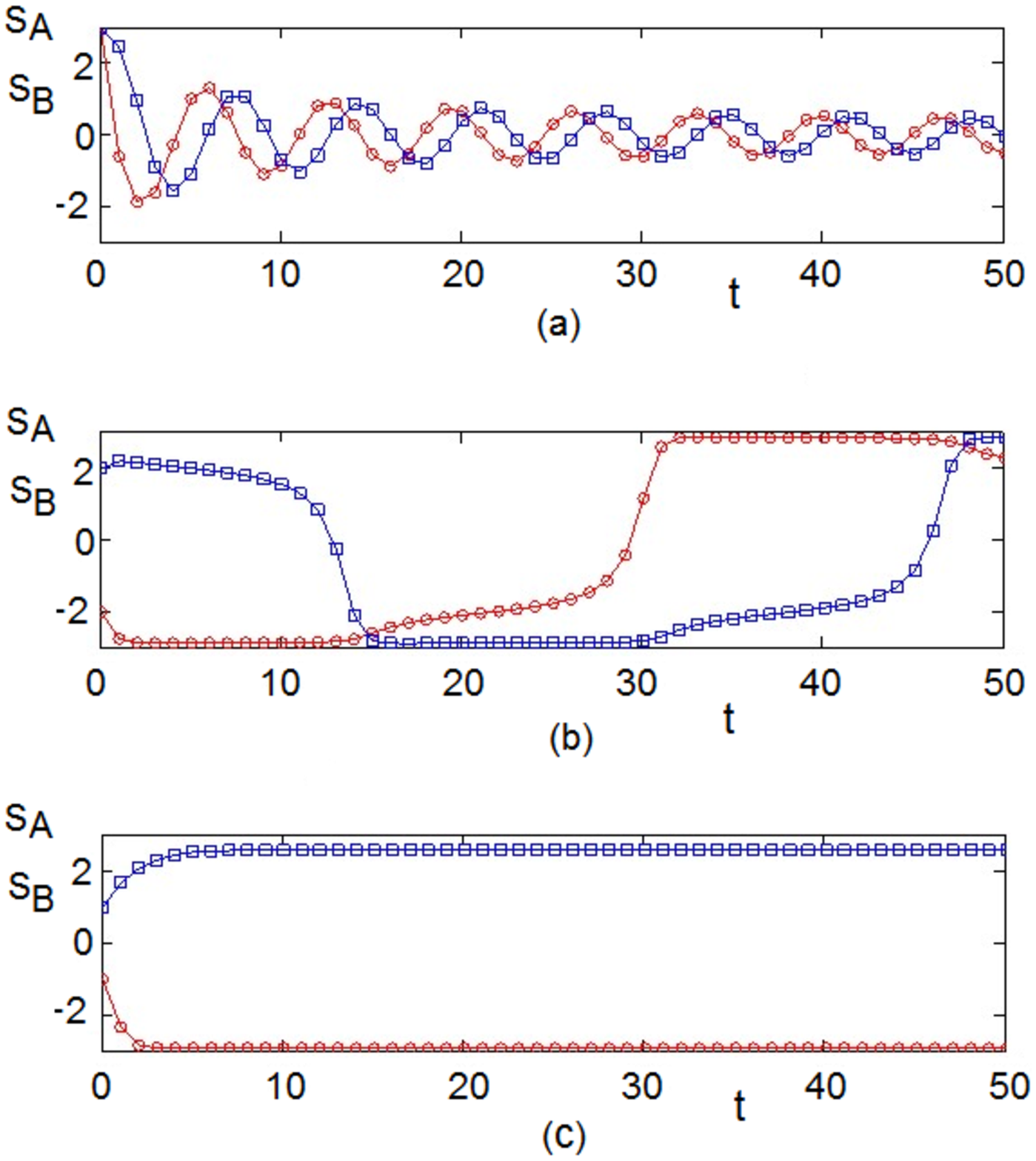}
\caption{All graphs are for $-K_{12} = K_{21} = 0.2$.  For $J_1 = J_2 < 0.15$ the time evolution is decaying oscillations as in Fig. 2a. In Fig. 4a $J_1= J_2 =0.15$ the state is critical and the amplitude of oscillations does not decay in time. In Fig. 4b $J_1 = J_2 = 0.6$ the period of oscillations is long. In Fig. 4c $J_1 = J_2 = 0.7$ the $S_A$, $S_B$ evolve in time to non-zero steady state values. The transition from Fig. 4b to Fig. 4c is discontinuous.\label{ffig4}}
\end{figure}

In Fig. \ref{ffig4} we consider the time evolution of preferences $S_A$ and $S_B$ in the presence of inter-group and intra-group interactions. First we consider the inter-group couplings of Fig. \ref{ffig2}a,  $-K_{12} = K_{21} = 0.2$ and include intra-group interactions. As we increase the strength of $J_1$, $J_2$ the time evolution evolves from decaying oscillations (as in Fig. \ref{ffig2}a); it continuously changes to sustained oscillations of increasing period; it discontinuously changes to steady non-zero values for $S_A$, $S_B$.

\begin{figure}[ht!]
\centering
\includegraphics[width=8cm,angle=0]{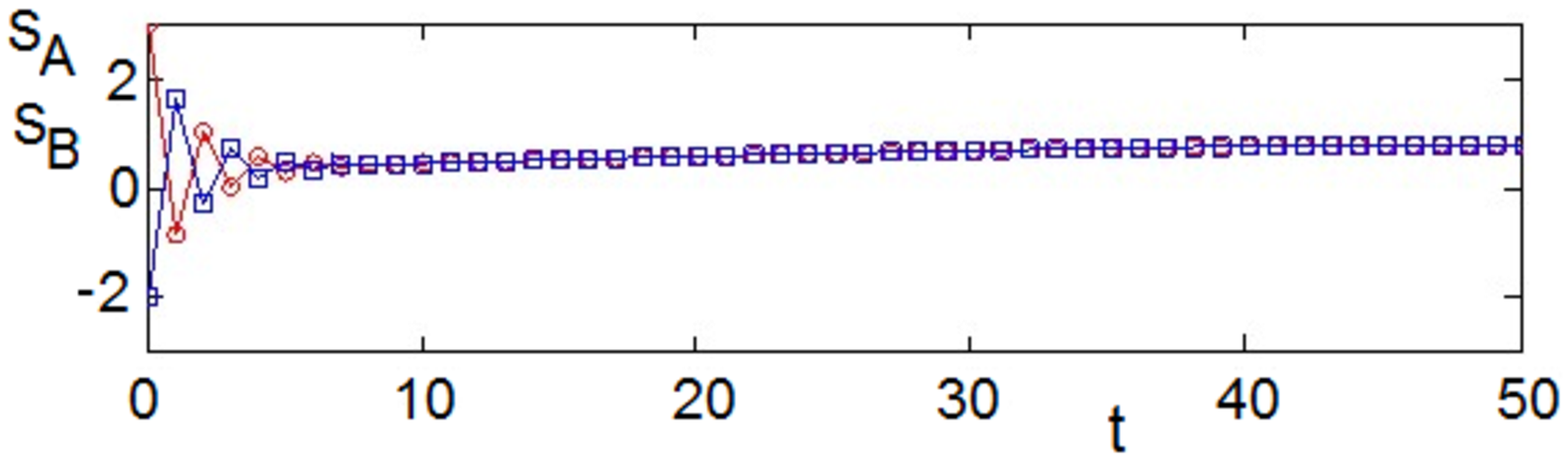}
\caption{$K_{12} = K_{21} = 0.2$.  For $J_1 = J_2 < 0.05$, $S_A$ and $S_B$ go to zero for long times, as shown in Fig. 2c. For $J_1 = J_2 = 0.06$, $S_A$ and $S_B$ approach a non-zero value for long time.\label{ffig5}}
\end{figure}

\begin{figure}[ht!]
\centering
\includegraphics[width=8cm,angle=0]{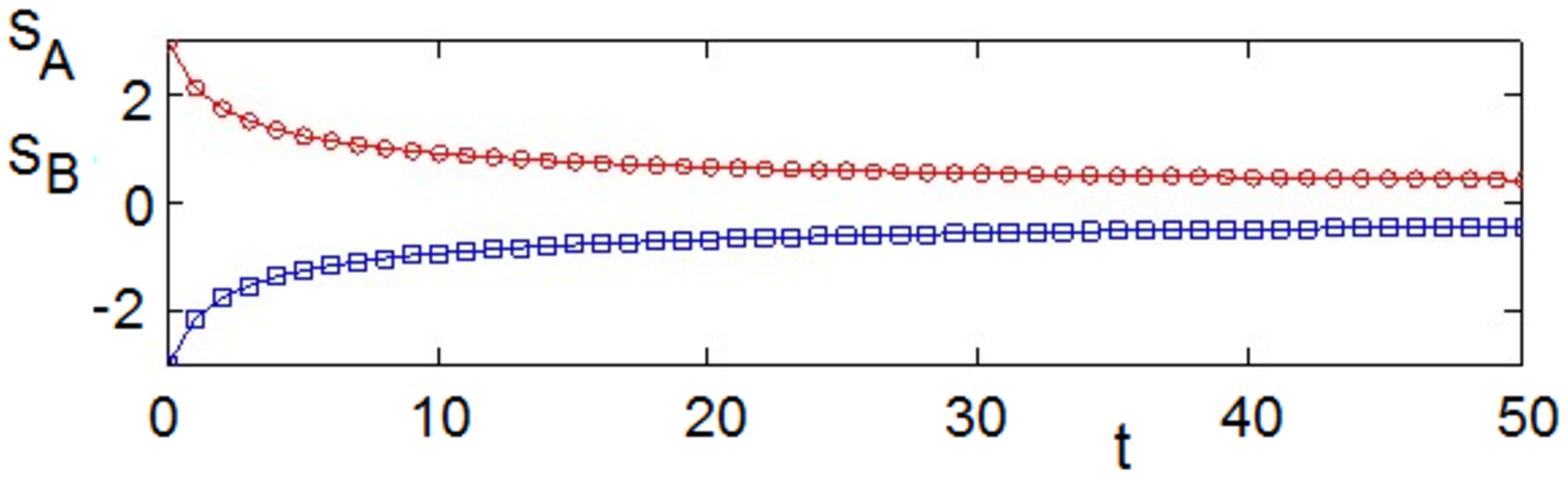}
\caption{$K_{12} = K_{21} = -0.2$.  For $J_1 = J_2 < 0.05$, $S_A$ and $S_B$ go to zero for long times, as shown in Fig. 2c. For $J_1 = J_2 = 0.05$, $S_A$ and $S_B$ approach a non-zero value for long time.\label{ffig6}}
\end{figure}

In Figure 2c we have shown that the two groups preferences $S_A$, $S_B$ evolve to zero in the absence of intra-group couplings and for weak inter-group couplings $K_{12} = K_{21}$.  Turning on the inner couplings, for weak values of $J_1$ and $J_2$ this situation is preserved (Fig. \ref{ffig5}).  However for sufficiently large intra-group couplings one gets a time evolution to non-zero $S_A$ and $S_B$ as shown in Fig. \ref{ffig6}. The transition between the two situations is continuous.

In Fig. 2e we have shown that the two groups preferences $S_A$, $S_B$ evolve to zero in the absence of intra-group couplings and for weak inter-group couplings $K_{12} = K_{21}= -0.20$.  For small values of $J_1$ and $J_2$ this situation is preserved.  For sufficiently large intra-group couplings $J_1 = J_2 = 0.05$ one gets a time evolution to non-zero $S_A$ and $S_B$ as shown in Fig. 5.  The transition between the two situations is discontinuous.

\section{Exploring applications to real conflicts}\label{MFD}

We illustrate below a series of possible trajectories of the two groups' willingness to engage in negotiations, when the individuals' preferences for resolving the conflict range between -3 and 3. Individuals with $s = 3$ show extreme adherence to their group's values, with a consequent lack of willingness to enter negotiations and make concessions. $s = -3$ for individuals lacking ideological adherence (independents), with a consequent openness to be persuaded to an opponent's views. Individuals with $s$ in the mid-range are willing to negotiate settlements consistent with their own values.

In Fig. \ref{ffig1} there is no inter-group communication. This can occur in very protracted political disputes where even discussing matters with individuals from an opposing group is discouraged or considered a sign of wavering values. This is reminiscent of the 2009 debates surrounding the Affordable Care Act, which have barely abated since passage of the Act.
In Fig. 1, where the two networks are "decoupled," individuals within each group act weakly in their efforts to persuade each other to their own points of view. In time the two groups converge to disorder. This does not mean that they would necessarily reach an agreement.

In Fig. \ref{ffig2}, the two groups are still decoupled. While Group $A$ has the same weak intra-group interactions as before, individuals in Group $B$ interact with each other more vigorously. As a result, Group $A$ slowly converges to 0 (disorder, open to negotiations or conflict) as before, but Group $B$ heads toward polarization, becoming more attached to its own core values and less open to compromises. This asymmetry in the intensity of activism within groups can occur if the contentious policy under consideration is of more importance to one group than the other. In our political example, gun control policies are more tied to Republican identity than for Democrats for whom it may be a matter of practical interest. The result (for the time period considered here) is that in the mid-term the groups may converge to a willingness to negotiate (e.g, negotiate some background check measures), but in the long run the group for whom opposing any gun control policy is more important (here Group $B$) will gravitate toward intransigence.

In Figs. \ref{ffig3}-\ref{ffig6} we explore situations where individuals within each networks interact not only with each other but also with those from the opposing network. In Fig. \ref{ffig3}, the two groups have weak intra-group interactions as in Fig. \ref{ffig1}, but inter-group communications produce oscillations of the average group opinions before they converge in time to a centrist stance for both, as in Fig. 1. Thus the possibility of interacting with opponents gives each individual food for thought and even moves the stance of each group before eventually causing openness to negotiations. Interactions between Democrats and Republicans regarding national security issues may exhibit this pattern.
The interaction pattern of Fig. 4a is the result of the fact that each group reacts negatively to the stances of individuals in the other group $K_{12}*K_{21} < 0$. As a result, instead of converging the stances of the groups are locked in a long-run, lagged oscillation from belligerence to conciliatory stances. Such dynamics occur sometimes in Democrat-Republican debates over social issues such as raising the minimum wage.

\begin{figure}[ht!]
\centering
\includegraphics[width=8cm,angle=0]{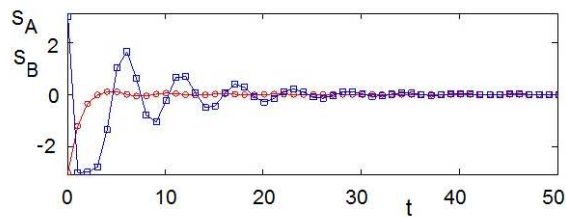}
\caption{Coupled networks with $J_1=J_2=0.25$, $K_{12}=-0.01$, $K_{21}=4$: $S_A$ and $S_B$ approach zero, oscillate in long run, $S_A$ monotonically, and $S_B$ through oscillations between conflict and conciliation.\label{ffig7}}
\end{figure}

Figure \ref{ffig7} shows a return to convergence, with one group (with strong intra-group interactions) arriving  monotonically at a willingness to negotiate with the other group (with weak intra-group connections) oscillating between some turning points before stabilizing, because individuals are swayed more by the other group than by their peers. In our political example, such asymmetric patterns may emerge around issues that are cross-cutting across party lines, such as unwillingness on both sides to intervene militarily in international conflicts ("boots on the ground").

The difference between Fig. \ref{ffig5} and Fig. \ref{ffig6} is the heightened effect of Group $B$ on members of Group $A$. This asymmetry produces wide, long-term oscillations in both groups' average stances, which may be due to the lagged mutual reactions to dialog. As well, Group $A$ has swings of lower amplitude than Group $B$. The temporary hold at each extreme value for Group $B$ reflects the lesser influence of Group $A$ on it. This pattern might correspond to how Democratic and Republican candidates in the upcoming elections react both to pressures from their own constituencies and to opponents' proposals for responding to immigration and refugees challenges.

\begin{figure}[ht!]
\centering
\includegraphics[width=8cm,angle=0]{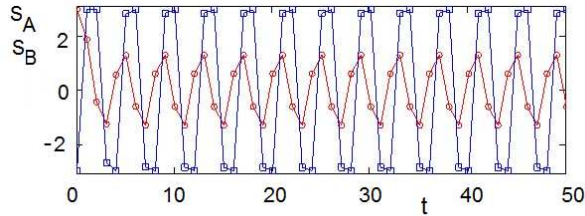}
\caption{As in Fig. \ref{ffig7} with more negative $K_{12}$, namely $J_1=J_2=0.25$, $K_{12}=-0.1$, $K_{21}=4$: both $S_A$ and $S_B$ oscillate in the long run.\label{ffig8}}
\end{figure}

Figures \ref{ffig1} to \ref{ffig7} illustrate long-term effects of the multiplex interactions between networked individuals in two groups, each of whom has a preference for how the inter-group conflict should be addressed, depending on their degrees of adherence to group values. They show that different assumptions about the intensity of intra- and inter-group interactions yield qualitatively different long-term patterns. Convergence, polarization or oscillatory patterns may correspond to real situations in which two conflicting groups are involved.

\section{Monte Carlo simulations}\label{MC}
While the mean field model is exact \cite{Kaufman1989},\cite{Cohen} for the static (equilibrium, infinite time limit) of an infinitely large system on the equivalent neighbor lattice, we are interested to model finite size systems and to expand the model to finite range interactions.  To this end we next perform Monte Carlo simulations on the dynamic model.
\subsection{Method}
We consider two coupled systems, called $A$ and $B$, having the same number of individuals $N$.  As in the previous sections, each individual interacts with all other individuals of his or her group and with the average of the other group. The energy at time $t+1$ of an individual $s_i$  of Group $A$ and that of  an individual $s_k$ of Group $B$ are written as

\begin{eqnarray}
E_i^A(t+1)&=&- J_A s_i(t+1)\frac{1}{N}\sum_{j\in A } s_j(t) -K_{AB}<S_B>(t) s_i(t+1) \label{eqn:hamil0}\\
E_k^B(t+1)&=&- J_B s_k(t+1)\frac{1}{N}\sum_{m\in B } s_m(t) -K_{BA}<S_A>(t) s_k(t+1)
 \label{eqn:hamil1}
\end{eqnarray}
where $J_{A}$ is the interaction between individual $i$ of group $A$ at time $t+1$ with other individuals $j$ belonging to the same group at time $t$ and $K_{AB}$ denotes the interaction between an individual of Group $A$ at time $t+1$ with the average $<S_B>(t)$ of Group $B$ at the previous time $t$.   $J_{B}$ and $K_{BA}$ are defined in the same manner for Group $B$. The sums are performed over all individuals belonging to a group as indicated.

We have performed Monte Carlo (MC) simulations using the above equations at various "social temperatures" $T$ for $N=1296$ individuals for each group.  Most of simulations have been done with this group size, but we have checked the validity of our results for twice and three times larger groups. The results do not vary with sizes of this order of magnitude. 
As we know in statistical physics, the finite-size effect manifests at and near the transition temperature $T_c$ defined in the next subsection.
Far from $T_c$, the finite-size effect is not significant in the collective behavior of the system. 
That is the reason why the theory of finite-size effect (see for example Ref. \onlinecite{Binder})
relates various physical quantities to the system size only around $T_c$. This theory helps determine critical properties of a system by using several sizes in MC simulations.  In our simulations, we did not focus on the critical 
properties, so we did not use systematically many sizes. The reason why we used $N=1296$ spins is that at this size 
the error (relative uncertainty) on $T_c$ is $\simeq 1/\sqrt(N)=3\%$ which is sufficient for our purpose. Note that in opinion surveys, populations of about one thousand individuals are often used to have results with this error (larger sizes yield smaller errors but this is not necessary because there are other factors in opinion surveys which alter the results at this error magnitude such as the rate of opinion change, the non-rigorous population representative composition in the sample etc).

Note that in the mean-field sections presented above, the "social temperature" has been included in the definitions of $J_1$, $J_2$, $K_{12}$ and $K_{21}$, namely $J_1=J_A/T$, $J_2=J_B/T$, $K_{12}=K_{AB}/T$
and $K_{21}=K_{BA}/T$.

The algorithm, called Metropolis, is the following:

(i) we generate independently two groups each of which has an initial condition at $t=0$ for its members;

(ii) we equilibrate each group at a fixed $T$ before turning on the interaction between them;

(iii) At time $t+1$, we calculate the energy $E_o$ of an individual $s_i$ of Group $A$ with Eq. (\ref{eqn:hamil0}) using the average of Group $B$ at the previous time, namely $<S_B>(t)$.  We change the state of $s_i$ at random and calculate its new energy $E_n$. If the new energy $E_n$ is lower than the old one $E_o$, the new state is accepted. Otherwise it is accepted only with a probability proportional to $\exp[-(E_n-E_o)/T]$. We go next to another individual of Group $A$ and update its state until all are considered. We then calculate the average $<S_A>(t+1)=\sum_{i\in A}s_i(t+1)/N$ of Group $A$ to be used for the next time. We do this for all members of Group $B$.

As mentioned earlier, the two groups may have different intrinsic characteristics defined by their own parameters which are the intra-group interaction $J_A$ or $J_B$, the number of individual states $q_A$ or $q_B$, and the individual stance $M_A$ or $M_B$. These parameters may differ among the two groups. For simplicity, we will take $M_A=M_B=3$ as in the mean-field sections. An individual of  Group $A$ has $q_A$ states ranging from $-M_A$ to $+M_A$, and an individual of Group $B$ has  $q_B$ states ranging from $-M_B$ to $+M_B$. The meaning of these states has been discussed in the mean-field sections.

\subsection{Thermodynamic properties at equilibrium}

To facilitate the understanding of the dynamic behavior of the two groups upon interaction, let us show first the thermodynamic properties of each group before the interaction. We use the word "thermodynamic" borrowed from statistical physics to describe the properties of a system in which we introduce the social temperature (or noise) to represents a societal perturbation. The term "equilibrium" indicates the state where there is no time-dependence in contrast to the dynamics presented below. In the absence of social perturbation, $T=0$, each group is in its base state: its energy, given by the $J$ term of Eqs. (\ref{eqn:hamil0})-(\ref{eqn:hamil1}), is at a minimum, with all individuals aligned with each other. When $T \neq 0$ each individual fluctuates between the $q_A$ (or $q_B$) states.  We define a measure which expresses the average strength of  Group $\ell$ ($\ell=A,B$) as
\begin{equation}
S_{\ell}=\frac{1}{N}\langle \sum_{i\in \ell}s_i(t)\rangle
\end{equation}
where $\sum_{i\in \ell}s_i(t)/N$ is the spatial average at time $t$ and $\langle...\rangle$ denotes the long-term average. Then $|S_{\ell}|=M_{\ell}$ is at the base state but decerases as the individuals fluctuates.
There exists a so-called "critical temperature" $T_c$ above which $S_{\ell}=0$ due to fluctuations of individuals between $-M_{\ell}$ and $M_{\ell}$. As will be seen below, $T_c$ plays an important role in understanding the dynamic behavior of two groups $A$ and $B$ when their interaction is turned on.

Note that while studying the group dynamics, the time average will not be computed. The time dependence of $S_A$ and $S_B$ are computed with the mean-field approach in Eqs. (\ref{eq1})-(\ref{eq2}). For MC simulations they are computed with the algorithm described in the previous subsection.

\subsection{Case of similar groups with opposite attitudes: dynamic properties}

\begin{figure}[ht!]
\centering
\includegraphics[width=6cm,angle=0]{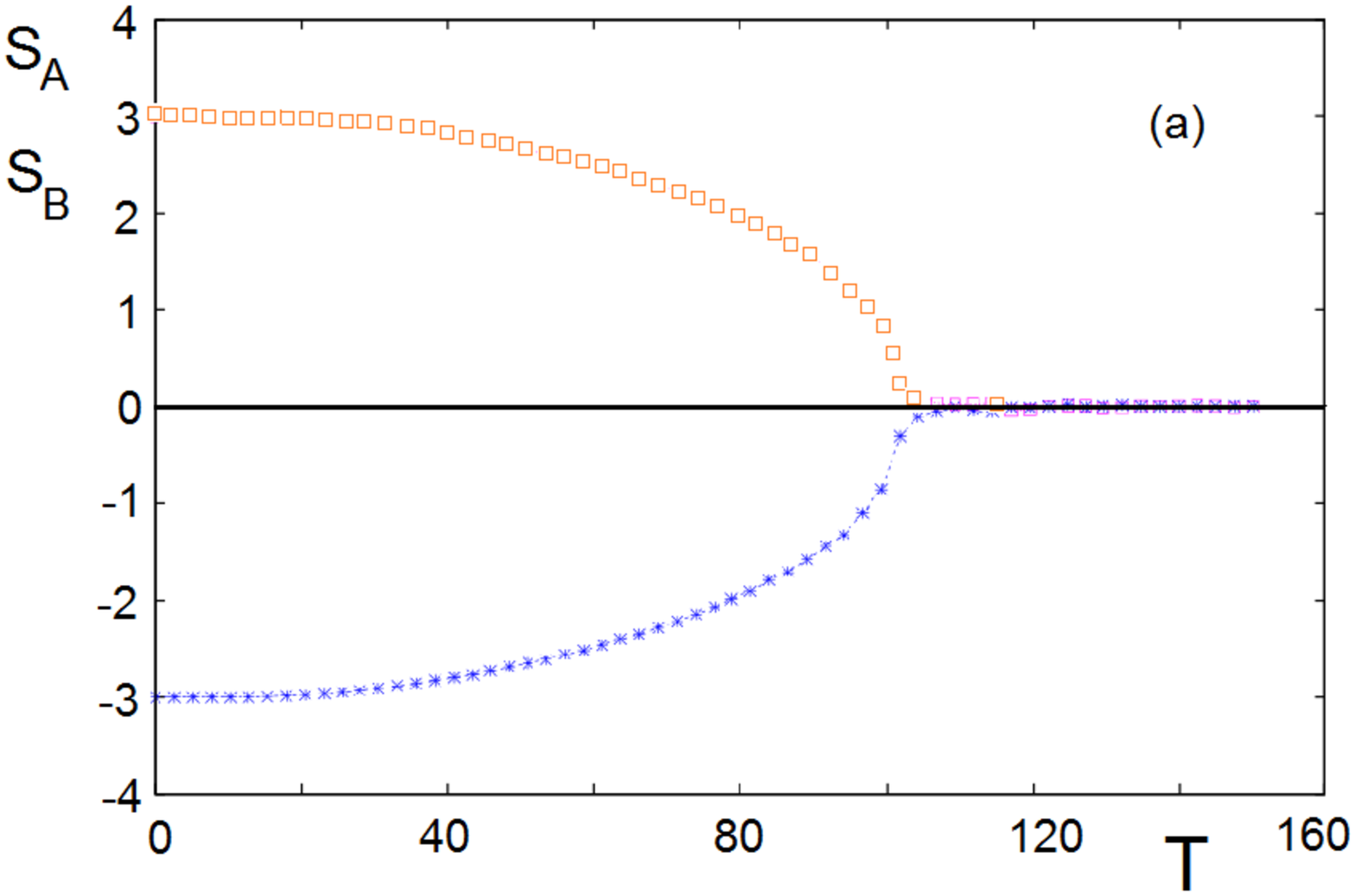}
\includegraphics[width=6cm,angle=0]{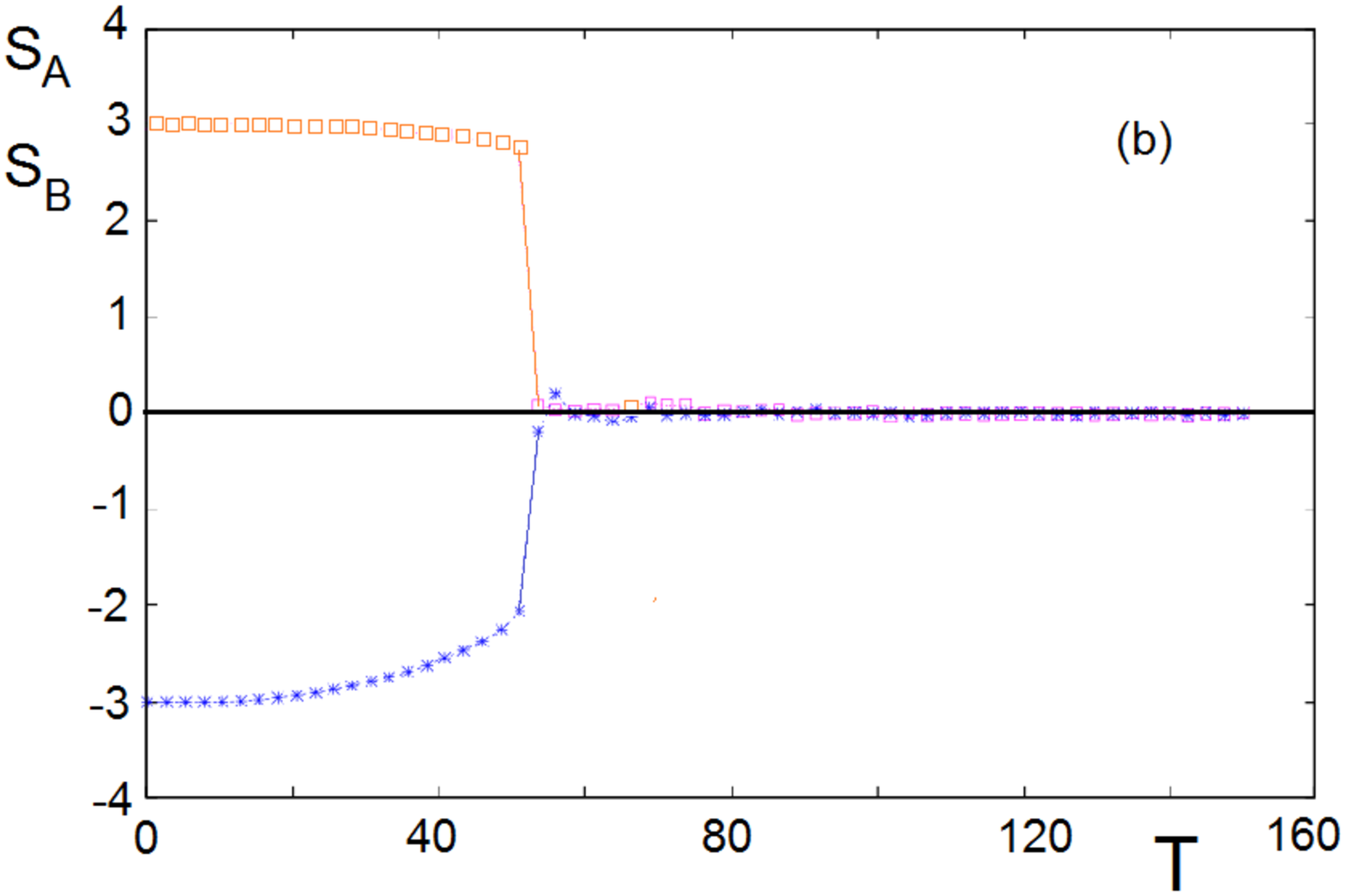}
\caption{(a) Before interaction $J_A=J_B=0.02$, $q_A=q_B=7$, initial conditions $S_A=-S_B=3$, both groups become "disordered" at $T_{c}^0\simeq 102$ (arbitrary unit), (b) With inter-group interaction $-K_{AB}=K_{BA}=0.005$, both groups become "disordered" at $T_{c}=53$. \label{ffig9}}
\end{figure}

\begin{figure}[ht!]
\centering
\includegraphics[width=6cm,angle=0]{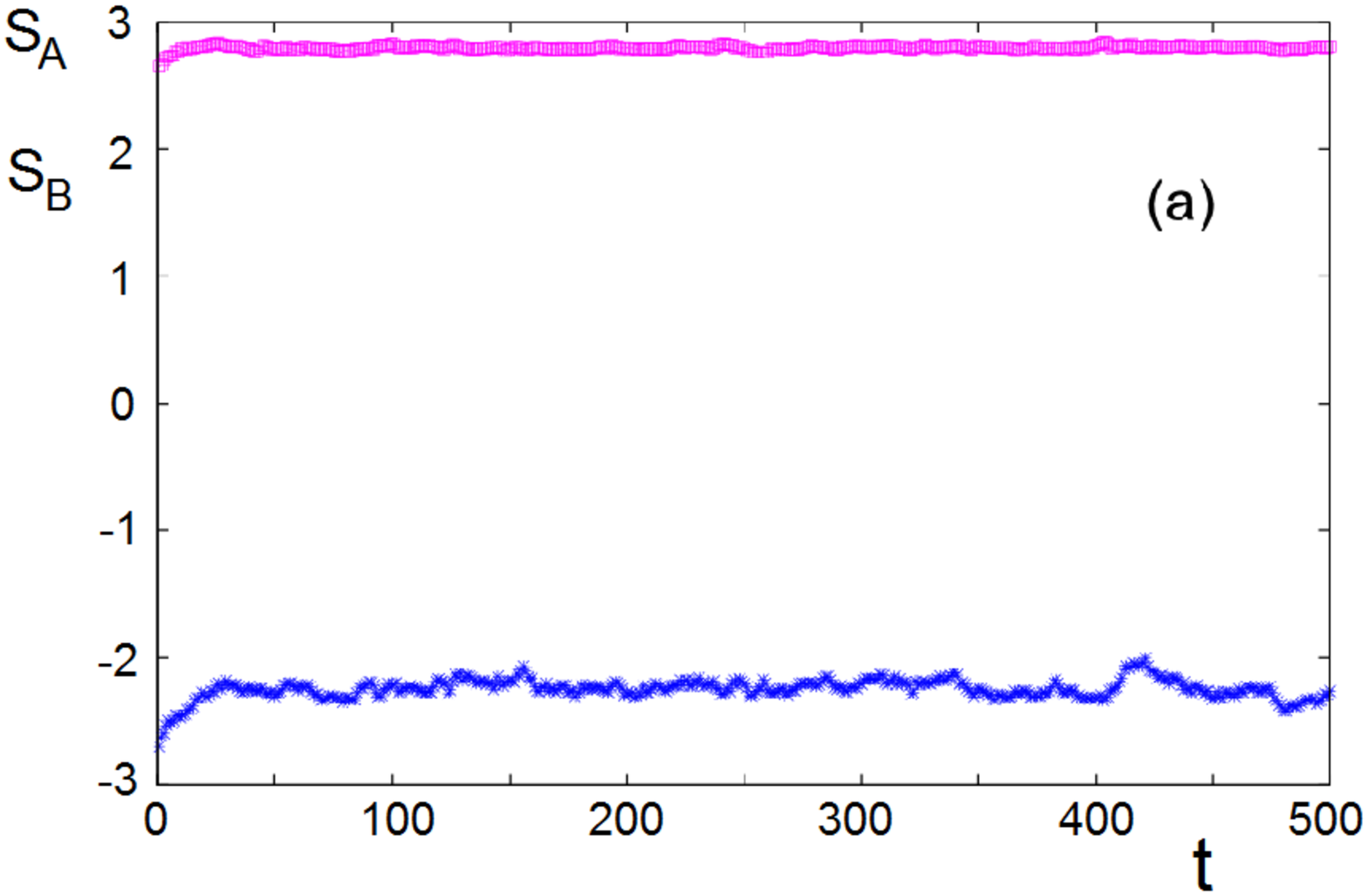}
\includegraphics[width=6cm,angle=0]{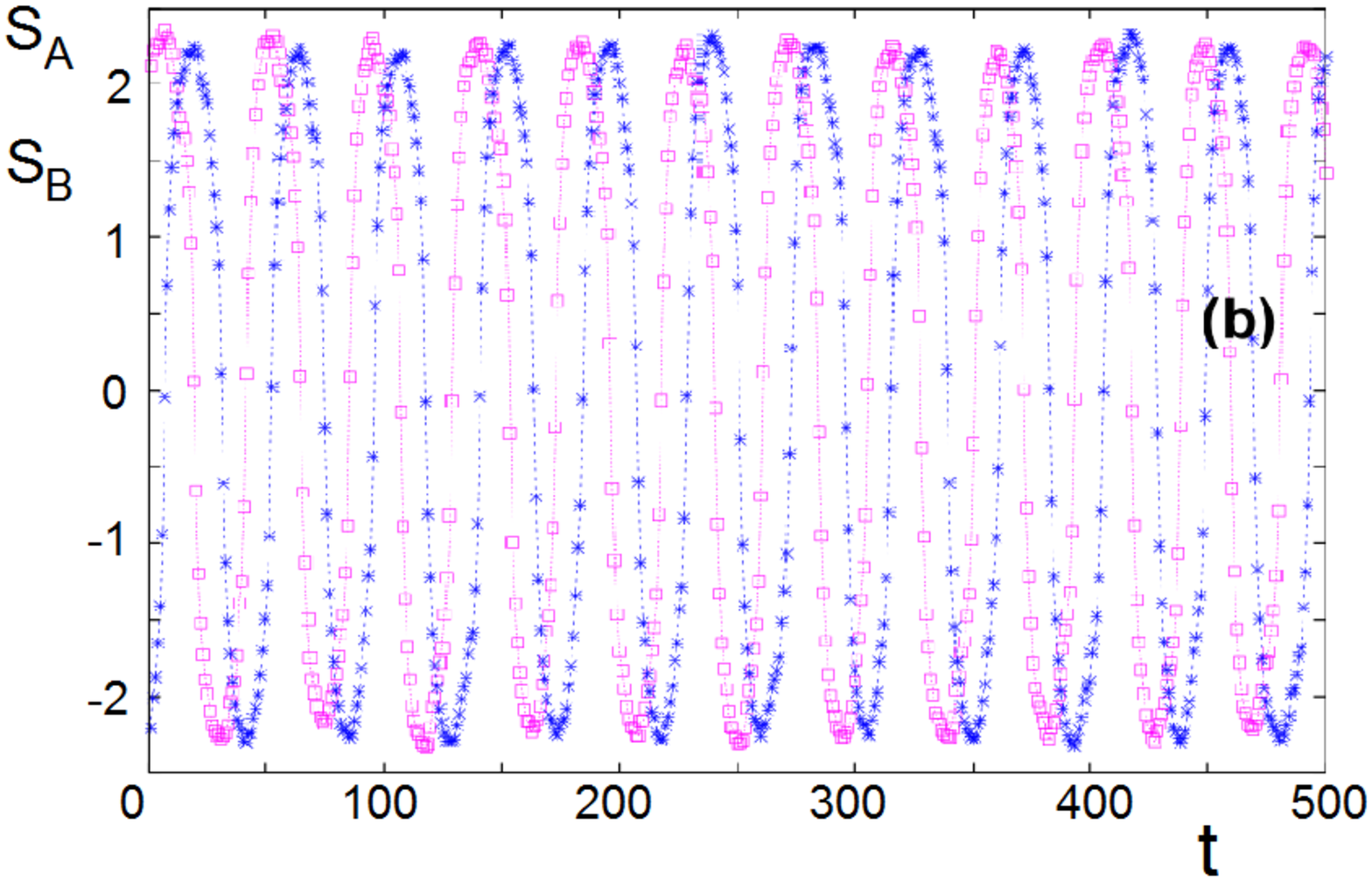}
\includegraphics[width=6cm,angle=0]{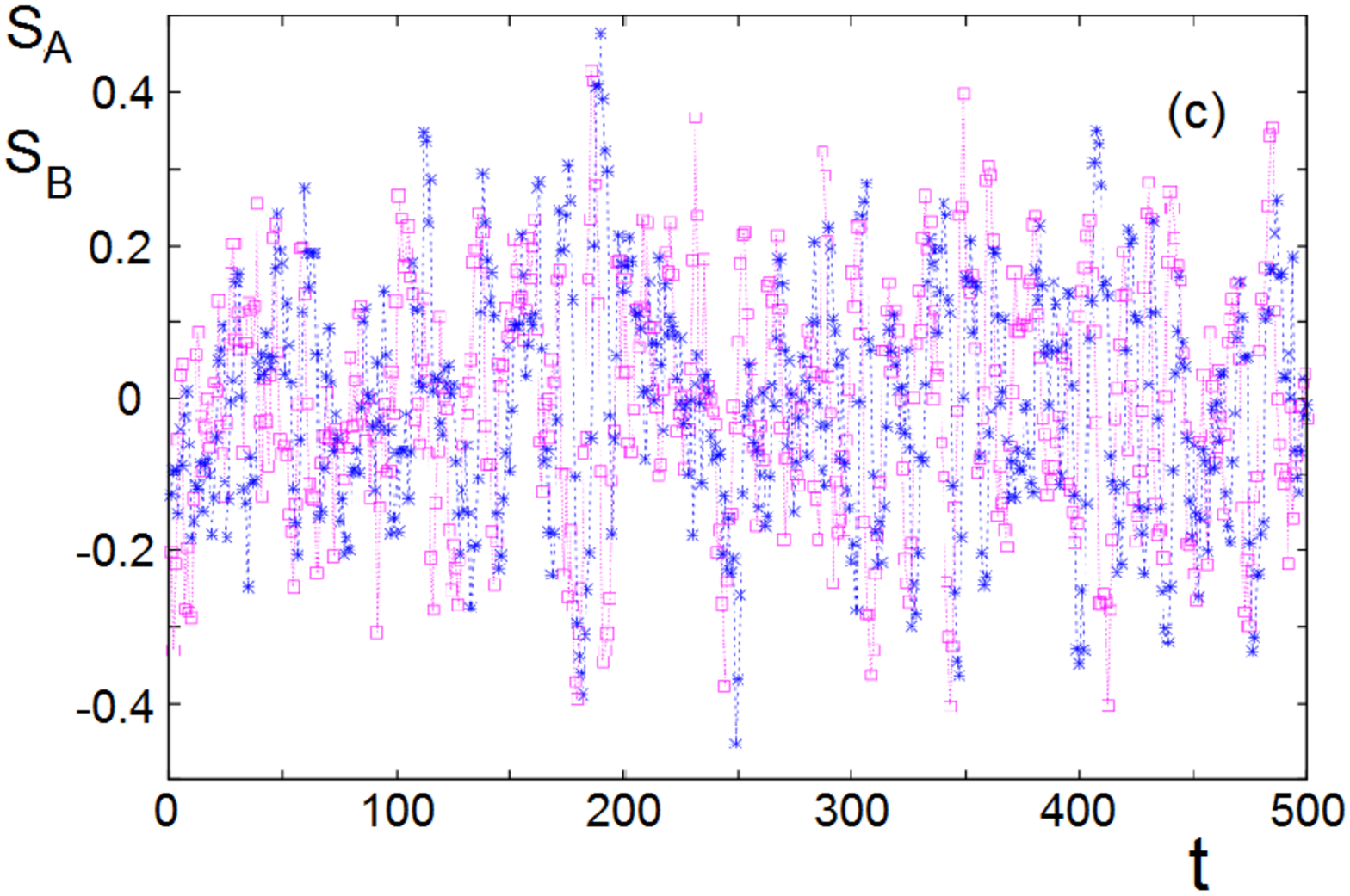}
\caption{Dynamics of two groups upon interaction $-K_{AB}=K_{BA}=0.005$ with $J_A=J_B=0.02$, $q_A=q_B=7$, initial conditions $S_A=-S_B=3$: (a) at "social temperature" $T=48$ below  $T_{c}=53$ where both groups are "ordered", (b) at "social temperature" $T=74$ between  $T_{c}$ and $T_{c}^0$ (values given in the caption of Fig. \ref{ffig9}), (c) at $T=125$ above $T_{c}^0$ in the initial disordered phase of both groups. See text for comments. \label{ffig10}}
\end{figure}

We show the simplest example in Fig. \ref{ffig9} where $J_A=J_B$, $M_A=M_B=3$, $q_A=q_B=7$ with the initial condition $S_A=-S_B=3$, namely Group $A$ does not want to negotiate while Group $B$ wants to negotiate.  Without interaction, the two groups are stable up to $T_c^0\simeq 102$ as shown in Fig. \ref{ffig9}a.  When the two groups interact with each other, the critical temperature is reduced to $T_c\simeq 53$.  The temperature zone between $T_c$ and $T_c^0$ seen belowis interesting: it is in this zone that each group, though disordered upon interaction, has the "memory" of its initial "ordered" state when there was no interaction with the other. Indeed, it is in this zone that the stances of two groups oscillate periodically in time as seen in Fig. \ref{ffig10}b. Below this zone, each group keeps its own stance (Fig. \ref{ffig10}a) and beyond this zone the stances of two groups fluctuate in a "chaotic" manner with small oscilations around zero, as seen in Fig. \ref{ffig10}c.

\subsection{Case of two groups with different strengths: dynamic properties}
\begin{figure}[ht!]
\centering
\includegraphics[width=6cm,angle=0]{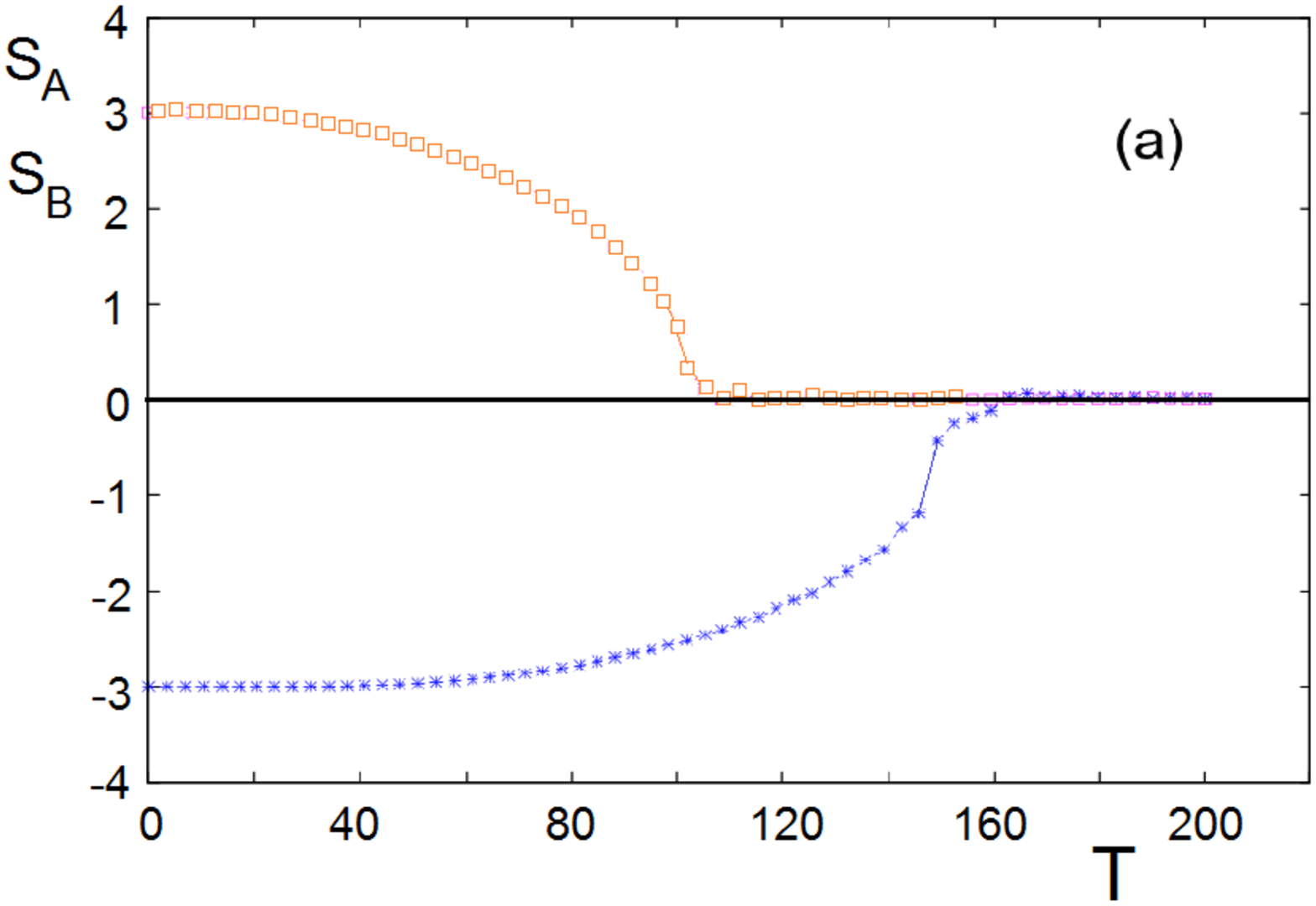}
\includegraphics[width=6cm,angle=0]{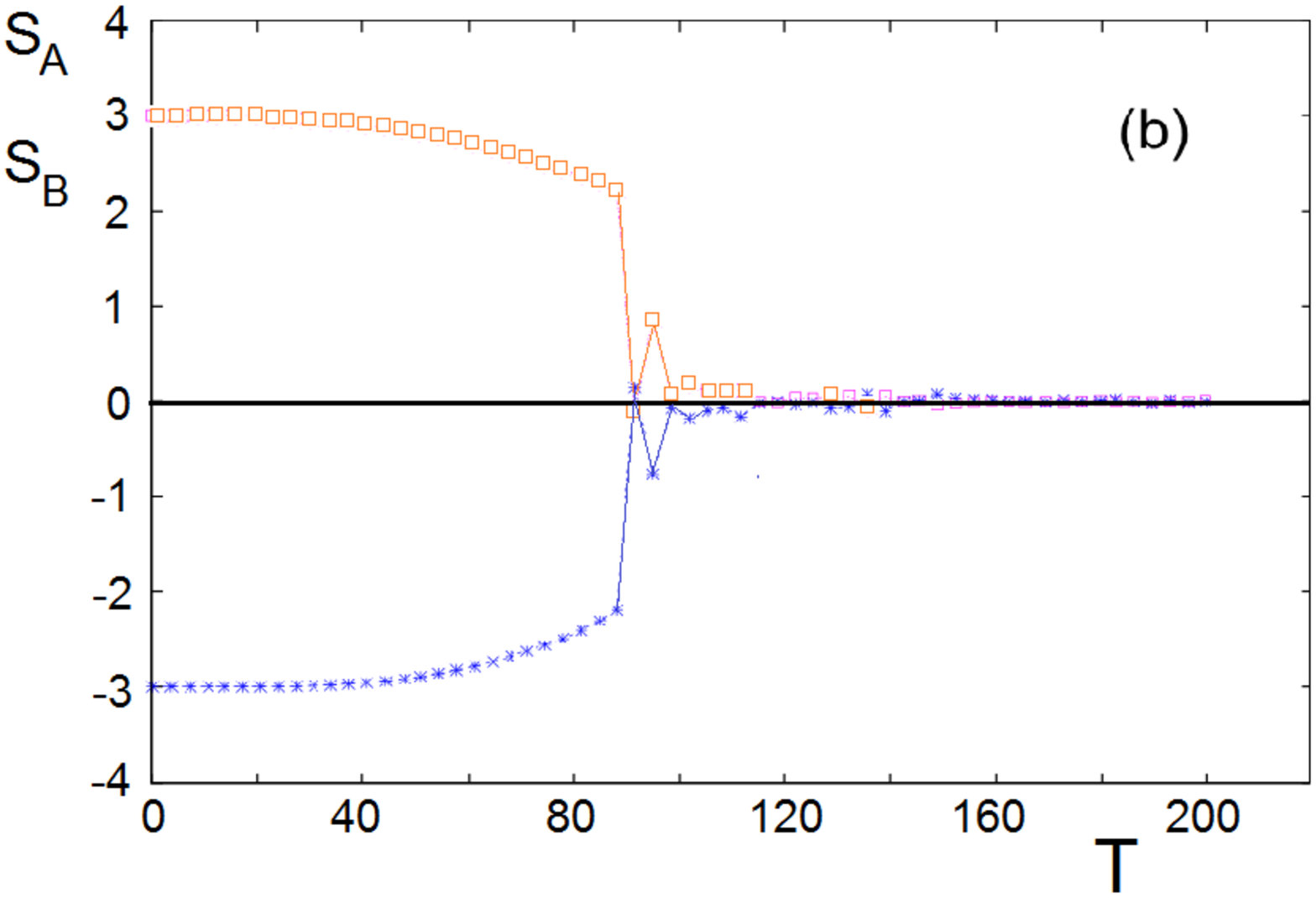}
\caption{(a) Before interaction $J_A=J_B=0.02$, $q_A=7$, $q_B=3$, initial conditions $S_A=-S_B=3$, Group $A$ becomes "disordered" at $T_{c}^0(A)\simeq 102$ (arbitrary unit) while Group $B$ has $T_{c}^0(B)\simeq 151$, (b) With interaction $-K_{AB}=K_{BA}=0.005$, both groups become "disordered" at $T_{c}\simeq92$. \label{ffig11}}
\end{figure}

Let us show now a second example where the numbers of states in the two groups are different: we take $q_A=7$ and $q_B=3$. As it is known in statistical physics, the larger the number of states the lower the critical temperature. We see this in Fig. \ref{ffig11}a when the two groups do not interact: Group $A$ has $T_c^0(A)\simeq 102$ and Group $B$ has $T_c^0(B)\simeq151$.  When the groups interact with each other, they become disordered at the same critical "social temperature" $T_c\simeq 92$ which is much lower than their initial non-interaction critical temperatures.

\begin{figure}[ht!]
\centering
\includegraphics[width=6cm,angle=0]{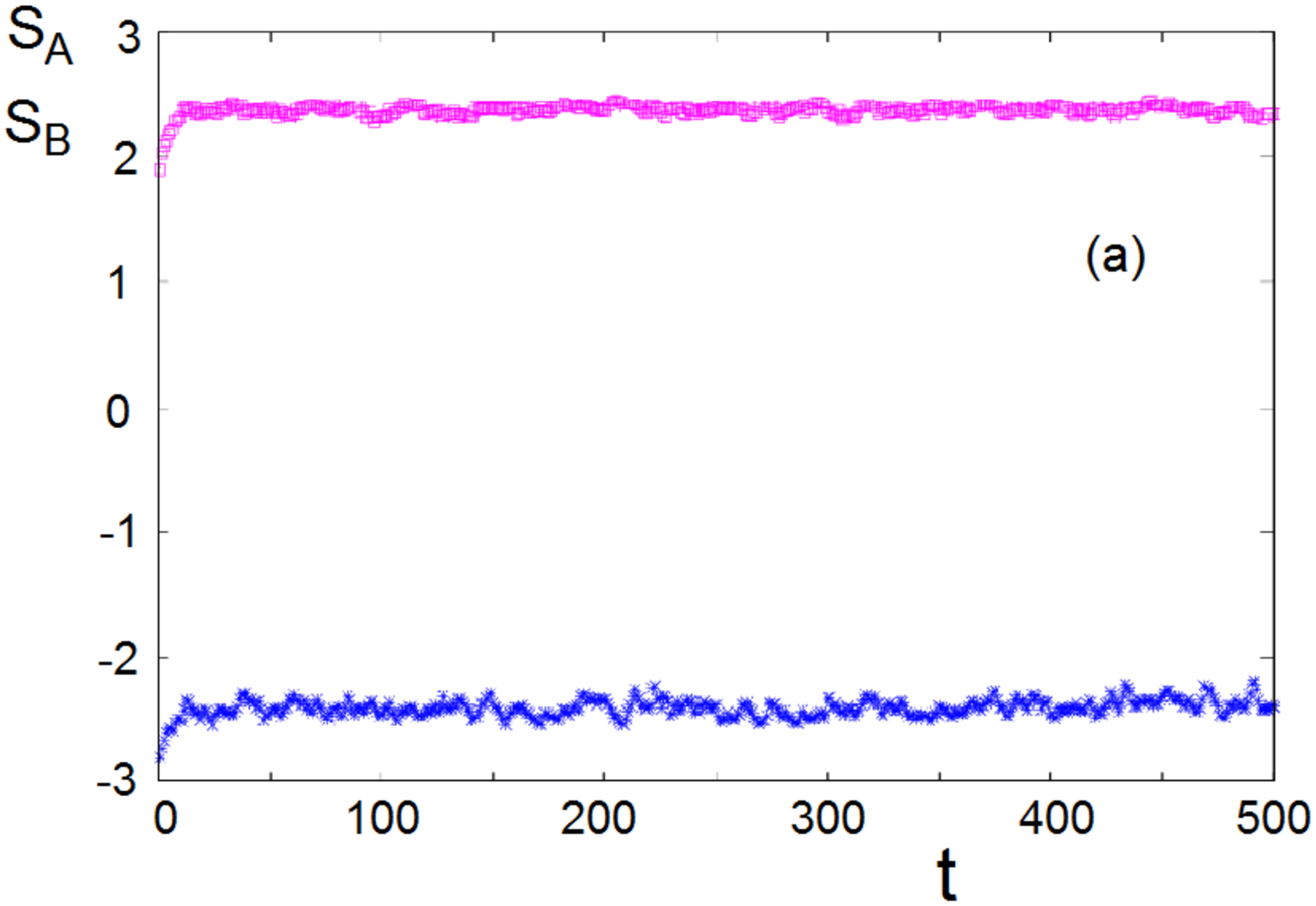}
\includegraphics[width=6cm,angle=0]{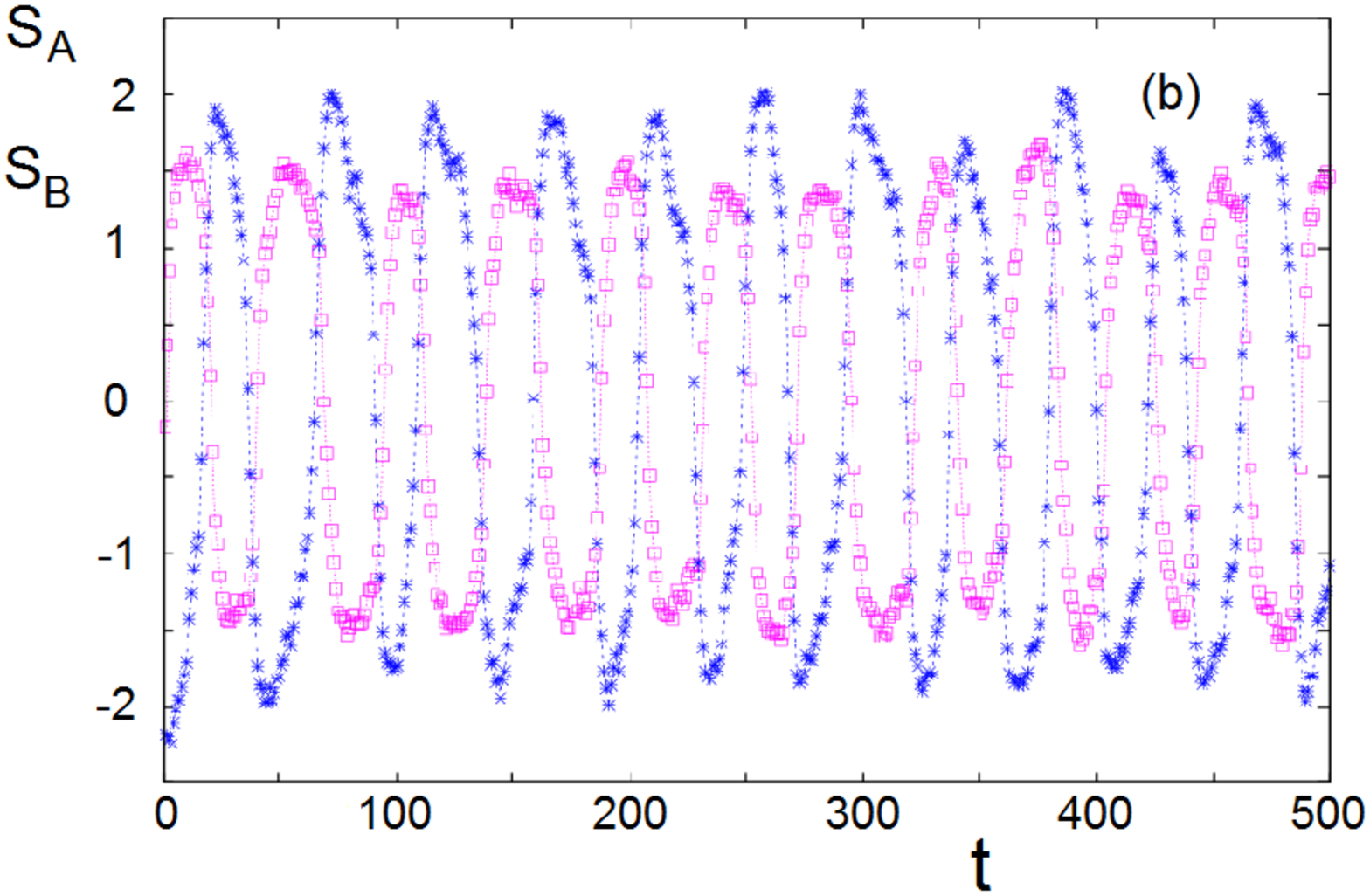}
\includegraphics[width=6cm,angle=0]{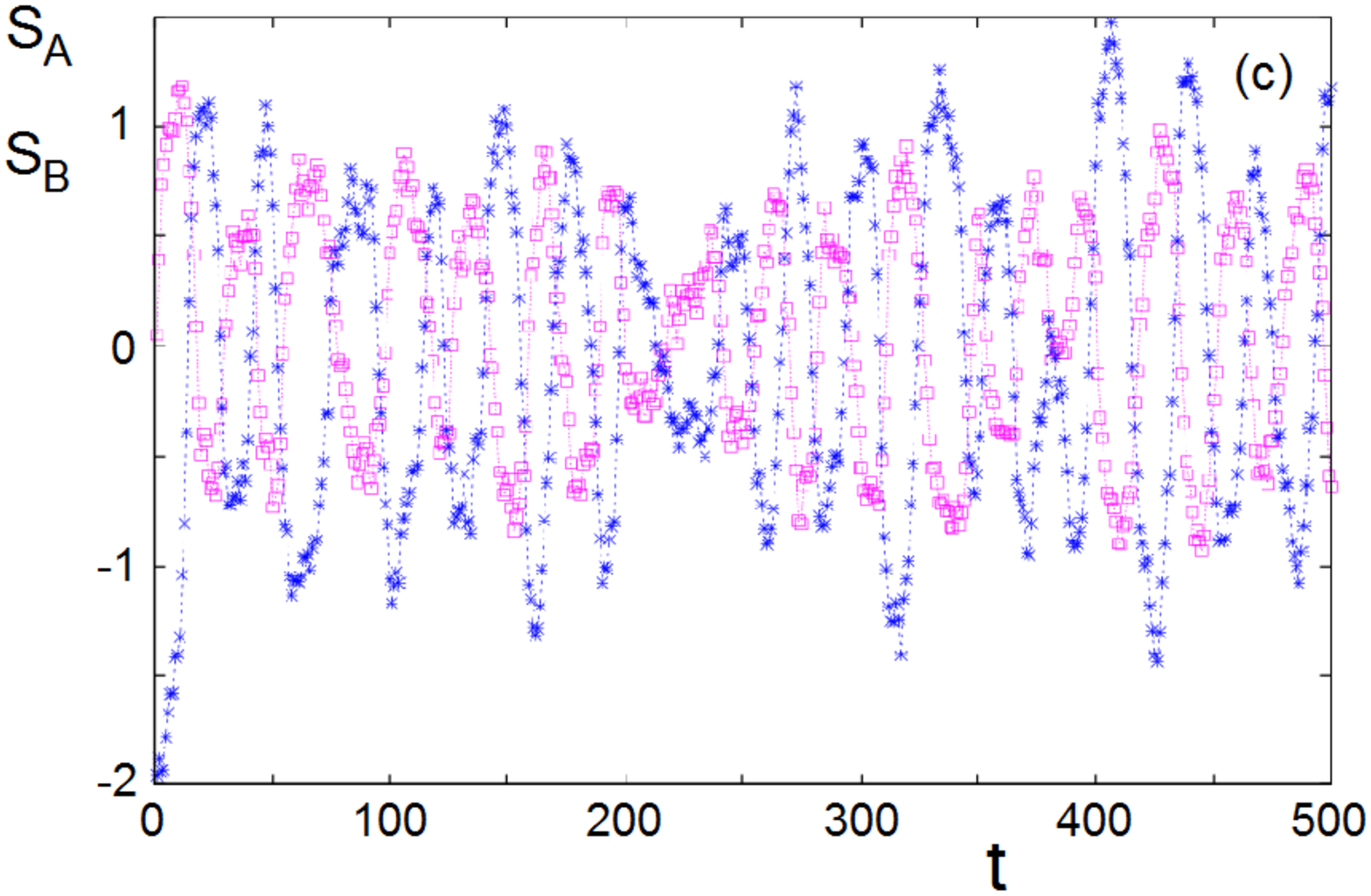}
\includegraphics[width=6cm,angle=0]{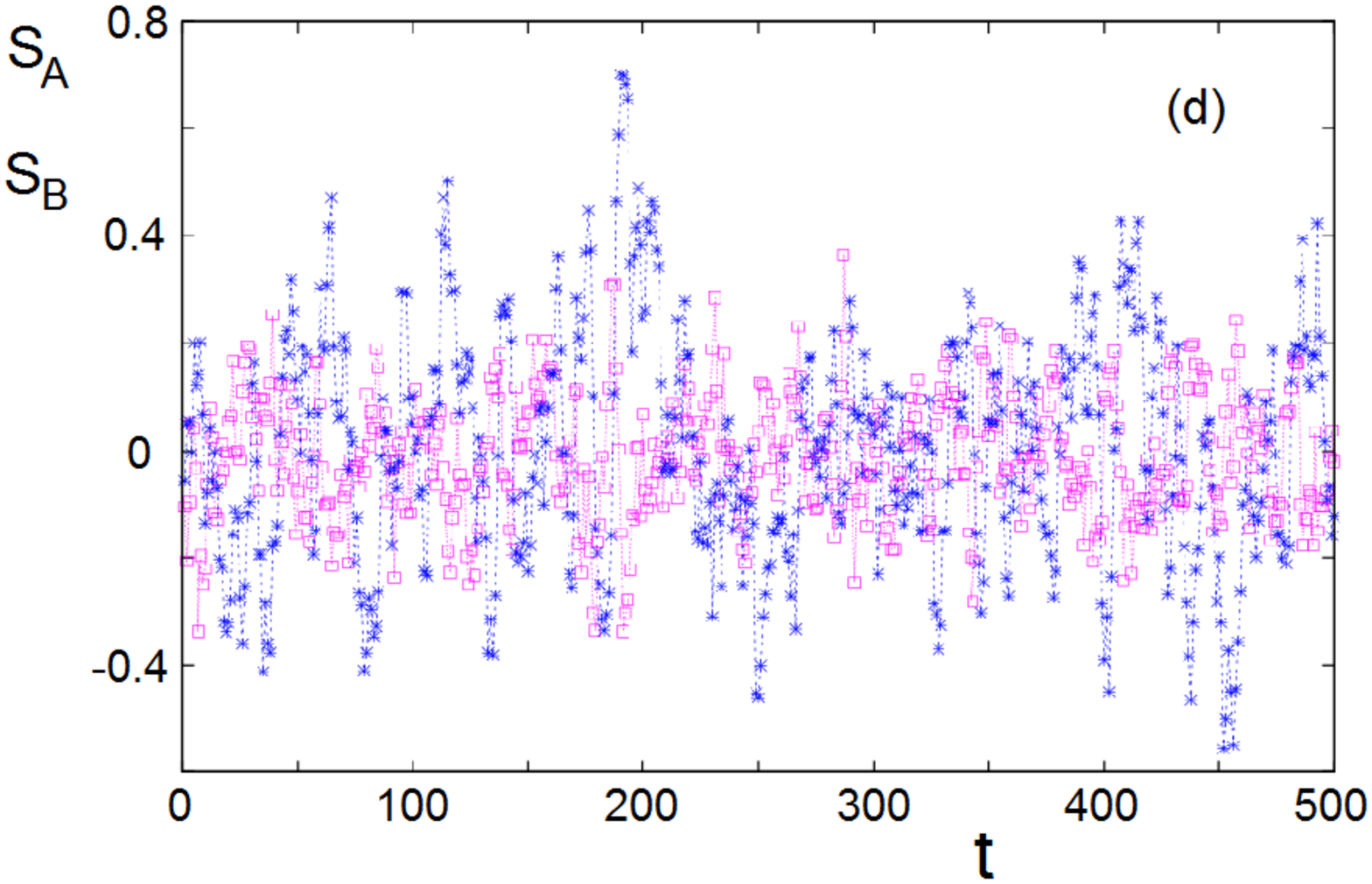}
\caption{Dynamics of two groups upon interaction $-K_{AB}=K_{BA}=0.005$ with $J_A=J_B=0.02$, $q_A=7$, $q_B=3$, initial conditions $S_A=-S_B=3$: (a) At "social temperature" $T=81$ below  $T_{c}=92$ where both groups are "ordered", (b)-(c) at "social temperatures" $T=115$ and 132 between  $T_{c}^0(A)=102$ and $T_{c}^0(B)=151$ (cf. Fig. \ref{ffig11}), (d) at $T=166$ above $T_{c}^0(A)$ and $T_{c}^0(B)$. See text for comments. \label{ffig12}}
\end{figure}

As in the previous case, the two groups with different initial critical temperatures $T_{c}^0(A)$ and $T_{c}^0(B)$ show stability of their stances below $T_c$ but an oscillation of the stances between  $T_{c}^0(A)$ and $T_{c}^0(B)$ as seen in Fig. \ref{ffig12}a and Fig. \ref{ffig12}b.  At temperatures beyond  $T_{c}^0(B)$ the stances strongly fluctuate around their zero value (see Fig. \ref{ffig12}c).

\subsection{Case of two groups with inter-group conflict interaction}

\begin{figure}[ht!]
\centering
\includegraphics[width=6cm,angle=0]{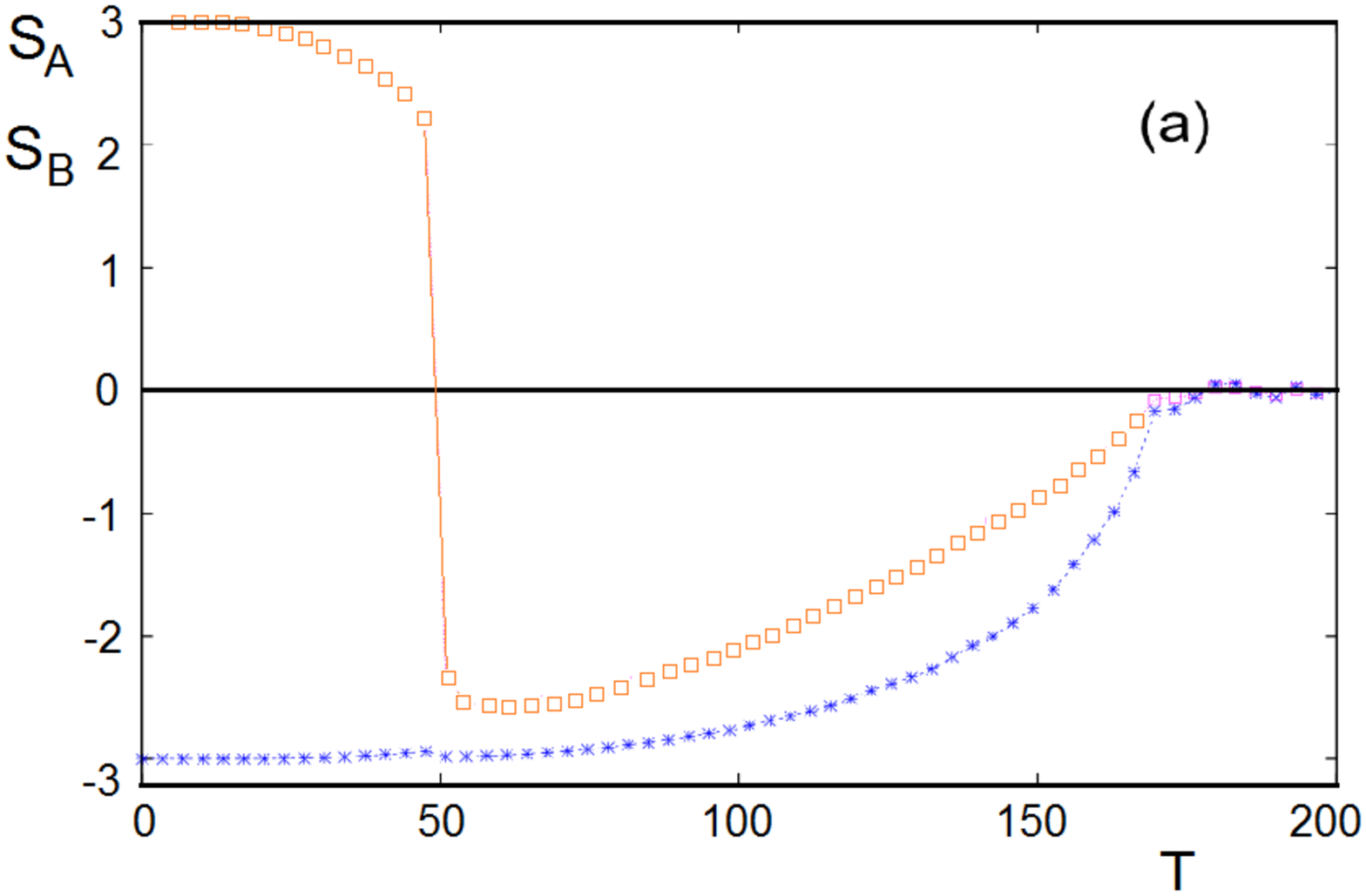}
\includegraphics[width=6cm,angle=0]{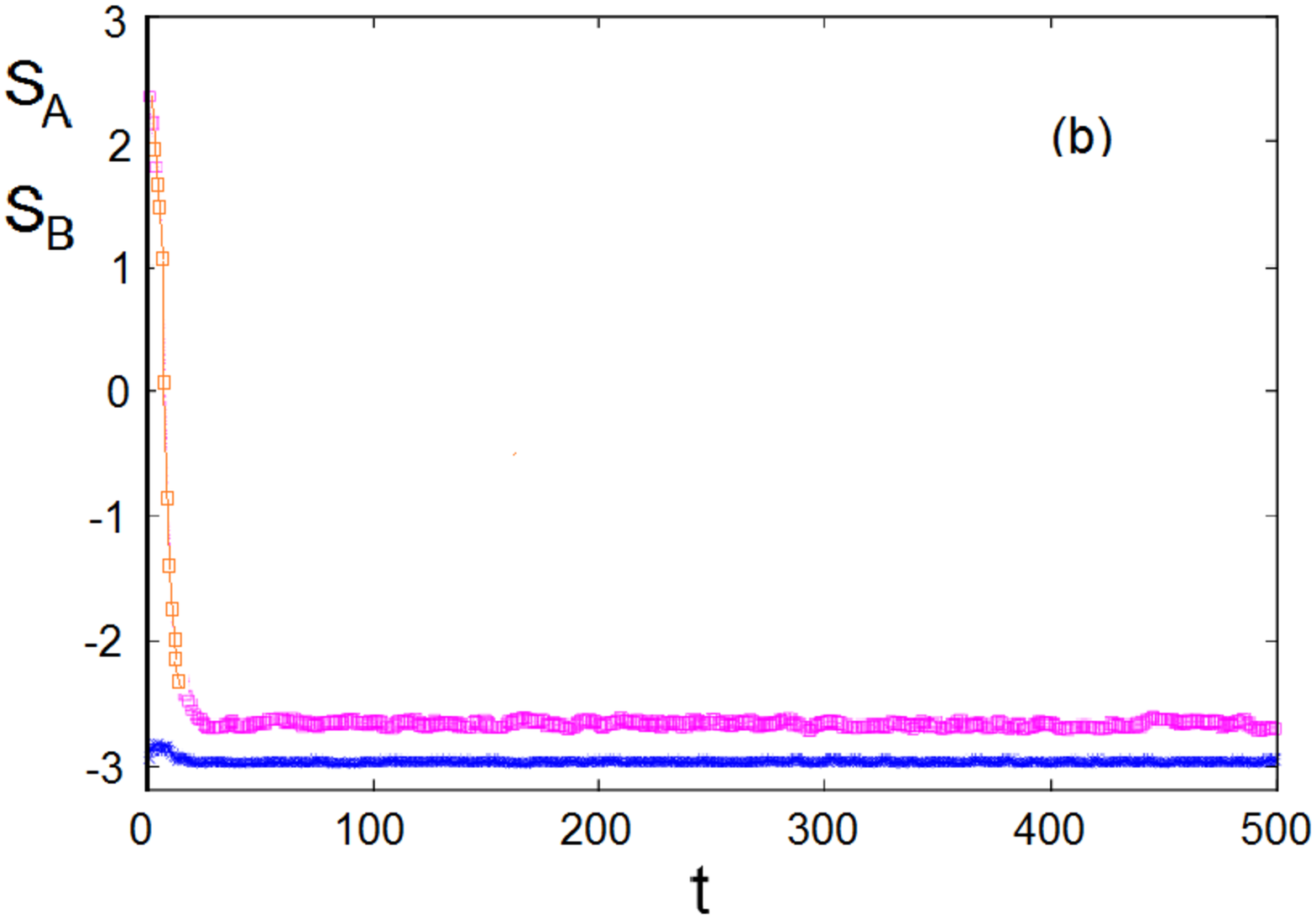}
\caption{Before interaction, Groups $A$ and $B$ with $J_A=J_B=0.02$, $q_A=7$, $q_B=3$ and initial conditions $S_A=-S_B=3$, are shown in  Fig. \ref{ffig11}a.
Curve (a) shows $S_A$ and $S_B$ vs $T$ with interaction $K_{AB}=K_{BA}=0.005$.
 Group $A$ changes sign (attitude) at $T_R\simeq 51$. Curves (b) show time dependence of $S_A$ and $S_B$ at $T=64$, above $T_R$ but below $T_c=169$. \label{ffig13}}
\end{figure}

In the examples above, we considered cases where $-K_{AB}=K_{BA}>0$. If we look at the second term in Eqs. (\ref{eqn:hamil0})-(\ref{eqn:hamil1}) we see that for the signs of these inter-group interactions the  energy of each individual is lowered by the interaction, taking into account the initial conditions. As a consequence, although the interaction softens their stance, each group keeps its stance, namely its sign, up to the critical social temperature (see Figs. \ref{ffig9}b, \ref{ffig11}b).
shows
We show now another case where the interaction of one group with the other group destabilizes its stance, causing a change in its attitude with respect to negotiation.  For this purpose, we take the same sign of $K_{AB}$ and $K_{BA}$, $K_{AB}=K_{BA}>0$. Choosing the initial condition as before, namely $S_A=-S_B=3$, we see from Eqs. (\ref{eqn:hamil0})-(\ref{eqn:hamil1}) that the second term of Group $A$ is positive. This increases the energy of individuals in $A$. As a consequence, we expect that Group $A$ is destabilized and changes its stance (its sign) at some temperature. Group $B$ with negative interaction energy "succeeds" in persuading Group $A$ to negotiate. Let us show the case where $J_A=J_B=0.02$, $q_A=7$, $q_B=3$ and initial conditions $S_A=-S_B=3$. The non-interacting groups have been shown in  Fig. \ref{ffig11}a. The groups in interaction with  $K_{AB}=K_{BA}=0.005$ are shown in Fig. \ref{ffig13}a: Group $A$ changes its mind at $T_R\simeq 51$ from positive to negative. Both become disordered at $T_c\simeq 169$. Note that this value is larger than the value of non-interacting $B$ which is $T_c^0(B)\simeq 151$ (see Fig. \ref{ffig11}a): this means that interaction with the opposite group in this case yields a stability benefit.

\subsection{Asymmetric interaction amplitudes}
\begin{figure}[ht!]
\centering
\includegraphics[width=6cm,angle=0]{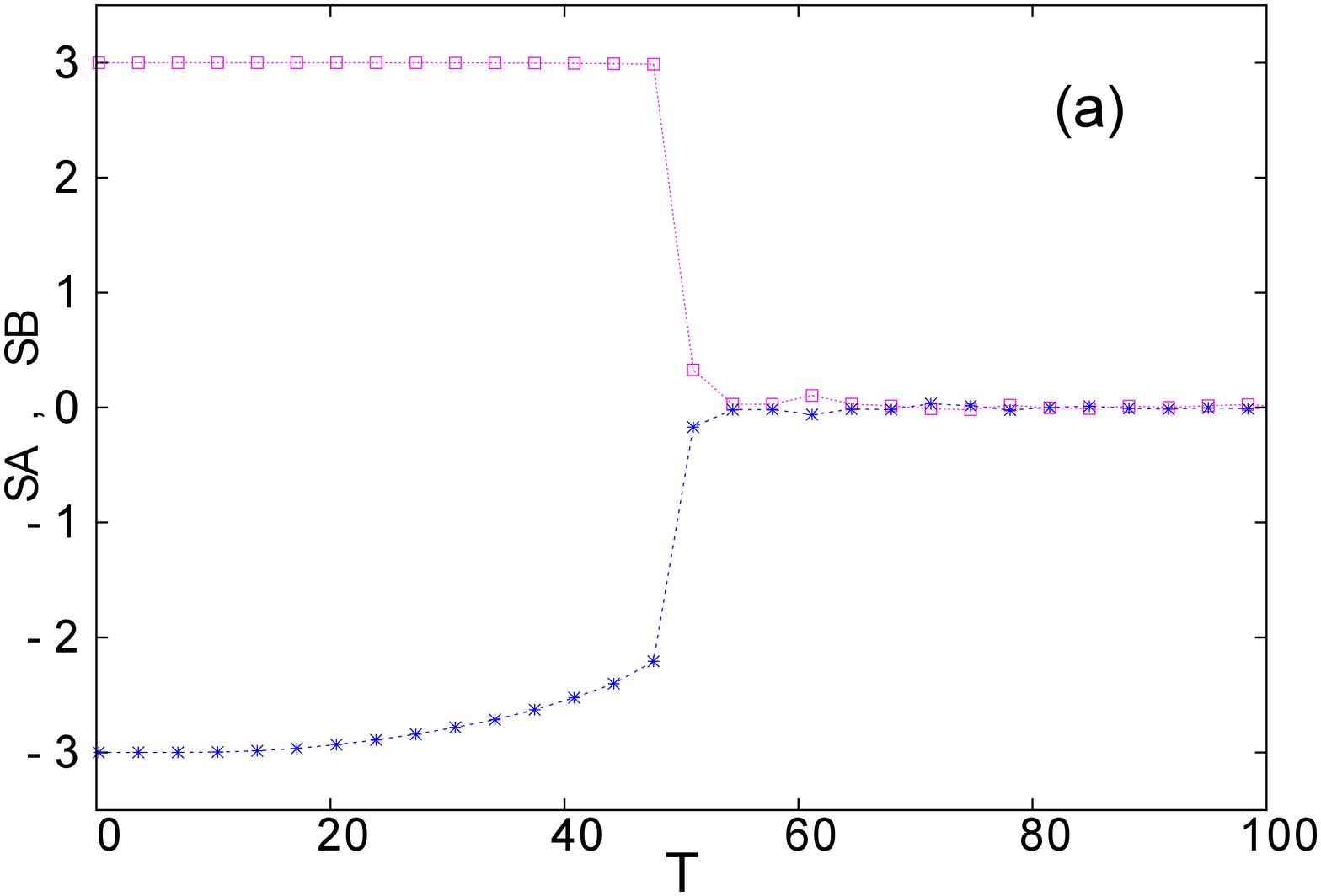}
\includegraphics[width=6cm,angle=0]{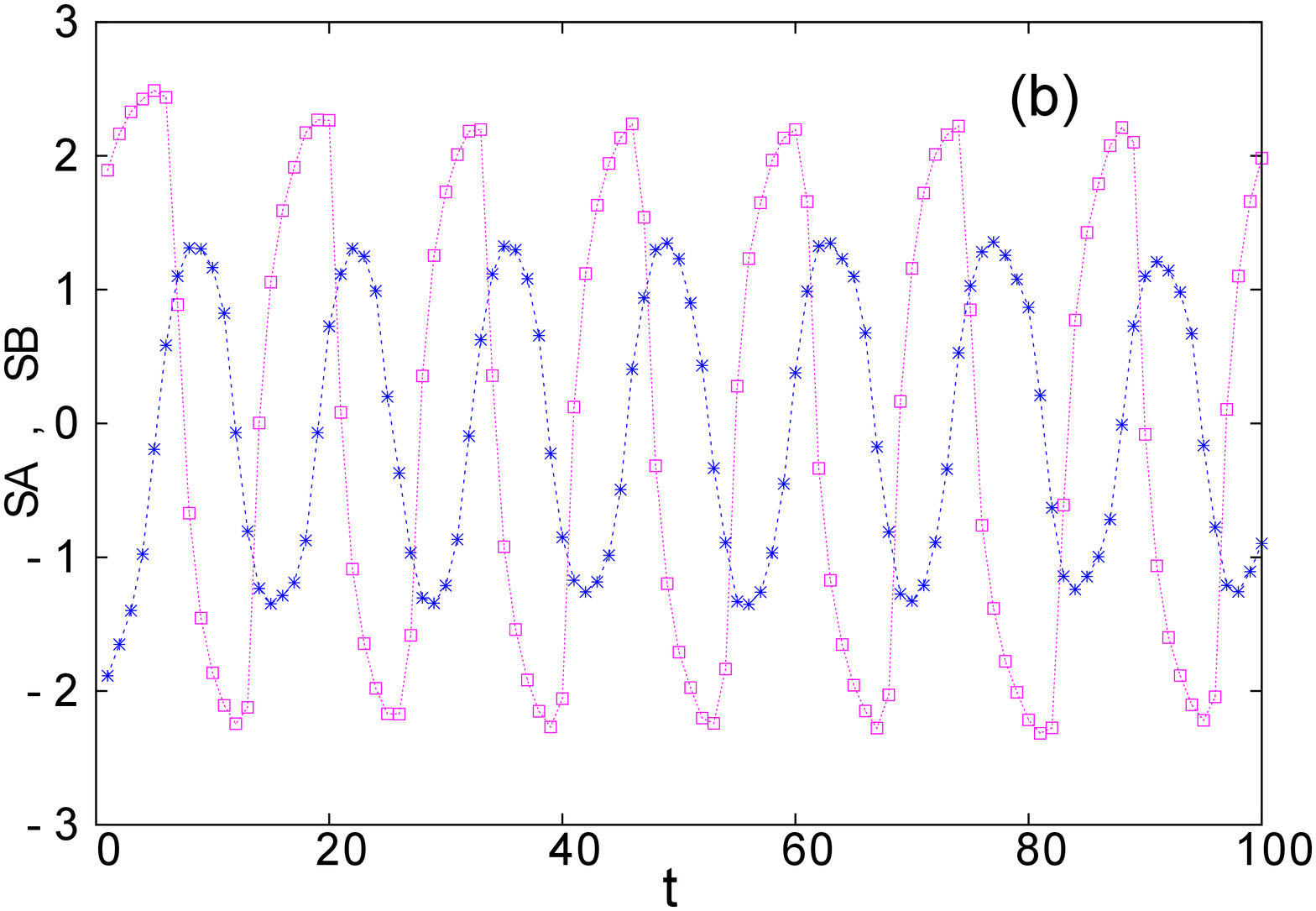}
\includegraphics[width=6cm,angle=0]{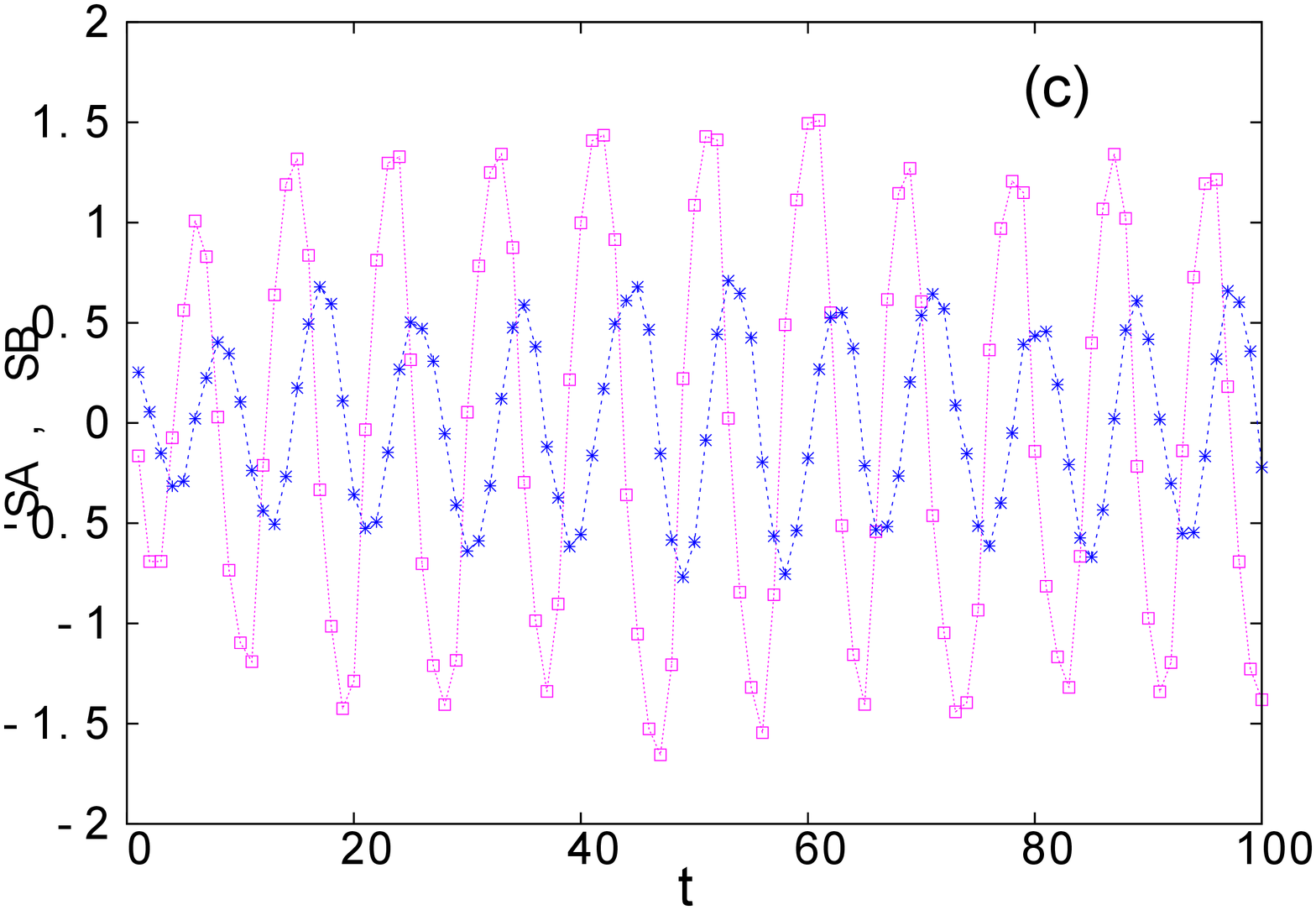}
\includegraphics[width=6cm,angle=0]{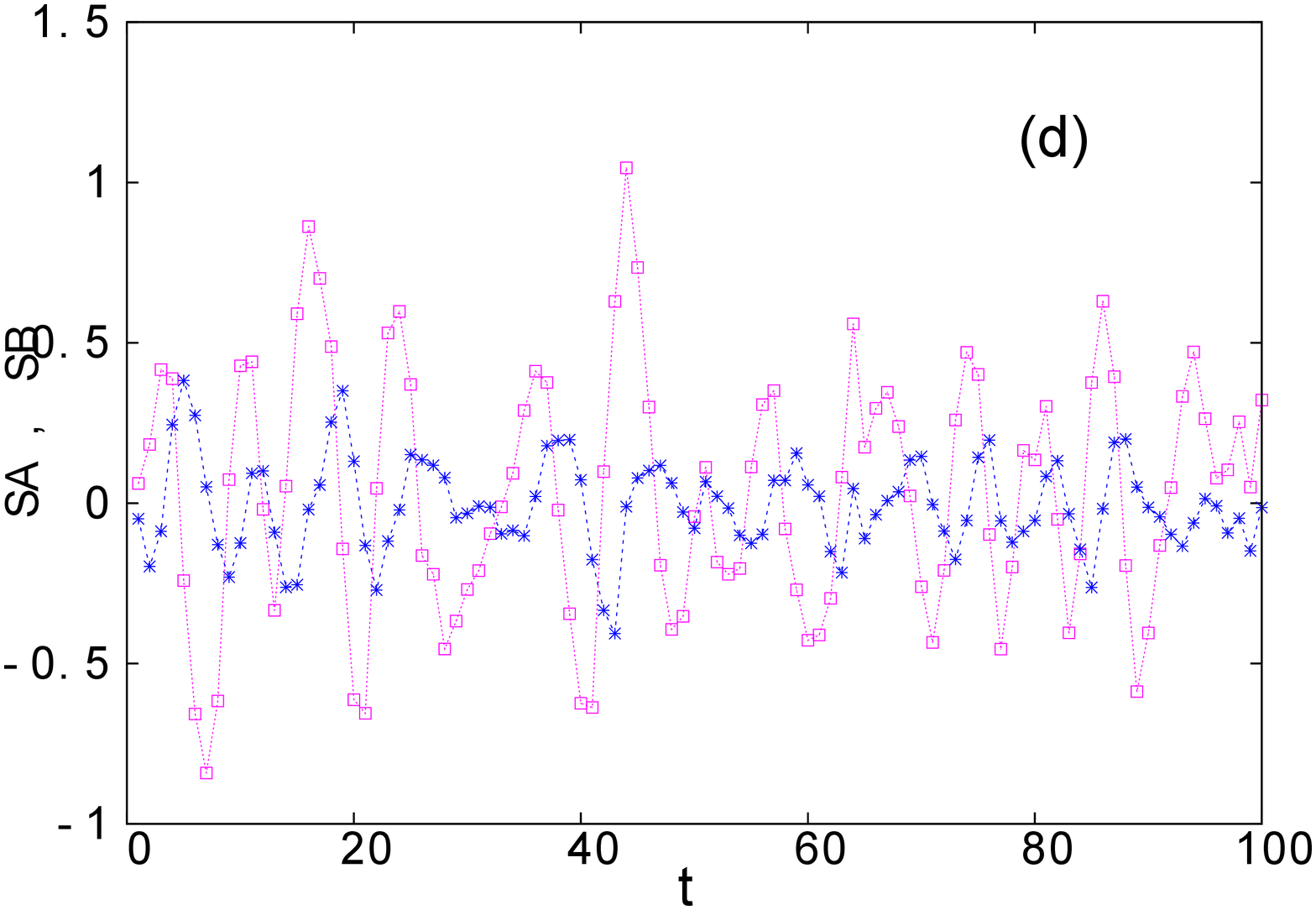}
\caption{Before interaction, Groups $A$ and $B$ with $J_A=J_B=0.02$, $q_A=7$, $q_B=7$ and initial conditions $S_A=-S_B=3$, are shown in  Fig. \ref{ffig9}a with $T_c^0(A,B)=102$. The present figure shows the effect of the asymmetric interactions $K_{AB}=-0.05$, $K_{BA}=0.005$:
 (a) $S_A$ and $S_B$ vs $T$. One has $T_c=51$.
 (b)-(c) time dependence of $S_A$ and $S_B$ at $T=81$ and $T=115$, respectively.
 (d) time dependence of $S_A$ and $S_B$ at $T=166$ above $T_c^0(A,B)$. See text for comments. \label{ffig14}}
\end{figure}

\begin{figure}[ht!]
\centering
\includegraphics[width=6cm,angle=0]{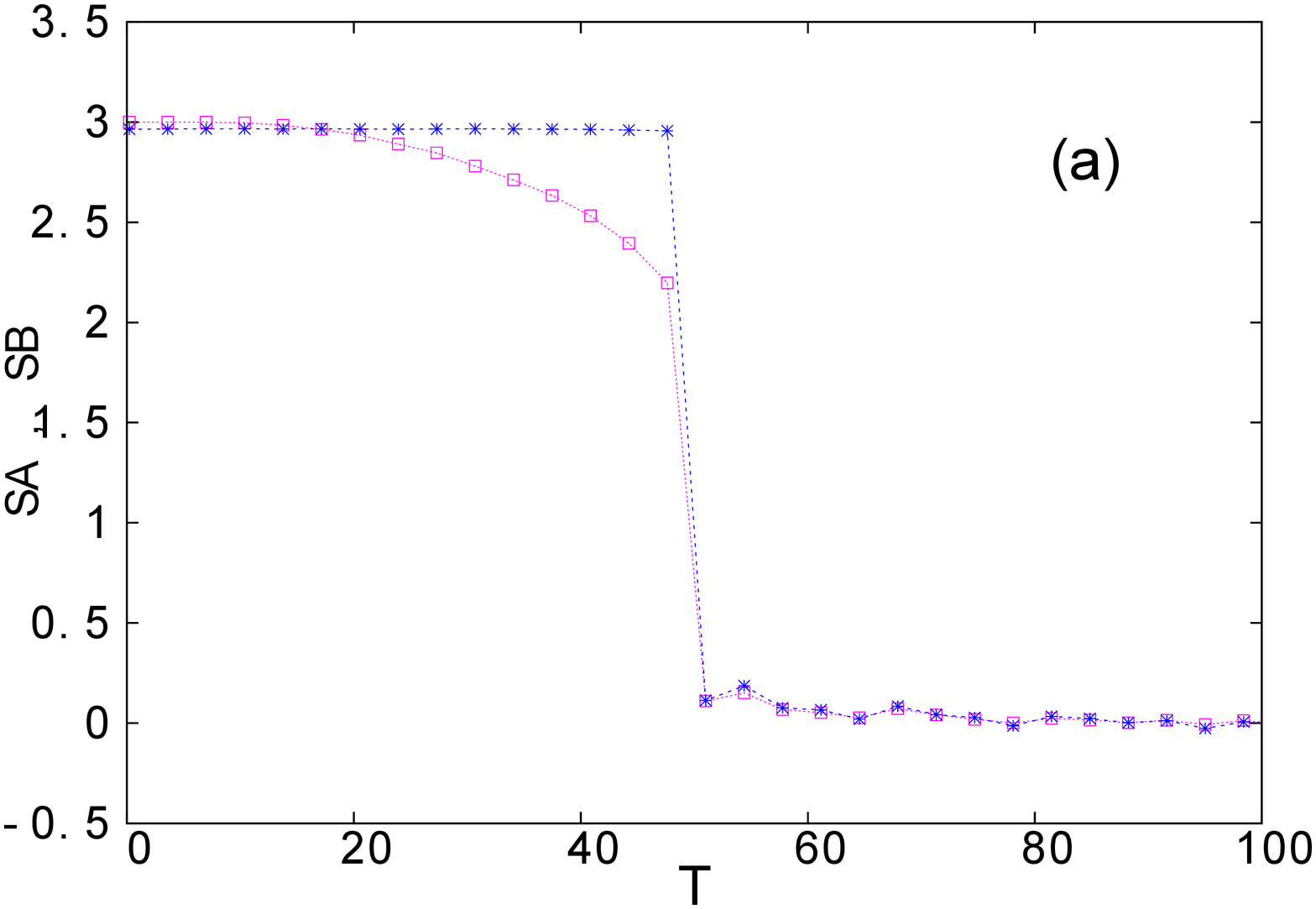}
\includegraphics[width=6cm,angle=0]{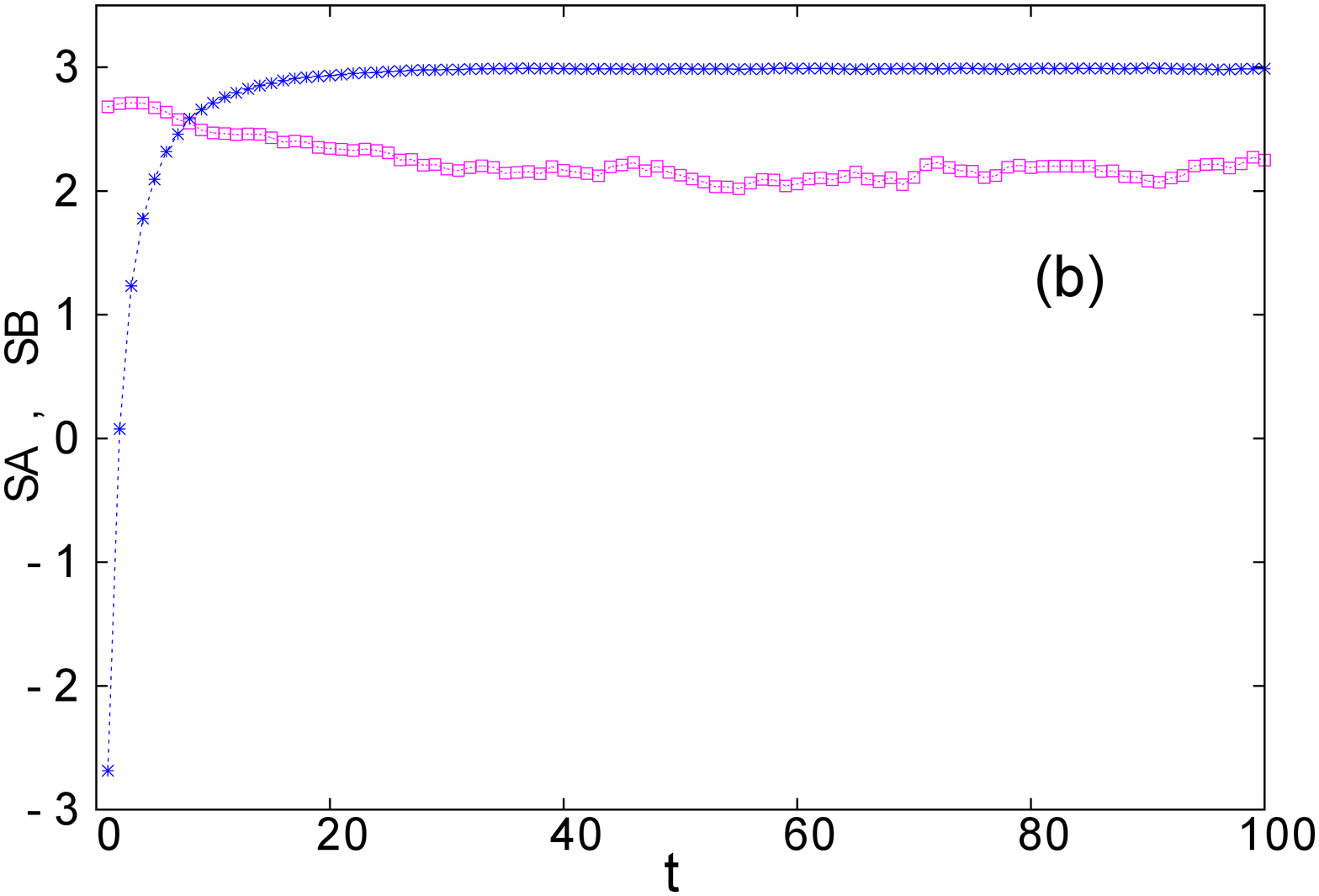}
\includegraphics[width=6cm,angle=0]{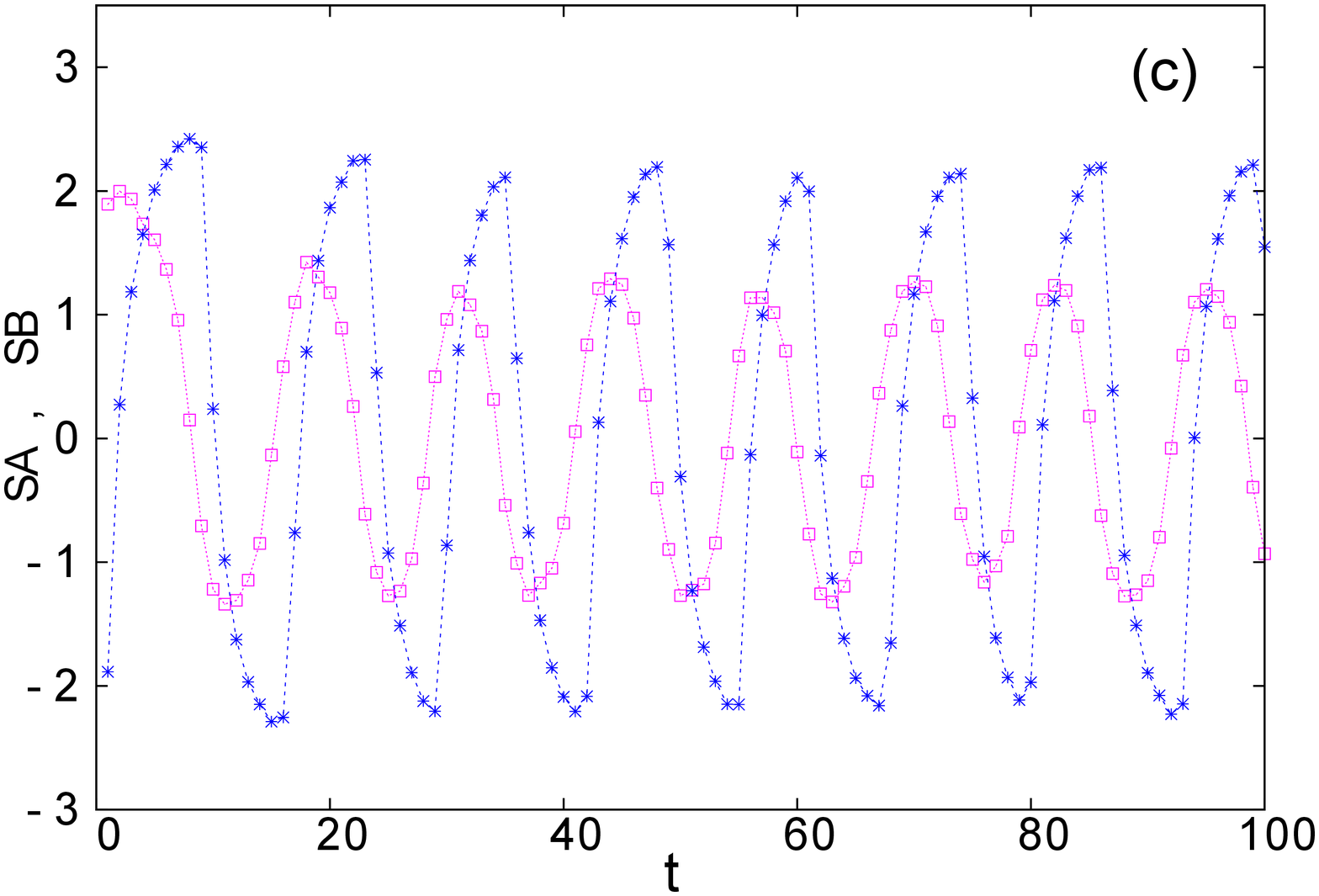}
\includegraphics[width=6cm,angle=0]{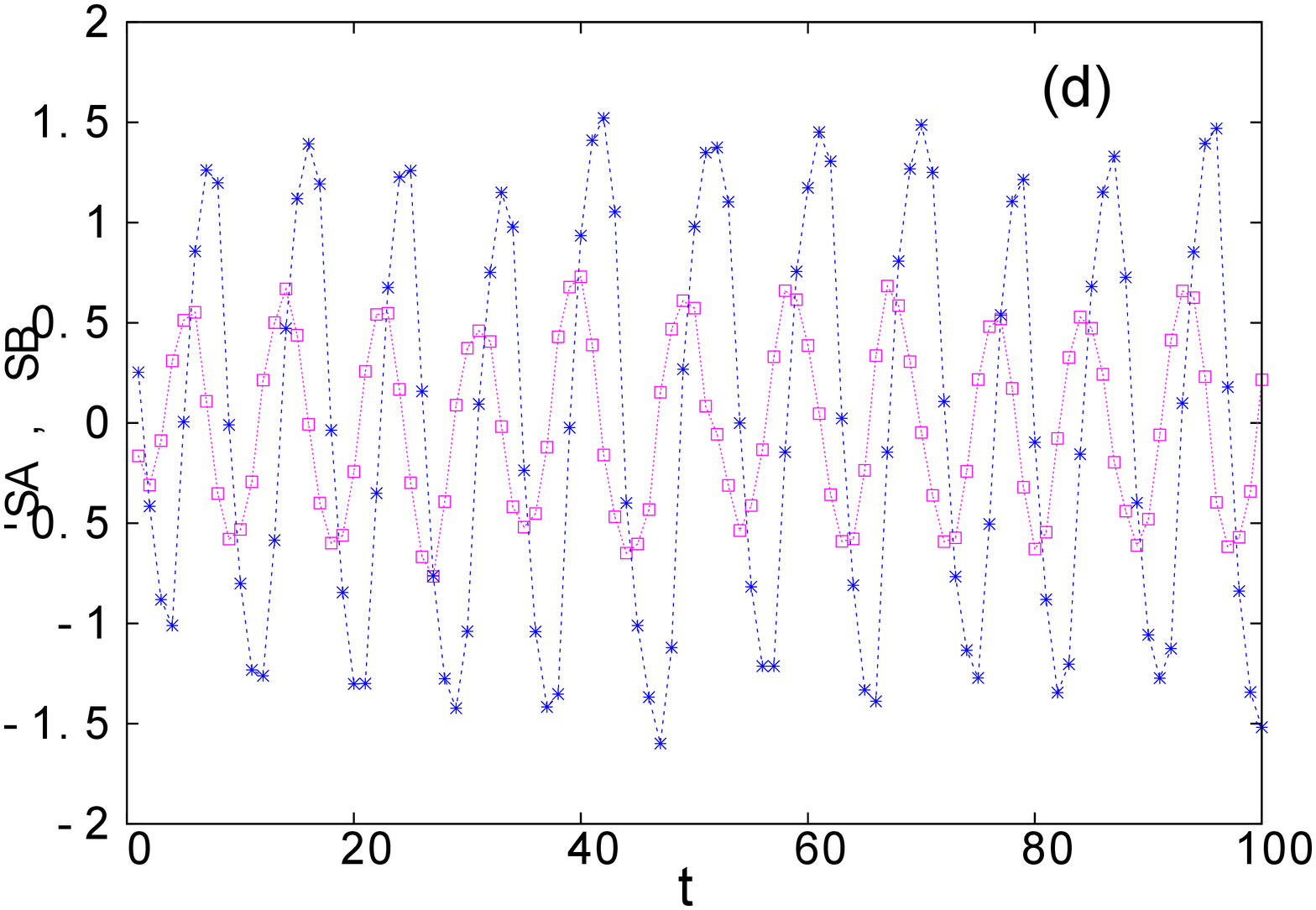}
\caption{The same parameters as those of Fig. \ref{ffig14} except for interactions $K_{AB}=-0.005$, $K_{BA}=0.05$:
 (a) $S_A$ and $S_B$ vs $T$, the critical temperature is $T_c\simeq 51$.
 (b) time dependence of $S_A$ and $S_B$ at $T=47$ below $T_c$.
 (c)-(d) time dependence of $S_A$ and $S_B$ at $T=81$ and  $T=115$, respectively.
 See text for comments. \label{ffig15}}
\end{figure}

Figures \ref{ffig14} and \ref{ffig15} show the case where $K_{AB}$ and $K_{BA}$ have different magnitudes
in addition to their different signs.

In Fig. \ref{ffig14}a, we show a case where $K_{AB}=-0.05$, $K_{BA}=0.005$. The two groups do not change their attitude upon interaction but the critical temperature is reduced from 102 to 51.  Both groups have strong periodic oscillations in time in the temperature range between their new critical temperature $T_c=51$ and their "old" critical temperature $T_c^0(A,B)=102$ before interacting, as seen in Fig. \ref{ffig14}b at $T=81$. This dynamic behavior has been observed above in Fig. \ref{ffig10}b and Fig. \ref{ffig12}b in the respective temperature regions between their $T_c$ and $T_c^0(A,B)$. Note that the oscillations are still present above $T_c^0(A,B)$ but they lose progressively their periodic character both in amplitude and periodicity as seen in Fig. \ref{ffig14}c at $T=115$ and Fig. \ref{ffig14}d at $T=166$.

We reverse now the strength of $K$, namely $K_{AB}=-0.005$, $K_{BA}=0.05$. In Fig. \ref{ffig15}a and Fig. \ref{ffig15}b, Group $B$ changes its attitude from the initial state -3 to positive value upon interaction. The critical temperature remains as that in the previous case, namely $T_c= 51$. As before, between $T_c$ and $T_c^0(A,B)$ we expect perfect periodic oscillations of $S_A$ and $S_B$. This is indeed observed in Fig. \ref{ffig15}c at $T=81$. Above $T_c^0(A,B)$ we lose the regularity of the oscillations progressively as seen in Fig. \ref{ffig15}d at $T=115$, as in previous cases examined so far.

\subsection{Effect of intra-group interactions}

\begin{figure}[ht!]
\centering
\includegraphics[width=6cm,angle=0]{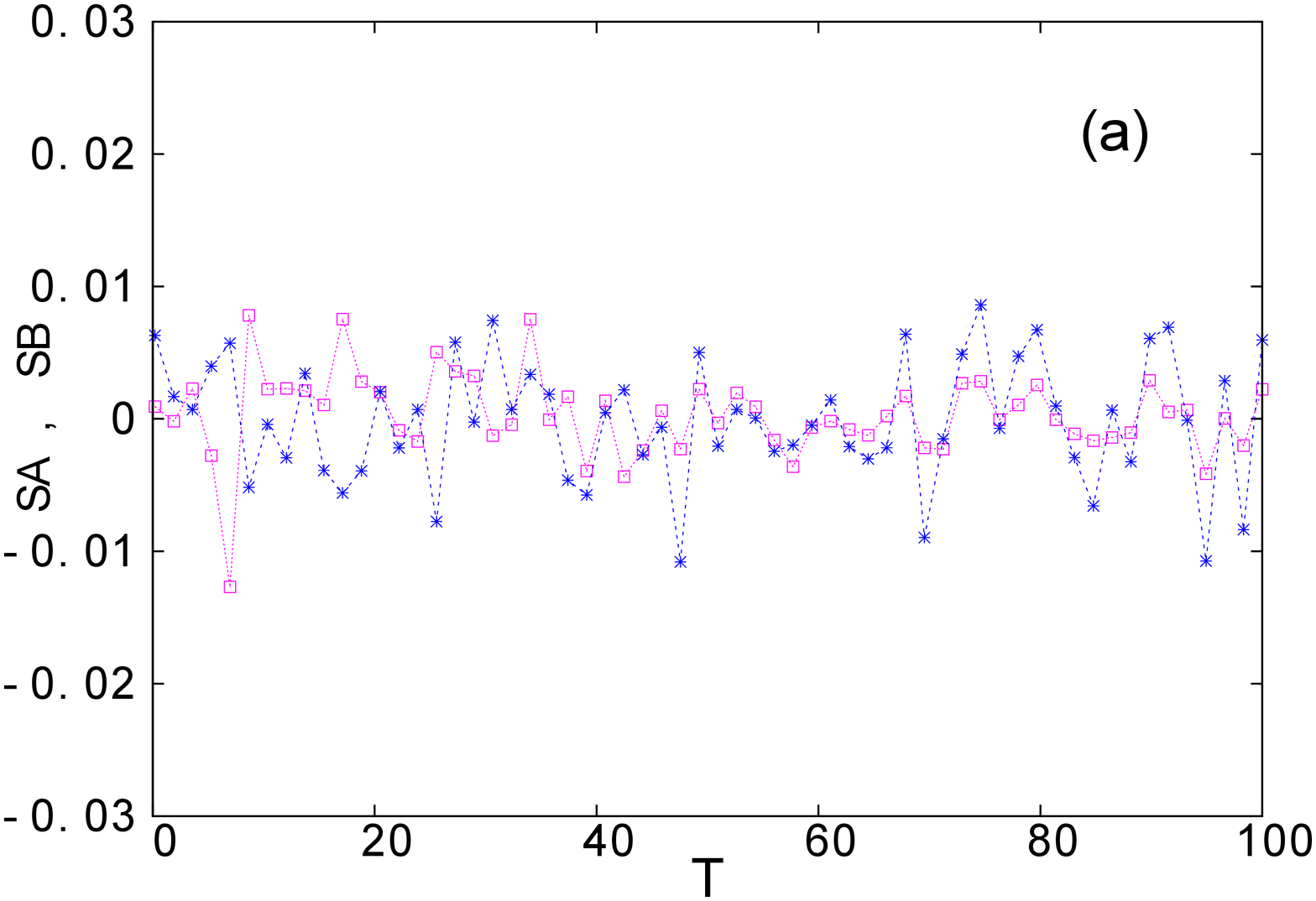}
\includegraphics[width=6cm,angle=0]{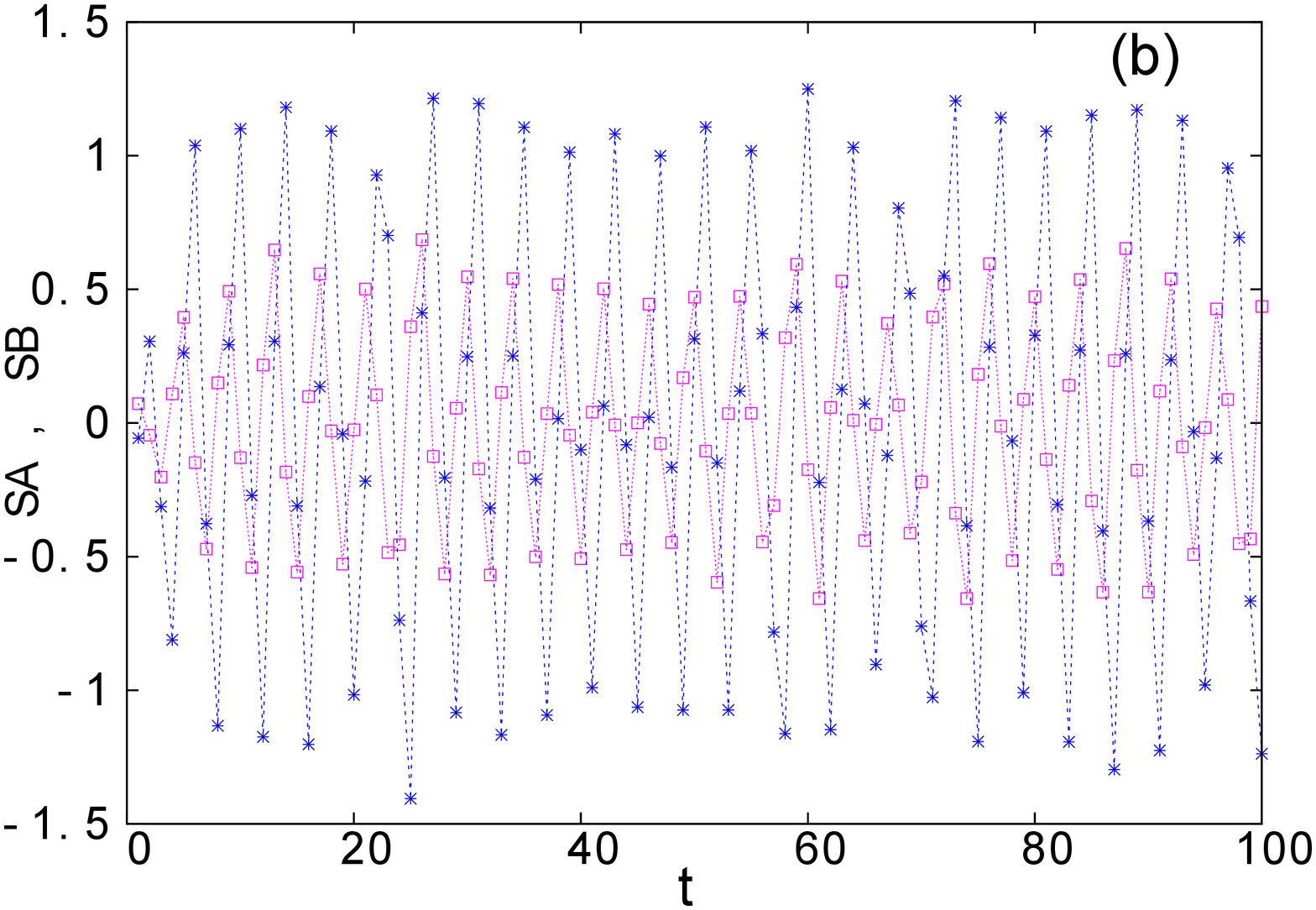}
\includegraphics[width=6cm,angle=0]{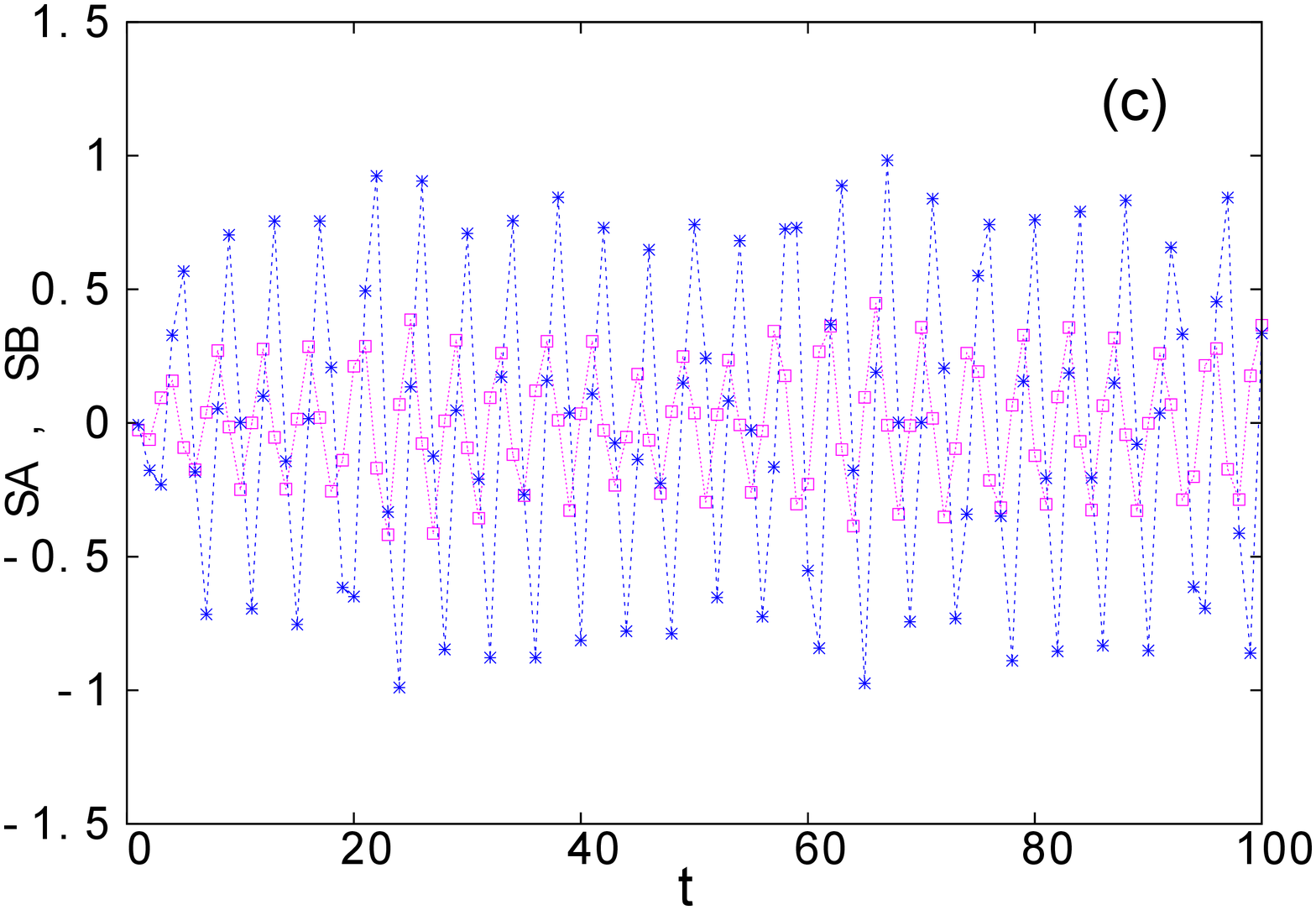}
\includegraphics[width=6cm,angle=0]{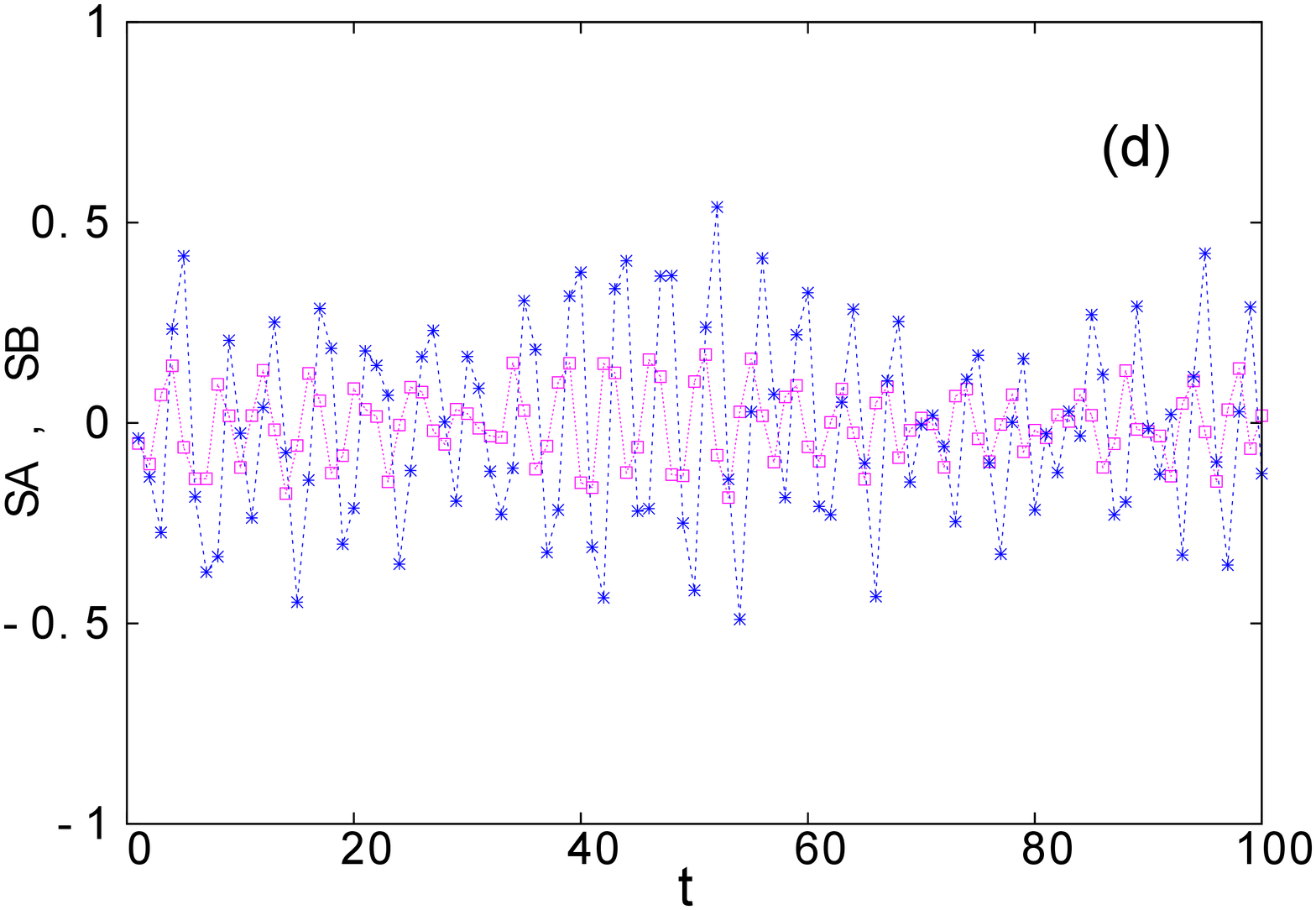}
\caption{Before interaction, Groups $A$ and $B$ with $J_A=J_B=0$, $q_A=7$, $q_B=7$ and initial conditions $S_A=-S_B=3$, have no order at any $T$.
 (a) $S_A$ and $S_B$ vs $T$ with interaction $K_{AB}=-0.005$, $K_{BA}=0.05$, there is no order at any $T$.
 (b)-(c)-(d) $S_A$ and $S_B$ vs time $t$ at $T=41$, 58 and 83, respectively. See text for comments. \label{ffig16}}
\end{figure}

In the cases shown above We have assumed the same value for $J_A$ and $J_B$. For completeness, we examine next the effect of $J_A$ and $J_B$.  If  $J_A$ and $J_B$ are both equal to  zero, then the group have no order at any $T$. When the interaction is turned on, they remain disordered as seen in Fig. \ref{ffig16}a. As seen in all cases above, in the disordered phase, $S_A$ and $S_B$ fluctuate strongly with time and progressively losing the periodic character with increasing $T$ as seen in Fig. \ref{ffig16}.

\begin{figure}[ht!]
\centering
\includegraphics[width=6cm,angle=0]{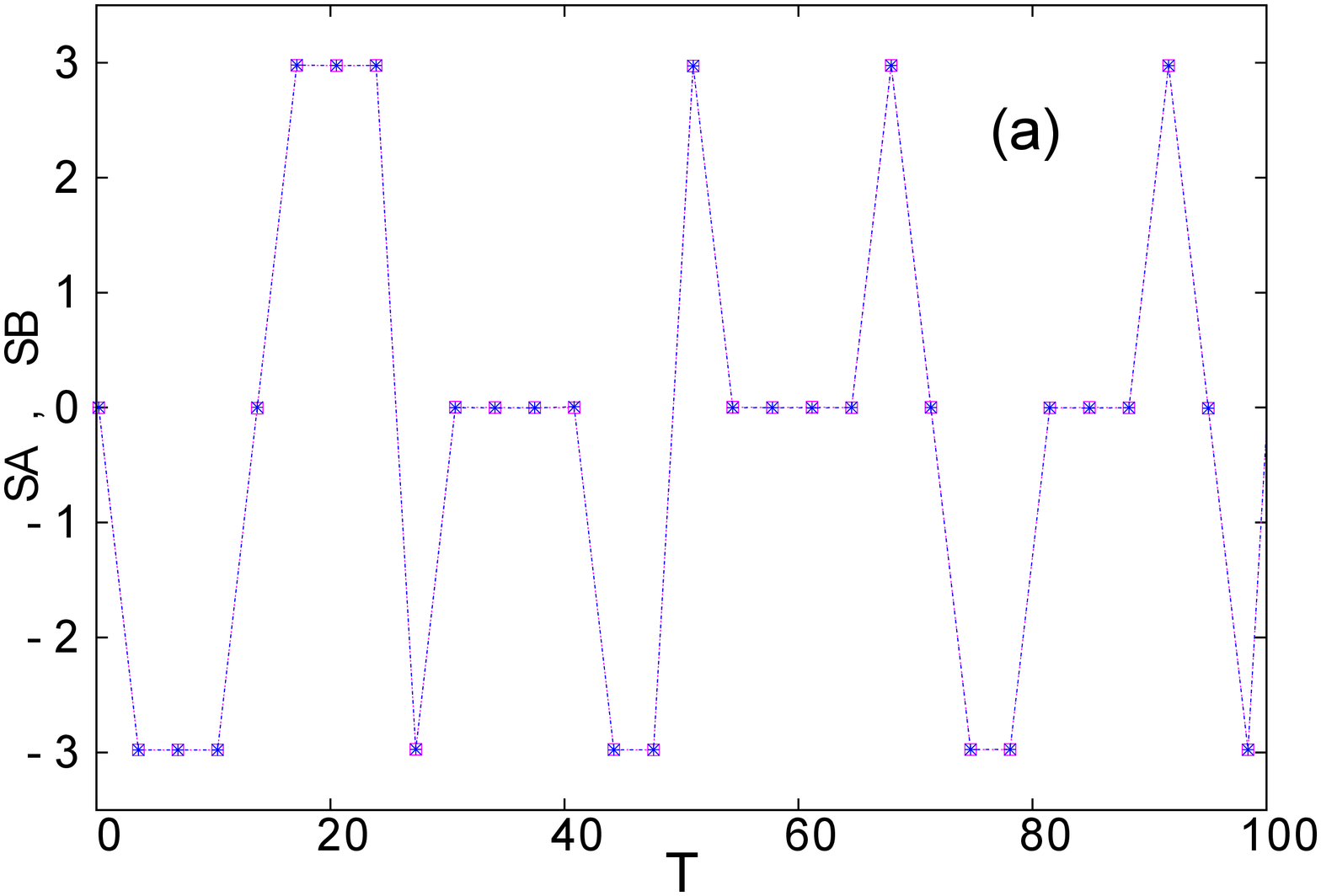}
\includegraphics[width=6cm,angle=0]{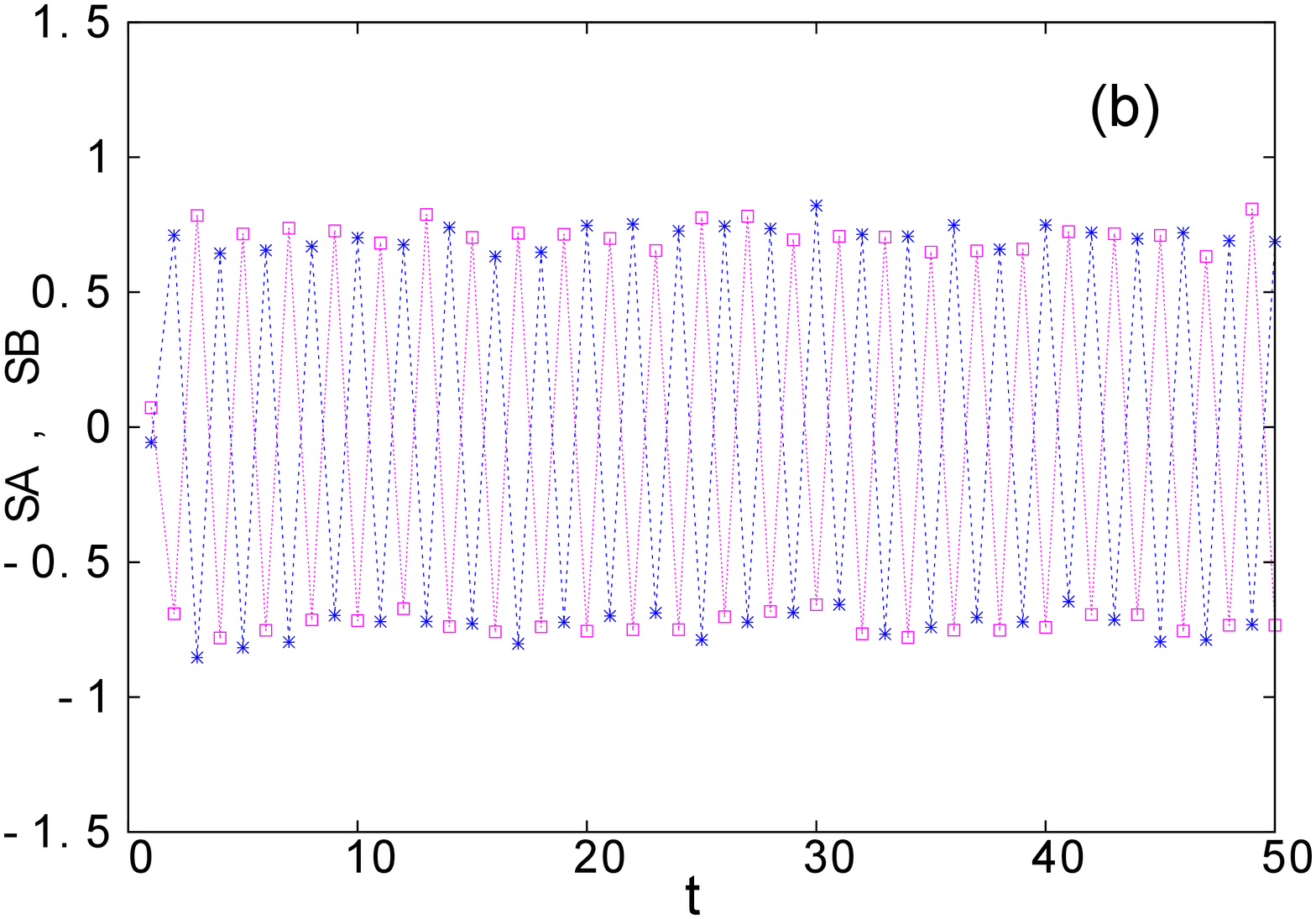}
\includegraphics[width=6cm,angle=0]{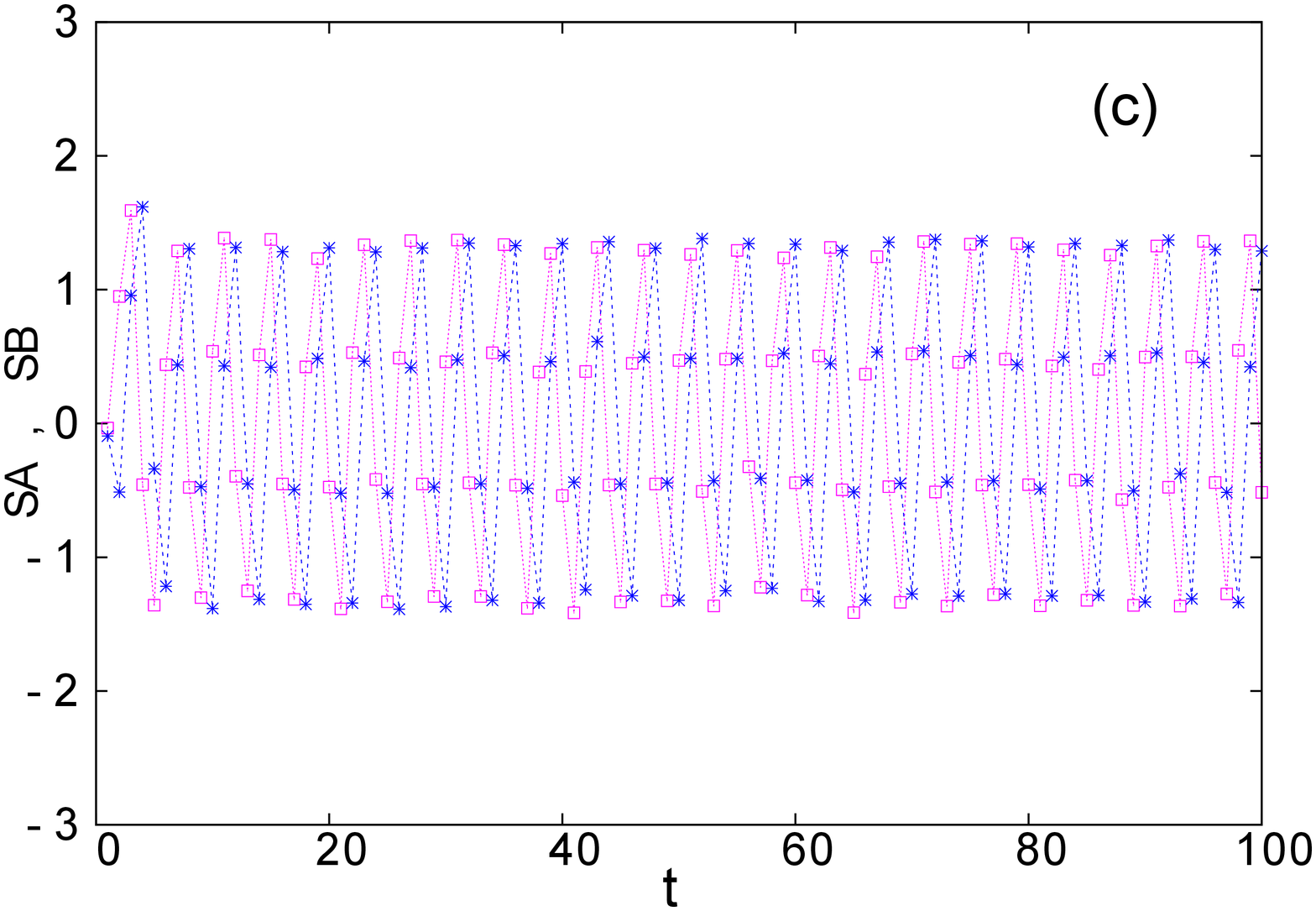}
\includegraphics[width=6cm,angle=0]{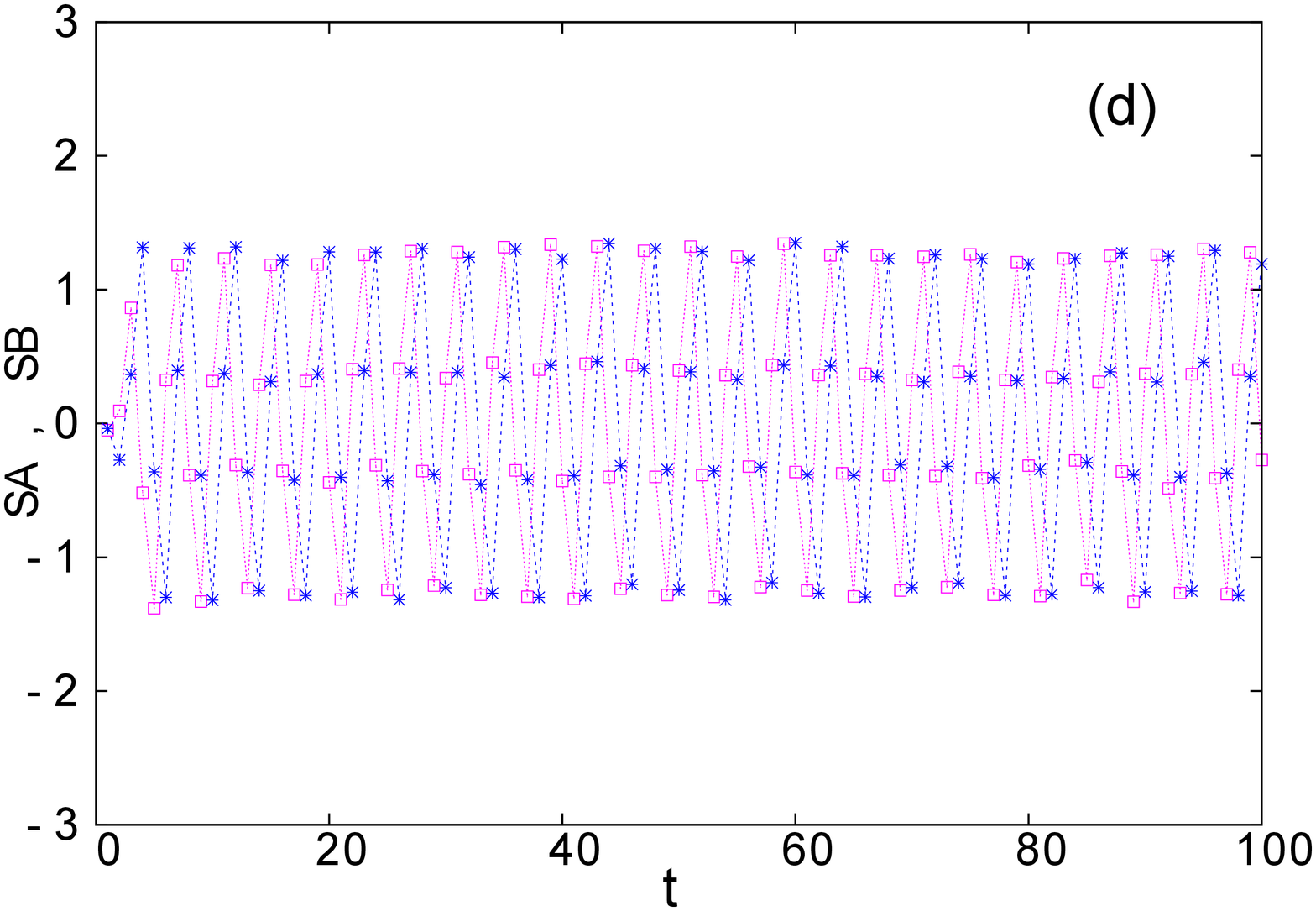}
\caption{Before interaction, Groups $A$ and $B$ with $J_A=J_B=0$, $q_A=7$, $q_B=7$ and initial conditions $S_A=-S_B=3$, have no order.
 (a) $S_A$ and $S_B$ vs $T$ with strong interaction $K_{AB}=K_{BA}=0.3$. There is no order at any $T$ but chaotic oscillations are seen between the two limits $\pm 3$ due to strong $K$.
 (b) $S_A$ and $S_B$ at $T=41$ for $K_{AB}=K_{BA}=0.3$. This oscillation behavior is observed for all $T$ unlike the weak $K$.
 (c)-(d) $S_A$ and $S_B$ vs $T$ with strong interaction $-K_{AB}=K_{BA}=\pm 0.2$ at $T=41$ and $T=166$. See text for comments. \label{ffig16bis}}
\end{figure}

We show in Fig. \ref{ffig16bis} the strong $K$ limit: with $K_{AB}=K_{BA}$ the groups oscillate with $T$ between $\pm 3$ (Fig. \ref{ffig16bis}a). At a given $T$, $S_A$ and $S_B$ oscillate periodically with time  (Fig. \ref{ffig16bis}b) as in the case of weak $K$ shown in Fig. \ref{ffig16}b.  Note however that these oscillations persist for all $T$ unlike the case of weak $K$.
With $-K_{AB}= K_{BA}$ (opposite signs), the two groups behave in the same manner with time (Fig. \ref{ffig16bis}c, Fig. \ref{ffig16bis}d) but $S_A$ and $S_B$ remain zero for all temperatures (not shown).

\begin{figure}[ht!]
\centering
\includegraphics[width=6cm,angle=0]{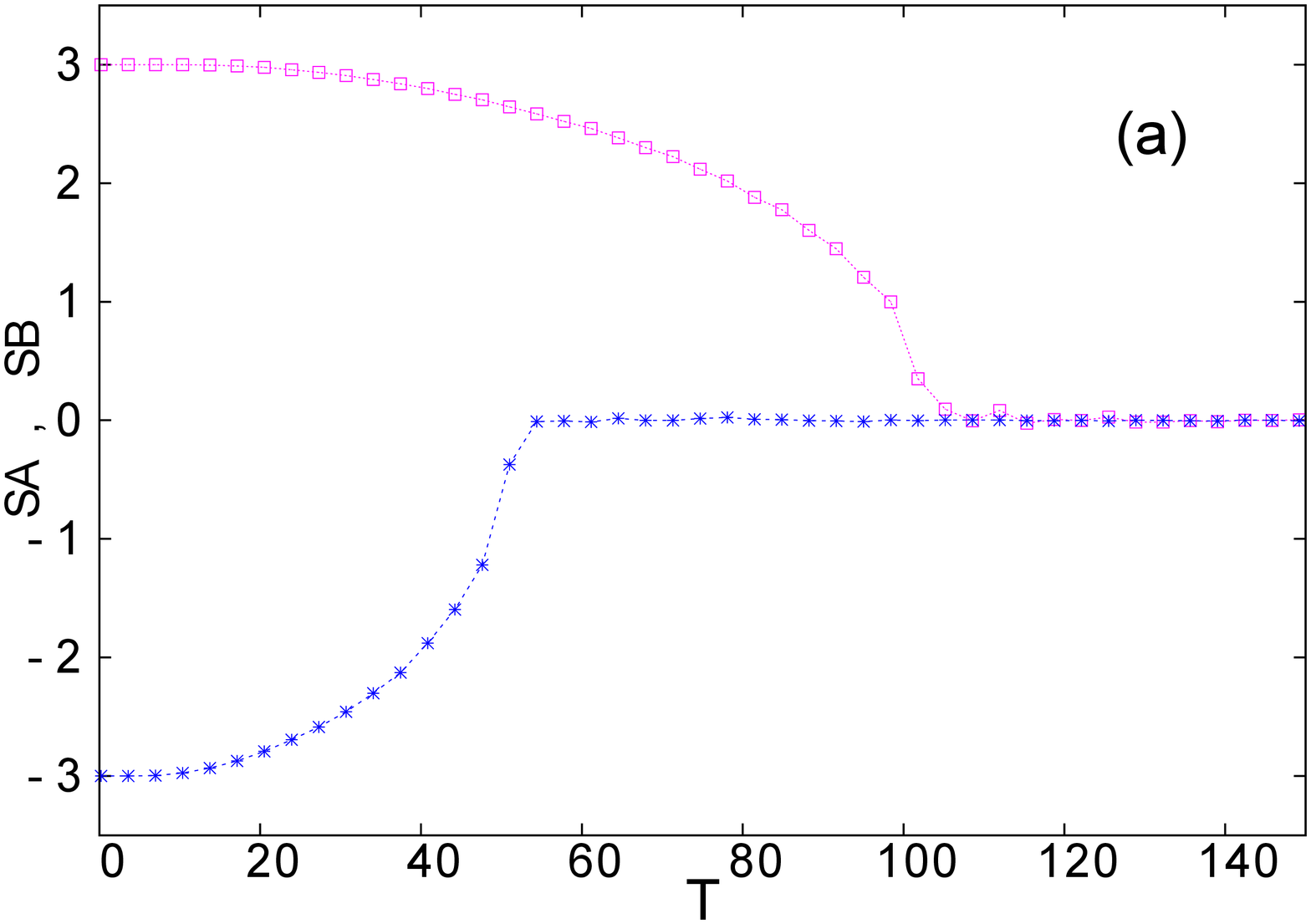}
\includegraphics[width=6cm,angle=0]{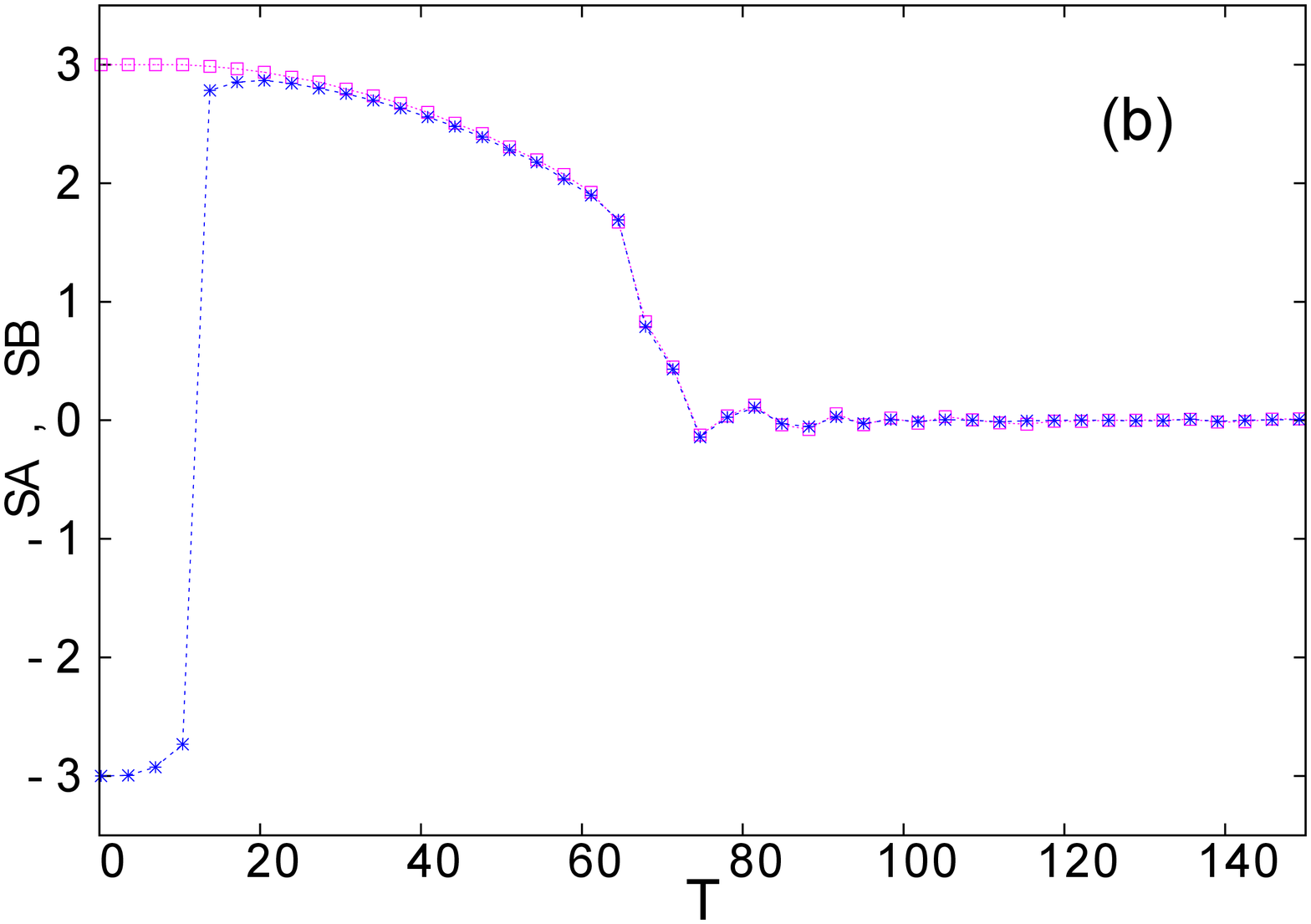}
\caption{Groups $A$ and $B$ with $J_A=0.02$, $J_B=0.01$, $q_A=7$, $q_B=7$,  and initial conditions $S_A=-S_B=3$.
(a) Without interaction: $S_A$ and $S_B$ vs $T$, one has $T_c^0(A)=102$, $T_c^0(B)=51$.
(b) With interaction $-K_{AB}=K_{BA}=0.005$: $S_A$ and $S_B$ vs $T$, one has $T_c\simeq 70$. Group $B$ changes its attitude.
 See text for comments. \label{ffig17}}
\end{figure}

\begin{figure}[ht!]
\centering
\includegraphics[width=6cm,angle=0]{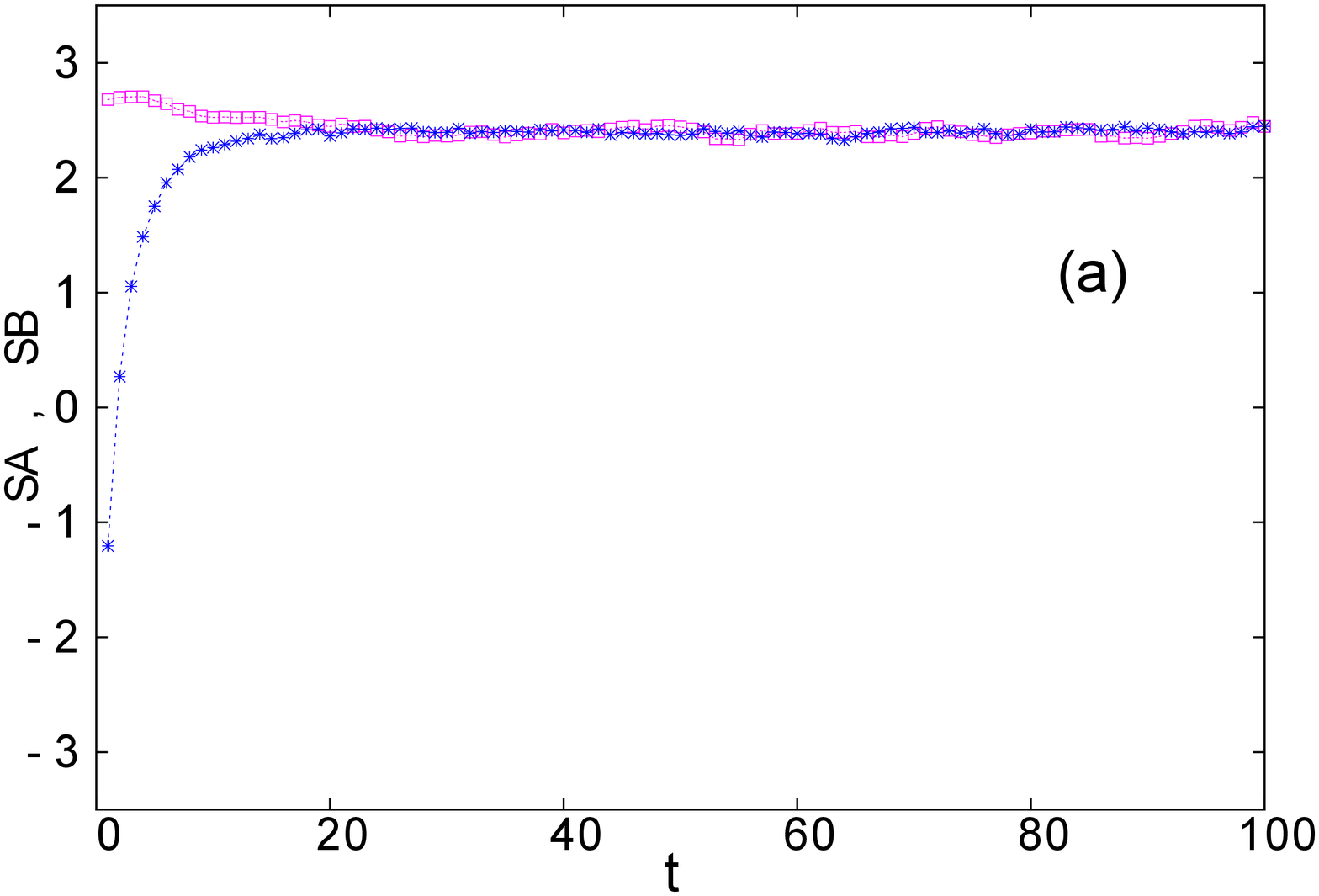}
\includegraphics[width=6cm,angle=0]{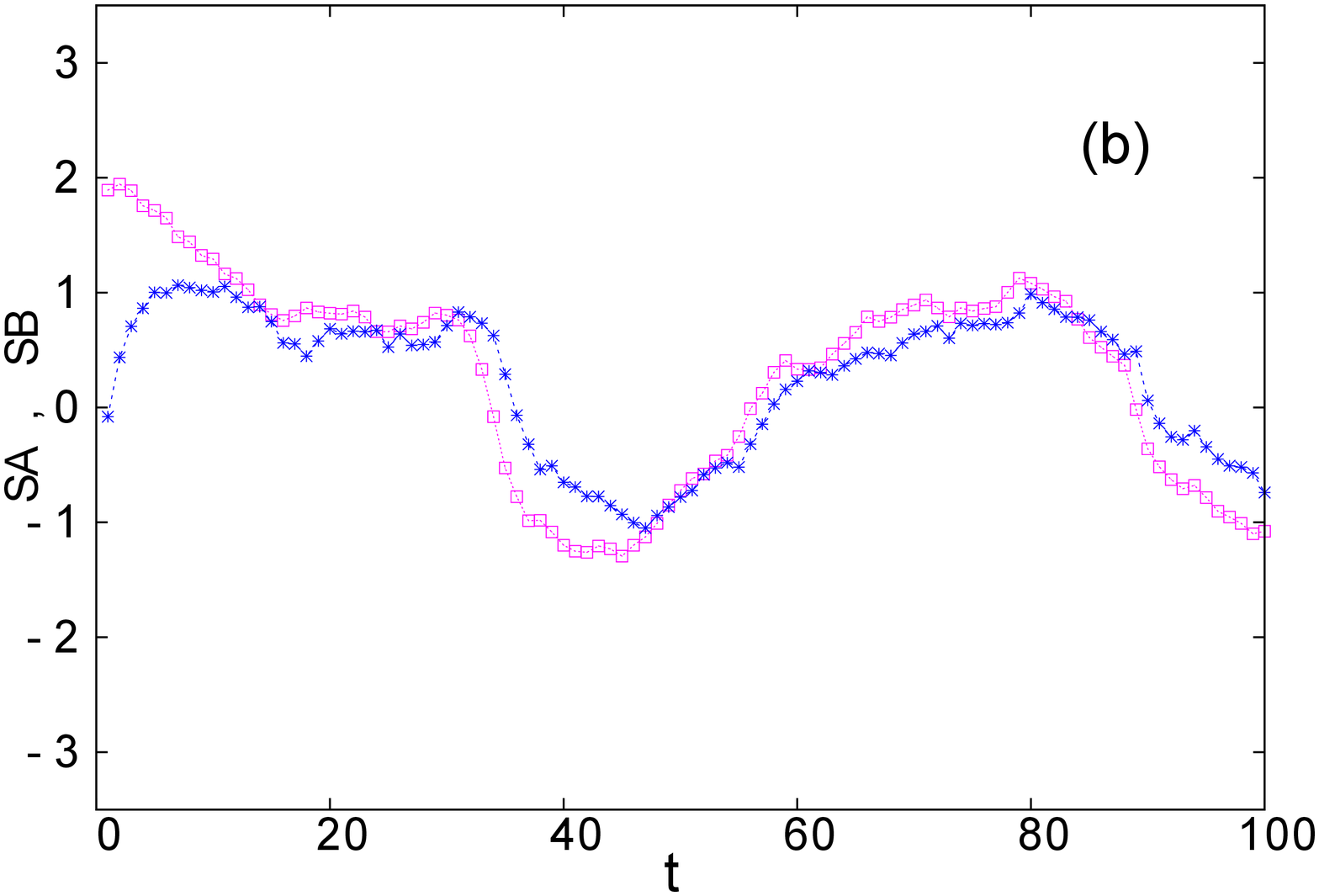}
\includegraphics[width=6cm,angle=0]{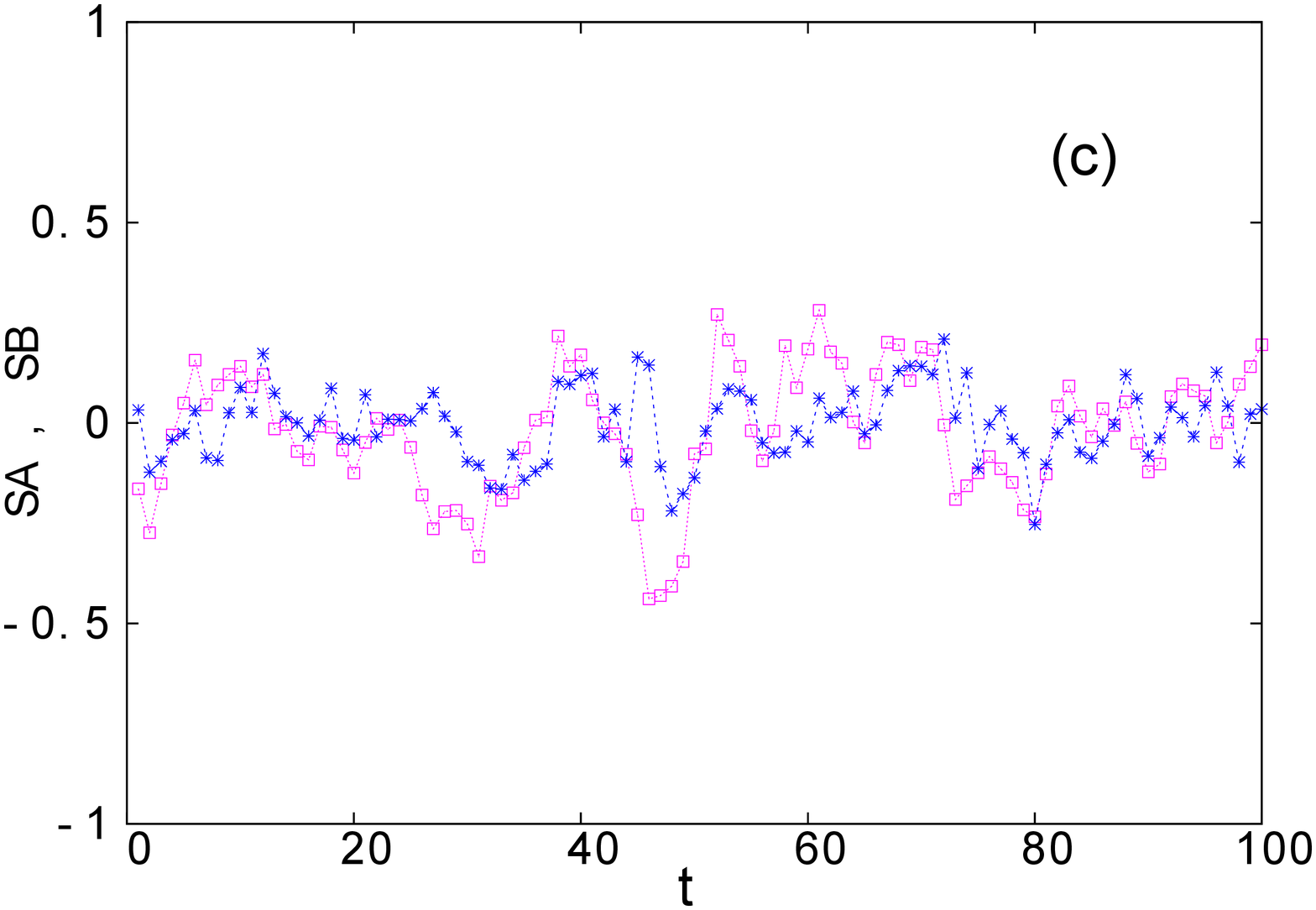}
\includegraphics[width=6cm,angle=0]{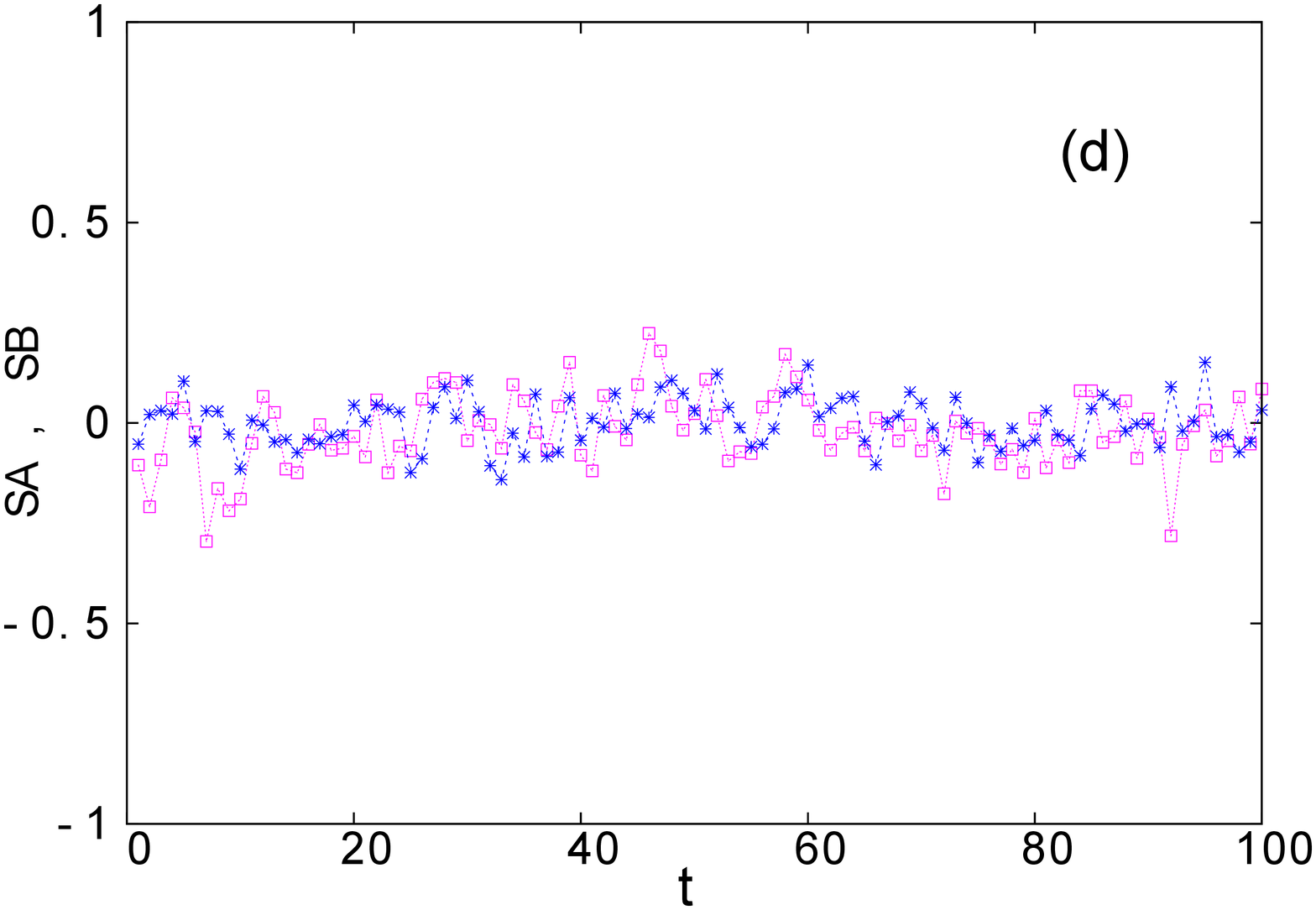}
\caption{Dynamics of the case shown in Fig. \ref{ffig17}:
(a)-(b)-(c)-(d) show $S_A$ and $S_B$ vs time $t$ at $T=47$ (below $T_c=70$), 81, 115 and 166, respectively.
 See text for comments. \label{ffig18}}
\end{figure}

Figure \ref{ffig17} shows the case where $J_A\neq J_B$. Without interaction $K$, the asymmetric interactions give rise to $T_c^0(A)\neq T_c^0(B)$ as seen in Fig. \ref{ffig17}. As expected, when the interaction is taken into account, both groups have a critical in-between temperature . Dynamics of $S_A$ and $S_B$ are shown in Fig. \ref{ffig18} in the case
$-K_{AB}=K_{BA}=0.005$ where there is a change of sign of $S_B$. It is interesting
to note that there are very slow but aperiodic oscillations at $T=81$ below $T_c^0(A)=102$ (Fig. \ref{ffig18}b).

\begin{figure}[ht!]
\centering
\includegraphics[width=6cm,angle=0]{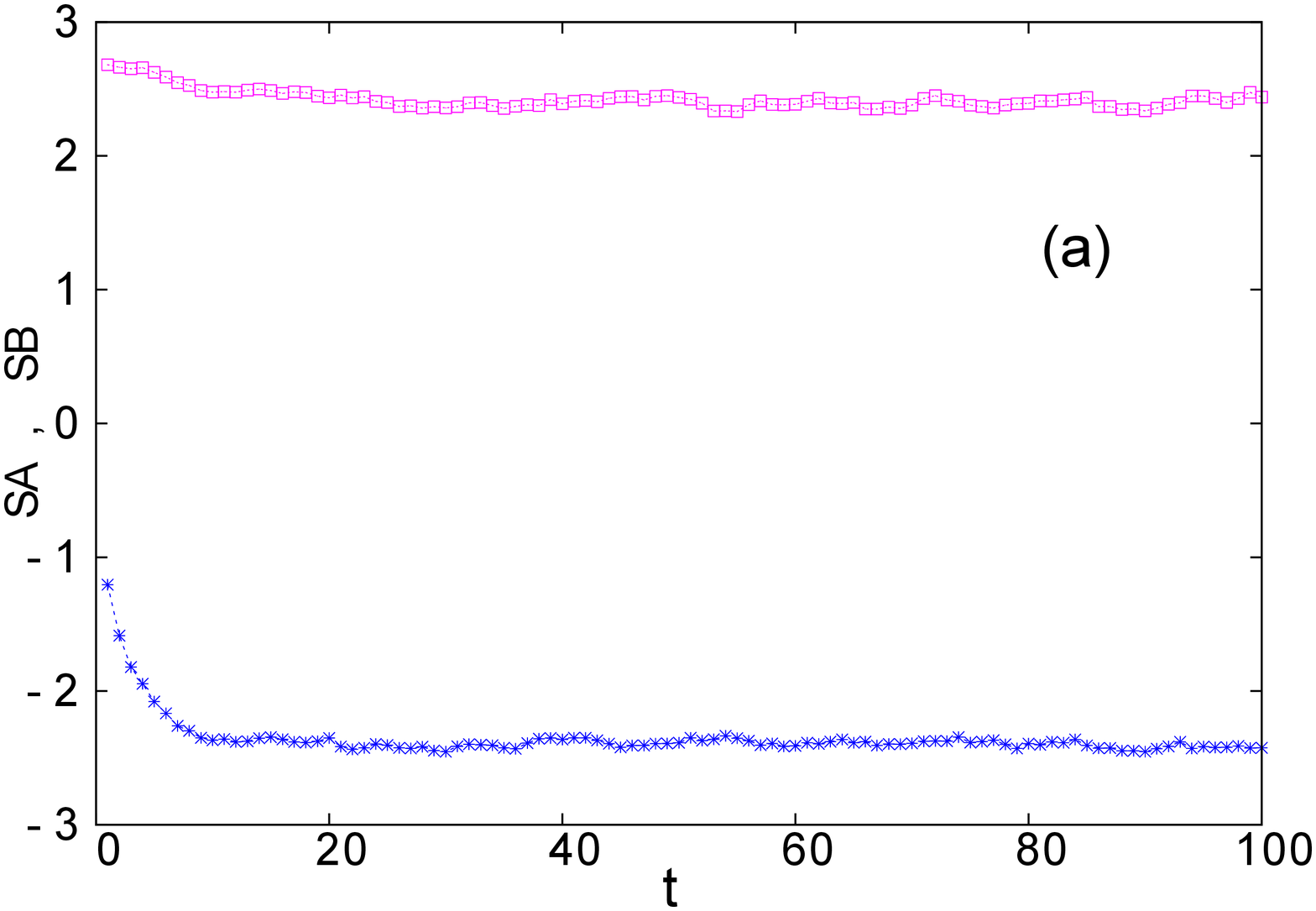}
\includegraphics[width=6cm,angle=0]{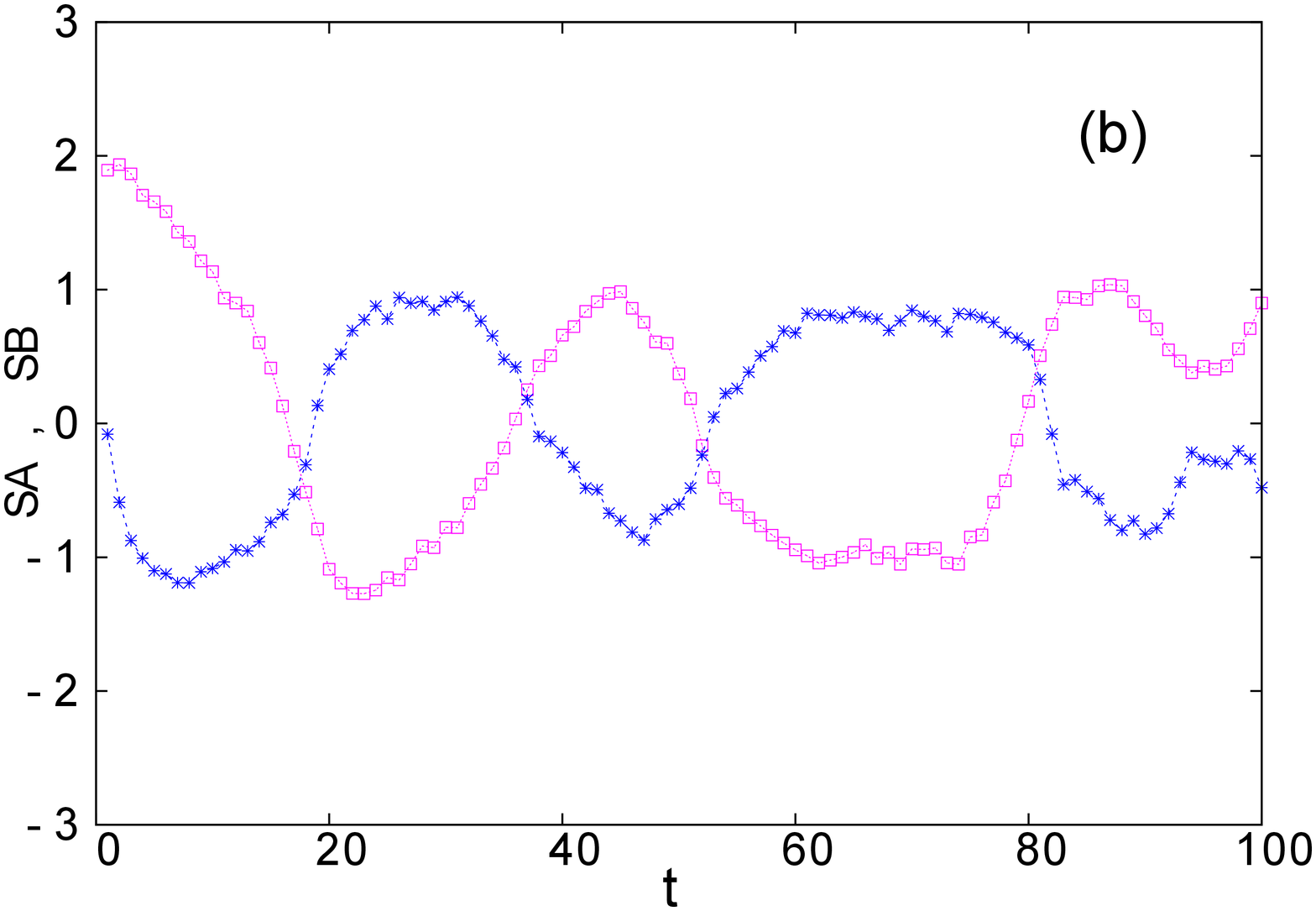}
\includegraphics[width=6cm,angle=0]{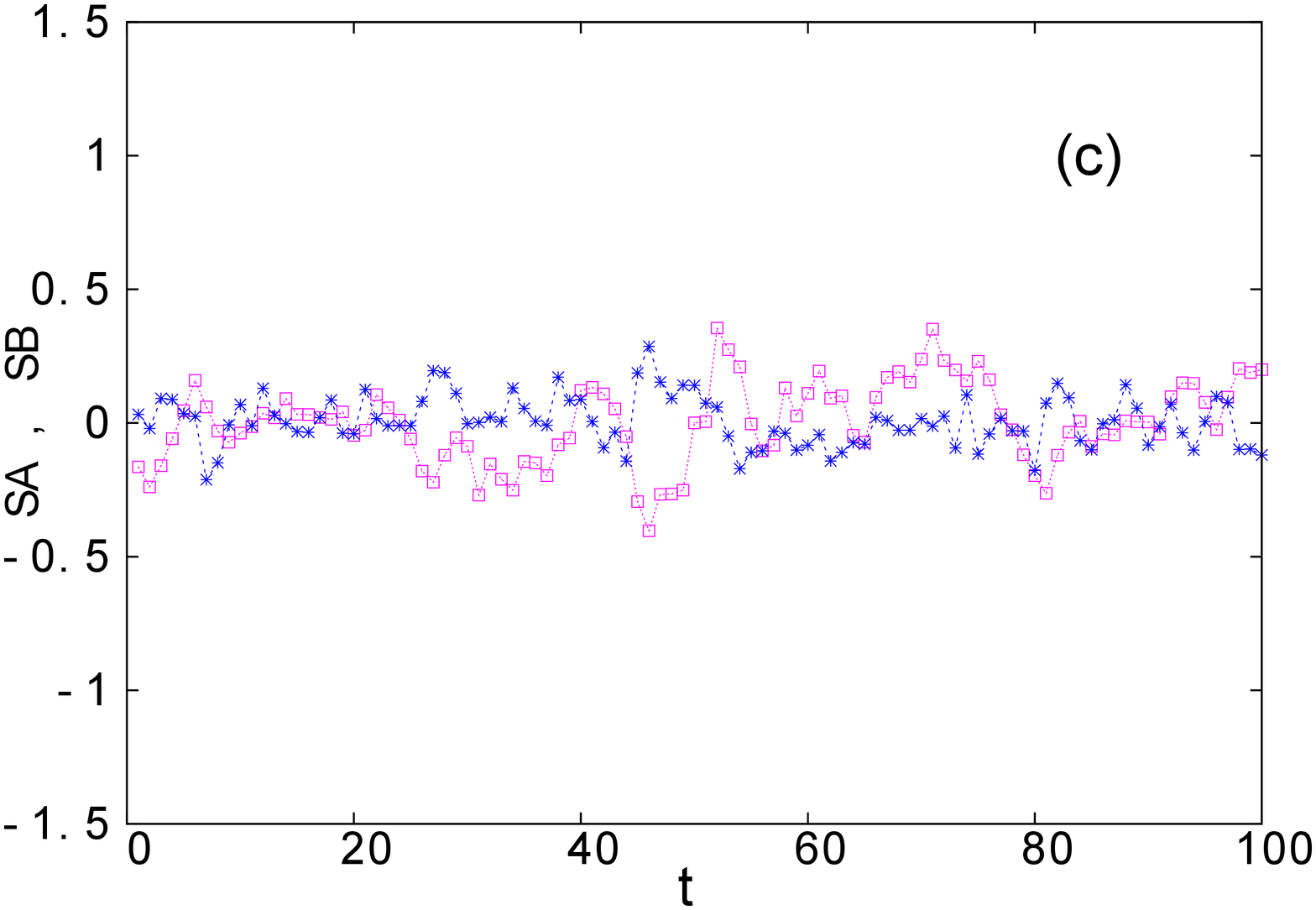}
\includegraphics[width=6cm,angle=0]{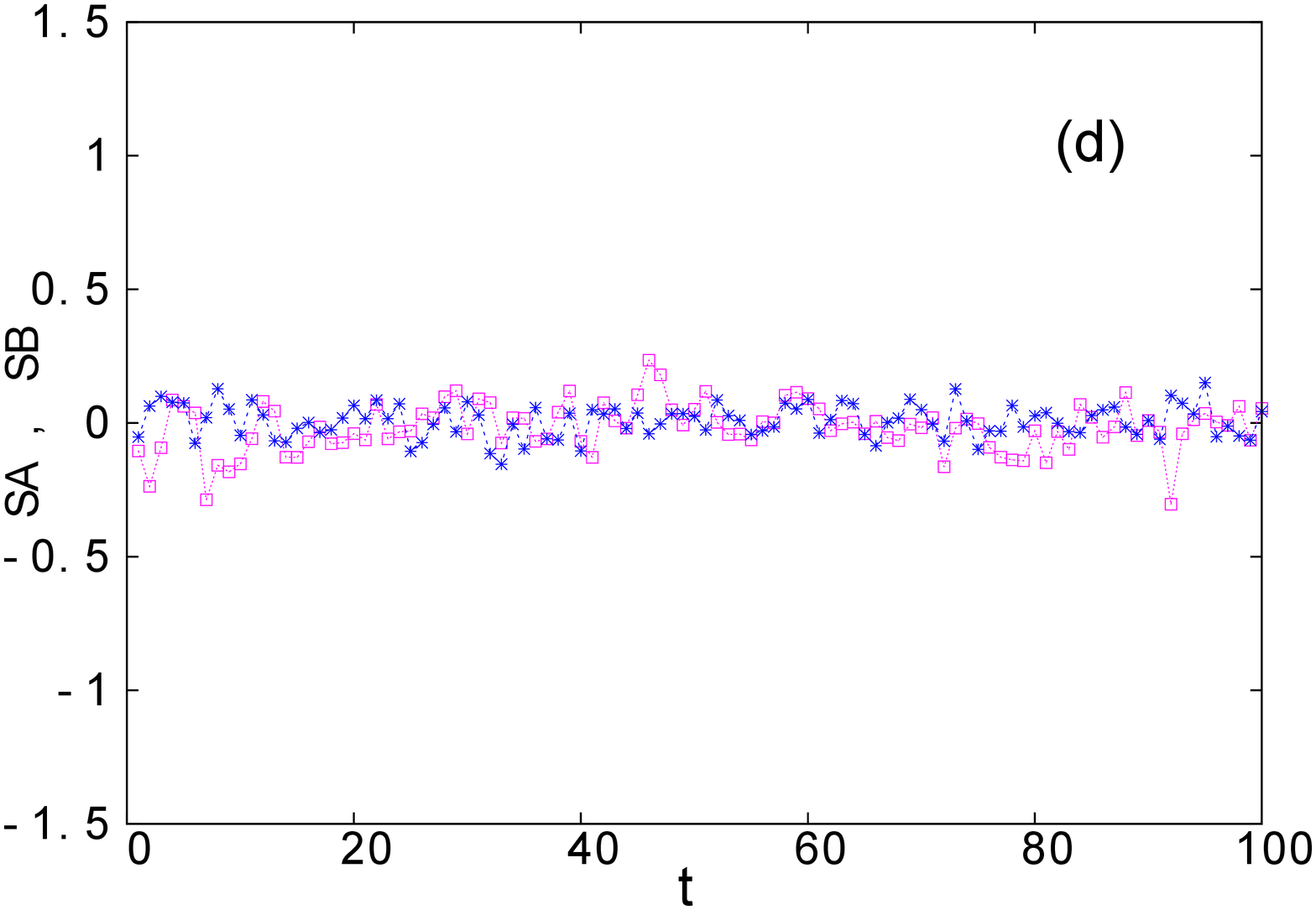}
\caption{Dynamics for the case $K_{AB}=-K_{BA}=0.005$ with $J_A=0.02$, $J_B=0.01$, $q_A=7$, $q_B=7$,  and initial conditions $S_A=-S_B=3$ (as in Fig. \ref{ffig17}): in this case $B$ does not change its sign and one has $T_c\simeq 70$ (not shown):
(a)-(b)-(c)-(d) show $S_A$ and $S_B$ vs time $t$ at $T=47$ (below $T_c$), 81, 115 and 166, respectively.
 See text for comments. \label{ffig19}}
\end{figure}

If we reverse the signs of $K$, namely $K_{AB}=-K_{BA}=0.005$, we do not obtain the change of sign of $B$ as
seen in Figs. \ref{ffig17} and \ref{ffig18}: this is shown in Fig. \ref{ffig19} where we also observe slow and non
periodic oscillations with time.

\section{Summary, comparison of Monte Carlo and mean-field results, discussion}\label{MCMF}

We have seen in MC simulations that qualitatively different dynamic behaviors of $S_A$ and $S_B$ occur in time for different regions of the parameter space. We summarize these findings and compare them to the mean-field results:
\begin{itemize}
\item When two groups interact with each other, whatever their respective critical temperatures $T_c^0(A)$ and $T_c^0(B)$ before the interaction, they have the same critical temperature $T_c$ upon interaction. If $T_c^0(A)= T_c^0(B)$ (similar systems with $J_A=J_B$, $q_A=q_B$ and $M_A=M_B$), $T_c$ lies below  $T_c^0(A,B)$ (Fig. \ref{ffig9}b, Fig. \ref{ffig14}a). If  $T_c^0(A)\neq  T_c^0(B)$, $T_c$ is lower than both (Fig. \ref{ffig11}b).  The transition at $T_c$ is of first order (see discontinuities in Fig. \ref{ffig9}b, Fig. \ref{ffig11}b, Fig. \ref{ffig14}a).

\item Periodic oscillations occur inside a temperature zone between the $T_c$ and the smaller of $T_c^0(A)$ and $T_c^0(B)$ (Fig. \ref{ffig10}b, Fig. \ref{ffig12}b, Fig. \ref{ffig14}b). For higher temperatures, $S_A$ and $S_B$ lose progressively the periodic character of their oscillations (Fig. \ref{ffig12}c, Fig. \ref{ffig14}c, Fig. \ref{ffig15}d). For $T$ higher than the larger of $T_c^0(A)$ and  $T_c^0(B)$, $S_A$ and $S_B$ exhibit small "chaotic" fluctuations around 0 (Fig. \ref{ffig10}c, Fig. \ref{ffig12}d, Fig. \ref{ffig14}d).
\item When $J_A>J_B$ or vice-versa, depending on the signs of $K_{AB}$ and $K_{BA}$, $S_A$ and $S_B$ keep or change their attitude at $T<T_c$. For higher temperatures, oscillations of attitude are very slow but they become chaotic with increasing $T$  (Fig. \ref{ffig18}, Fig. \ref{ffig19}).

\item In the absence of intra-group interactions, namely $J_A=J_B=0$, there is no order in either group. The dynamics of $S_A$ and $S_B$ depend on the signs and the amplitudes of $K_{AB}$ and $K_{BA}$:  if $K_{AB}$ and $K_{BA}$ are small, oscillations of $S_A$ and $S_B$ are regular but they do not have constant amplitudes (Fig. \ref{ffig16}). If $K_{AB}$ and $K_{BA}$ are large ($\simeq 0.2-0.3$ for example) oscillations are slow at low $T$ and chaotic at high $T$.
\end{itemize}

To compare the results from MC simulations to the results obtained by mean-field theory we recall that the equations of Section II include the temperature $T$ in the definition of the interactions. Hence the mean-field parameters are related to the MC parameters by $J_1=J_A/T$, $J_2=J_B/T$, $K_{12}=K_{AB}/T$ and $K_{21}=K_{BA}/T$. Increasing $J_1$ for example means decreasing $T$ or increasing $J_A$ in MC simulation and vice-versa. The critical value $J_{1c}$ means $J_A/T_c$ in MC parameters.

Mean-field results shown in Fig. \ref{ffig1}-Fig. \ref{ffig8} have the same main characteristics as those of MC simulations: oscillations of $S_A$ and $S_B$ within some ranges of the parameters and convergence toward different values within other ranges. We have interpreted those phenomena in terms of conflicts and willingness to negotiate. The mean-field model is an approximation valid in the limit of an infinitely large number of individuals. It captures the main features found in MC simulations on finite systems.  For the particular model used in this paper where each individual interacts with all others, the equilibrium (with no time dependence) state is exactly characterized by the mean-field theory\cite{Kardar},\cite{Cohen}. The oscillatory regime is also correctly captured in the mean field theory model. The chaotic time variations of  $S_A$ and $S_B$  observed in some MC simulations do not appear in the mean field model which does not exhibit deterministic chaos and ignores thermal fluctuations.
We note that the phenomenon which occurs in each case shown in this paper depends on the relative ratios and signs between values of parameters, but not on their absolute values.
Therefore, the choice of $J_1$ and $J_2$ of the order of 0.02 in MC simulations is to keep $K$ parameters in the same order of magnitude and $T$ in the range around 100.

The dynamics of each group depend on the sign of the total action on each individual: $J$ action from his or her partners in the same group and $K$ action from the other group, both at a previous time. The total action $H_A$ on an individual of group $A$ can be written using the energy given by Eq. (\ref{eqn:hamil0}):
\begin{eqnarray}
E_i^A(t+1)&=& -H_A s_i(t+1) \label{eqn:ha1}\\
\mbox {where}\ \ H_A&=&[J_A\frac{1}{N}\sum_{j\in A } s_j(t) +K_{AB}<S_B>(t)]\label{eqn:ha2}
\end{eqnarray}
$s_i(t+1)$ is such as to render $E_i^A(t+1)$ negative. To clarify the dynamic rule, let us imagine the following scenario. If $H_A>0$ then $s_i(t+1)$ should be $>0$; it means that if $s_i(t)$ is $<0$ at time $t$, individual i should change his/her preference sign at time $t+1$. This dynamic applies to other individuals of Group $A$ so that at the end all change their attitude. We can imagine other scenarios.  The rule is that   $s_i(t+1)$ should follow the sign of $H_A(t)$ which is, as seen from Eq. (\ref{eqn:ha2}), a function of $J_A$, $K_{AB}$ and the signs of $s_j(t)$ of $A$ and of $<S_B>(t)$.  The reader can verify the dynamic rule for each case presented above.  Of course, if $H_A$ and $s_i(t+1)$ have the same sign, then the individual keeps his or her stance although he or she can modify "a little bit" his or her attitude by choosing other "preference states" of the same sign.

Note that the above scenario describes the attitude dynamics at low $T$. However, in a society governed by a "high" temperature, there are always individuals who break the "negative-energy rule" mentioned above. As a consequence, the stance of each group is weakened as $T$ increases, until it breaks down at $T_c$, as shown in the previous sections.

\section{Concluding remarks}\label{Concl}

In this paper we have studied the dynamic behavior of two groups of individuals in conflict. Each group is defined by two parameters which express the intra-group strength of interaction among members and its attitude toward negotiations. The interaction with the other group is parameterized by a constant which expresses an attraction or a repulsion to the other group average attitude. In addition, the model includes a social temperature $T$ which acts on each group and quantifies the social noise.
For a given set of parameters, the results show that the dynamic behavior depends on $T$ as summarized and discussed in section \ref{MCMF}. One of the most striking features is the periodic oscillation of the attitudes towards negotiation or conflict for certain ranges of parameter values.

Clearly not all characteristics of a real two-group conflict can be captured in a model. We view this model to be a tool for anticipation rather than prediction. While prediction has to be right within a usefully narrow range, anticipation helps map the field of possibilities for strategizing purposes. The values/ranges chosen for the various model parameters  are illustrative and can be changed for exploratory purposes or if data show that it is warranted.

As a final remark, we believe that models in statistical physics \cite{Diep2015} are suitable for describing social phenomena thanks to the fact that the spatial average will retain only common macroscopic aspects of an ensemble. Personal atypical characters will be erased in the averaging. This is the reason why rules for particles may be applied to the study of human and animal behaviors. Opinion surveys for example classify in the same category people with similar backgrounds in education or psychologic or political profiles. These shared backgrounds limit the  number of macroscopic categories: it is unlikely that two millions of people have two millions of distinct preferences. The same is observed in physics: particles in the same system have each a different eigenstate, but observables or macroscopic properties  retain only their common characters after the averaging process.

\acknowledgments
MK is grateful to the UCP for a visiting professorship grant and for the hospitality extended to him during his working visit.

{}


\begin{thebibliography}{0}
\expandafter\ifx\csname natexlab\endcsname\relax\def\natexlab#1{#1}\fi
\expandafter\ifx\csname bibnamefont\endcsname\relax
  \def\bibnamefont#1{#1}\fi
\expandafter\ifx\csname bibfnamefont\endcsname\relax
  \def\bibfnamefont#1{#1}\fi
\expandafter\ifx\csname citenamefont\endcsname\relax
  \def\citenamefont#1{#1}\fi
\expandafter\ifx\csname url\endcsname\relax
  \def\url#1{\texttt{#1}}\fi
\expandafter\ifx\csname urlprefix\endcsname\relax\def\urlprefix{URL }\fi
\providecommand{\bibinfo}[2]{#2}
\providecommand{\eprint}[2][]{\url{#2}}

\end{thebibliography}


\begin{thebibliography}{9}
\bibitem{Coser}Coser L. A., {\it The functions of social conflict} (Vol. 9). Routledge  (1956).
\bibitem{Druckman} Druckman D., {\it Human Factors in International Negotiations: A Survey of Research on Social-psychological Aspects of International Conflict}  (1971).
\bibitem{Pruitt} Pruitt D. G., {\it Social conflict}, McGraw-Hill, New York  (1998).
\bibitem{Schelling} Schelling T. C.,{\it The strategy of confict}, Cambridge, Mass. (1960).
\bibitem{Simons} Simons H. W., Persuasion in social conflicts: A critique of prevailing conceptions and a framework for future research, Communications Monographs {\bf 39}(4), 227-247  (1972).
\bibitem{Kaufman-Kaufman2013}Kaufman S. and Kaufman M., Tipping points in the dynamics of peace and war, in: A. Colson, D. Druckman and W. Donohue (Eds.), International Negotiation: Foundations, Models and Philosophies (pp. 251-272). Dordrecht: RoL  (2013).
\bibitem{Kaufman-Kaufman2015} Kaufman S. and Kaufman M., Two-group dynamic conflict scenarios: "Toy model" with a severity index 2015. Negotiation and Conflict Management Research {\bf 8}(1), 41–55 (2015).
\bibitem{Galam}Galam S., {\it Sociophysics: A physicist's modeling of psycho-political phenomena}, Springer: Complexity, Berlin (2012).
\bibitem{Diep2014}Diep H. T., Ordre, d\'esordre dans des syst\`emes frustr\'es: Complexit\'e aux fronti\`eres des phases, hal-01090082 (2014).
\bibitem{Galam2016}Galam S., The Trump phenomenon an explanation from sociophysics, arXiv:1609.03933v1 (2016).
\bibitem{Gao} Gao J., Buldyrev S. V., Stanley E. H., Havlin S., Network formed from interdependent networks, Nature Physics, {\bf 8}, 40-48 (2012).
\bibitem{GalamGabay} Galam S. and Gabay M., Coupled spin systems and plastic crystals, Europhys. Lett. {\bf 8}, 167-171 (1989).
\bibitem{Alvarez} Alvarez-Zuzek L. G., La Rocca C. E., Braunstein L. A., and Vazquez F. Competing dynamical processes on two interacting networks, arXiv:1604.07444v1 (2016).
\bibitem{Kaufman1989} Kaufman M., Equilibrium Polymerization on the Equivalent-Neighbor Lattice, Phys. Rev. B{\bf 39}(10), 6898-6906 (1989).
\bibitem{Cohen}Cohen R., Dawid D. J., Kardar M., Bar-Yam Y., Unusual Percolation in Simple Small-World Networks, Phys. Rev. E{\bf79}, 066112 (2009).
\bibitem{Fernandez}Fernandez-Rosales I. Y, Liebovitch L. S., Guzman-Vargas L., The Dynamic Consequences of Cooperation and Competition in Small-World Networks, PLOS One 10(4):e0126234 (2015).
\bibitem{Binder} Binder K. and Heermann D. W., {\it Monte Carlo Simulation in Statistical Physics}, Springer-Verlag, Berlin  (1992).
\bibitem{Diep2015}Diep H. T., {\it Statistical physics : Fundamentals and application to condensed mater}, World Scientific, Singapore  (2015).
\bibitem{Kardar}Kardar M., Phase transitions in new solvable Hamiltonians by a Hamiltonian minimization, Phys. Rev. Lett. {\bf 51}, 523 (1983).







\end{thebibliography}
\end{document}